\documentclass[superscriptaddress,twocolumn,amsmath,amssymb,prd,floatfix,nofootinbib]{revtex4-2}

\usepackage[dvipdfmx]{graphicx}
\usepackage{dcolumn}
\usepackage[varg]{txfonts}
\usepackage{bm}
\usepackage{empheq}%
\usepackage{multirow}

\usepackage{color}

\newcommand{\sY}[2]{{}_{#1}\hspace*{-0.1ex}Y_{#2}}

\begin{document}

\title{Integrated perturbation theory for cosmological tensor
  fields. IV. Full-sky formulation
}

\author{Takahiko Matsubara} \email{tmats@post.kek.jp}
\affiliation{%
  Institute of Particle and Nuclear Studies, High Energy
  Accelerator Research Organization (KEK), Oho 1-1, Tsukuba 305-0801,
  Japan}%
\affiliation{%
  The Graduate Institute for Advanced Studies, SOKENDAI,
  Tsukuba 305-0801, Japan}%

\date{\today}

\begin{abstract}
  In Papers~I--III \cite{paperI,paperII,paperIII}, we use the flat-sky
  and distant-observer approximations to develop a formalism with
  which the correlation statistics of cosmological tensor fields are
  calculated by the nonlinear perturbation theory, generalizing the
  integrated perturbation theory for scalar fields. In this work, the
  formalism is extended to include the full-sky and wide-angle effects
  in evaluating the power spectra and correlation functions of
  cosmological tensor fields of any rank. With the newly developed
  formalism, one can evaluate the nonlinear power spectra and
  correlation functions to arbitrary higher orders in principle. After
  describing the general formalism, we explicitly derive and give
  analytic results of the lowest-order linear theory for an
  illustrative purpose in this paper. The derived linear formulas with
  full-sky and wide-angle effects are numerically compared with the
  previous formulas with flat-sky and distant-observer limits in a
  simple model of tensor bias.
\end{abstract}


\maketitle


\section{\label{sec:Introduction}%
  Introduction
}

The large-scale structure of the Universe is one of the most
invaluable sources of information about the nature of our Universe
itself. The spatial distributions of galaxies have been playing
important roles in unveiling the structure of the Universe, where
three-dimensional positions of galaxies are measured by various kinds
of modern galaxy surveys (e.g.,
Refs.~\cite{SDSS:2000hjo,2DFGRS:2001zay,BOSS:2012dmf,DES:2016jjg,LSSTScience:2009jmu,DESI:2016fyo,EUCLID:2011zbd}).
In order that maximal information can be extracted from such modern
surveys of large-scale structure, we can use not only the positions of
objects, but also various properties of objects, such as angular
momentum
\cite{Peebles:1969jm,White:1984uf,Catelan:1996hv,Catelan:1996hw} and
shape moments
\cite{Catelan:2000vm,Okumura:2008du,Joachimi:2015mma,Kogai:2018nse},
etc. The correlation statistics of the intrinsic alignment of galaxies,
characterized by second moments of shapes, have been extensively
studied in recent years, and they provide complementary information of
cosmology (for review, see, e.g., Ref.~\cite{Lamman:2023hsj} and
references therein).

Theoretical modeling of the large-scale structure of the Universe has
a long history. Among others, the perturbation theory of density field
plays an important role in analyzing the correlation statistics on
large scales, where the nonlinearity in gravitational evolutions of
density field is considered to be weak enough
\cite{Peebles1980,Bernardeau:2001qr}. As galaxy surveys become larger
and larger, methods using perturbation theory are expected to become
increasingly useful. Recently, methods of perturbation theory have
been actively applied to the correlation statistics of intrinsic
alignment by many authors. The intrinsic alignment is usually
characterized by the second moments of galaxy shapes
\cite{Bartelmann:1999yn}, and the rank-2 tensor is attached to each
galaxy. One can also consider higher moments of galaxy shapes as well
\cite{Kogai:2018nse}. When angular moments of galaxies can be
measured, a rank-1 tensor (i.e., a vector) is attached to each galaxy.

With the above examples in mind, we develop a generalized formalism to
predict correlation statistics of tensor fields in previous
Papers~I--III \cite{paperI,paperII,paperIII} in the series. The
formalism is a generalization of the integrated perturbation theory
(iPT)
\cite{Matsubara:2011ck,Matsubara:2012nc,Matsubara:2013ofa,Matsubara:2016wth},
which is originally formulated to describe the gravitational evolutions of
scalar-type biased fields. We generalize the theory so that
tensor-type biased fields can also be described by nonlinear
perturbation theory, using irreducible representations of tensor
fields throughout. In Paper~I, the basic formulations of the iPT for
tensor fields are described and some analytic results are derived in
lowest-order approximations. In Paper~II, methods of calculating
higher-order corrections of nonlinear evolutions are developed in the
formalism. In Paper~III, the previous formalism is extended to
predict statistics of projected tensors, because observed tensors
usually correspond to two-dimensional tensors projected onto the sky
from originally three-dimensional tensors.

In the developments of previous Papers~I--III, we derive most of the
results consistently assuming flat-sky and distant-observer
approximations. In this Paper~IV, we consider the corresponding
problem without assuming the last approximations, including full-sky
and wide-angle effects in power spectra and correlation functions for
tensor fields in the large-scale structure of the Universe.
Generalizing the method of rotationally invariant functions developed
in previous Papers~I--III, we derive various formulas which are
generally useful for full-sky analyses of cosmological tensor fields,
which is independent of the dynamics using the perturbation theory. 

The full-sky and wide-angle effects in power spectra and correlation
functions of density fields with redshift-space distortions have been
extensively investigated mostly using the linear perturbation theory
(for pioneering work, see
Refs.~\cite{Heavens:1994iq,Hamilton:1995px,Zaroubi:1996qt}). In this
paper, we develop a systematic way of evaluating the full-sky and
wide-angle effects in correlation statistics of tensor fields.
Applying the iPT, we explicitly derive analytic expressions in linear
theory, and extensions of the results to include nonlinear corrections
using the higher-order perturbation theory are straightforward by
using the present formalism. The expressions of the correlation
functions are natural extensions of the geometric method of evaluating
the linear correlation function in redshift space
\cite{Szalay:1997cc,Matsubara:1999du,Szapudi:2004gh,Papai:2008bd,Benabou:2024tmn},
which is recently generalized also to derive linear correlation
functions of galaxy ellipticity \cite{Shiraishi:2020vvj}. The
formalism of this paper generalizes the previous studies, and provides
a way to evaluate nonlinear correlation functions of tensor fields
with arbitrary ranks in redshift space by the iPT, and also to
evaluate full-sky effects in the power spectrum. For illustrative
purposes in particular, the numerical results of the power spectra and
correlation functions in linear theory are demonstrated in
Figs.~\ref{fig:2}--\ref{fig:4} of Sec.~\ref{subsec:LinearNumerical} in
this paper.

This paper is organized as follows. In Sec.~\ref{sec:TensorFullSky},
basic ingredients and notations are introduced to describe the
full-sky and wide-angle effects of the power spectrum and correlation
function of tensor fields, and rotationally invariant functions which
play central roles in this paper are introduced. Relations between the
full-sky, wide-angle formalism and the flat-sky, distant-observer
formalism are also derived in this section. In
Sec.~\ref{sec:iPTFullSky}, we explain how one can apply the iPT to
calculate the invariant functions in the general formalism, and
explicit results of applying the linear theory of gravitational
evolution are analytically derived. In Sec.~\ref{sec:ProjectedTensor},
we consider the tensor fields projected onto the sky, and
corresponding formulas for E/B power spectra and $+/\times$
correlation functions with full-sky and wide-angle effects are
derived. Analytic formulas of the linear theory are explicitly
derived, and they are numerically compared with the corresponding
formulas of the previous results in the flat-sky and distant-observer
limits. The conclusions are summarized in Sec.~\ref{sec:Conclusions}.
In Appendix~\ref{app:LargeScaleLimit}, the formulas of large-scale
limits of the power spectra and correlation functions are explicitly
given, where the scale dependencies of the bias functions of tensor
fields are neglected.

\section{\label{sec:TensorFullSky}%
  Irreducible tensor fields in full sky
}

\subsection{%
  Spherical bases in Cartesian and spherical coordinates
}

In this section, we first review the concept and conventions of
spherical basis in Cartesian coordinates introduced in Paper~I, and
then introduce the corresponding basis in spherical coordinates to
describe the tensors in the full-sky formulation of this paper. As
explained in Paper~I, any symmetric tensor of rank-$l$,
$T_{i_1i_2\cdots i_l}$, can be decomposed into traceless parts,
$T^{(s)}_{i_i\cdots i_s}$, where
$s = l, l-2, \ldots, (0 \mathrm{\ or\ } 1)$. A traceless part of the
tensor components forms an irreducible representation of the rotation
group SO(3). In Paper~I, a traceless tensor of rank-$s$ is decomposed
into the spherical basis as
\begin{equation}
  T^{(s)}_{i_1i_2\cdots i_s} =
  A_s
  \sum_{\sigma=-s}^s T_{s\sigma}
  \mathsf{Y}^{(\sigma)}_{i_1i_2\cdots i_s},
  \label{eq:1}
\end{equation}
where 
\begin{equation}
  A_s \equiv \sqrt{\frac{s!}{(2s-1)!!}}
  \label{eq:2}
\end{equation}
is a normalization factor of our convention, and $T_{s\sigma}$ is the
irreducible representation of the tensor, and
$\mathsf{Y}^{(\sigma)}_{i_1i_2\cdots i_s}$ with
$\sigma=0,\pm 1, \pm 2, \cdots, \pm s$ is the spherical basis of the
Cartesian coordinates.\footnote{In Paper~I, the indices $s$ and
  $\sigma$ are denoted by $l$ and $m$ for the irreducible tensors. In
  this paper, the letters like $s$ and $\sigma$ are used instead for
  indices of irreducible tensors. The letters like $l$, $m$ are saved
  to indices of multipoles of the angular positions $(\theta,\phi)$ of
  the field.} The spherical basis of the rank-$s$ tensor is
constructed from a set of vector bases in Cartesian coordinates,
\begin{equation}
  \mathbf{e}^0 = \hat{\mathbf{e}}_3, \quad
  \mathbf{e}^\pm =
  \mp \frac{\hat{\mathbf{e}}_1 \mp i \hat{\mathbf{e}}_2}{\sqrt{2}},
  \label{eq:3}
\end{equation}
where $\hat{\mathbf{e}}_i$ with $i=1,2,3$ are the unit base vectors of
the Cartesian coordinates $(x_1,x_2,x_3)$. For example, the spherical
bases of the rank-$0,1$ tensors are, respectively, given by
\begin{equation}
  \mathsf{Y}^{(0)} = 1, \quad
  \mathsf{Y}^{(\sigma)}_i = {\mathrm{e}^\sigma}_i,
  \label{eq:4}
\end{equation}
and those of the rank-$2$ tensors are given by
\begin{align}
  &
    \mathsf{Y}^{(0)}_{ij} = \sqrt{\frac{3}{2}}
    \left(
    {\mathrm{e}^0}_i\,{\mathrm{e}^0}_j - \frac{1}{3}\delta_{ij}
    \right),
  \label{eq:5}\\
  &
    \mathsf{Y}^{(\pm 1)}_{ij} = \sqrt{2}\,
    {\mathrm{e}^0}_{(i}\,{\mathrm{e}^\pm}_{j)},
  \label{eq:6}\\
  &
    \mathsf{Y}^{(\pm 2)}_{ij} =
    {\mathrm{e}^\pm}_i\,{\mathrm{e}^\pm}_j,
  \label{eq:7}
\end{align}
where ${\mathrm{e}^\sigma}_i = [\mathbf{e}^\sigma]_i =
\mathbf{e}^\sigma\cdot\hat{\mathbf{e}}_i$ with
$\sigma=0,\pm$ are Cartesian components of the spherical basis, and
round brackets in the indices of the right-hand side (rhs) of
Eq.~(\ref{eq:6}) indicate symmetrization with respect to the indices
inside the brackets, e.g., $a_{(i}b_{j)} = (a_ib_j + a_jb_i)/2$.

The spherical bases for rank-3 and -4 tensors are explicitly given in
Appendix~A of Paper~I, and how to systematically construct spherical
bases for arbitrary higher-order tensors is also explained there. A
relation of the spherical basis to the spherical harmonics is quite
useful. For an arbitrary unit vector
$\bm{n} = (\sin\theta\cos\phi, \sin\theta\sin\phi,\cos\theta)$, the
contraction of the indices in the spherical basis with those in unit
vectors is given by
\begin{equation}
  \mathsf{Y}^{(\sigma)}_{i_1i_2\cdots i_s}
  n_{i_1} n_{i_2}\cdots n_{i_s} =
  A_s\sqrt{\frac{4\pi}{2s+1}}\,
  Y_{s\sigma}^*(\theta,\phi),
  \label{eq:8}
\end{equation}
where $Y_{lm}^*(\theta,\phi)$ (with $l=s$ and $m=\sigma$) is the
complex conjugate of the spherical harmonics\footnote{The
  Condon-Shortley phase is included in our definition of the spherical
  harmonics $Y_{lm}(\theta,\phi)$ (see also Paper~I).},
$Y_{lm}(\theta,\phi)$. Using the above equation, explicit forms of
higher-rank spherical bases $\mathsf{Y}^{(\sigma)}_{i_1i_2\cdots i_s}$
can be uniquely determined (Appendix~A of Paper~I).

The spherical basis $\mathsf{Y}^{(\sigma)}_{i_1\cdots i_s}$ satisfies an
orthonormality relation,
\begin{equation}
  \mathsf{Y}^{(\sigma)*}_{i_1\cdots i_s}
  \mathsf{Y}^{(\sigma')}_{i_1\cdots i_s} = \delta_\sigma^{\sigma'},
  \label{eq:9}
\end{equation}
where $\delta_\sigma^{\sigma'}$ is the Kronecker's delta symbol, and
the indices $i_1,\ldots,i_s$ are summed over with omitting the
summation symbol $\sum_{i_1,\ldots,i_l}$, as we adopt the Einstein
summation convention for repeated indices of Cartesian coordinates.
From the construction of the spherical basis in Paper~I, the complex
conjugate of the basis is given by
\begin{equation}
  \mathsf{Y}^{(\sigma)*}_{i_1\cdots i_s} =
  (-1)^\sigma \mathsf{Y}^{(-\sigma)}_{i_1\cdots i_s} =
  \sum_{\sigma'}
  g^{(s)}_{\sigma\sigma'} \mathsf{Y}^{(\sigma')}_{i_1\cdots i_s},
  \label{eq:10}
\end{equation}
where, in the last equation, the spherical metric is defined by
\begin{equation}
  g^{(s)}_{\sigma\sigma'} \equiv (-1)^\sigma \delta_{\sigma,-\sigma'},
  \label{eq:11}
\end{equation}
for $\sigma,\sigma'=0,\pm 1,\ldots,\pm s$. The spherical metric of
Eq.~(\ref{eq:11}) is considered to lower the spherical index $\sigma$
of the spherical basis. The inverse metric $g_{(s)}^{\sigma\sigma'}$
which satisfies
$\sum_{\sigma''} g_{(s)}^{\sigma\sigma''}g^{(s)}_{\sigma''\sigma'} =
\delta_{\sigma'}^\sigma$ has the same matrix elements as the
rhs of Eq.~(\ref{eq:11}), i.e.,
$g_{(s)}^{\sigma\sigma'} = g^{(s)}_{\sigma\sigma'}$. With the above
notations, The orthonormality relation of Eq.~(\ref{eq:9}) is
equivalently represented by
\begin{equation}
  \mathsf{Y}^{(\sigma)}_{i_1\cdots i_s}
  \mathsf{Y}^{(\sigma')}_{i_1\cdots i_s}
  = g_{(s)}^{\sigma\sigma'}, \quad
  \mathsf{Y}^{(\sigma)*}_{i_1\cdots i_s}
  \mathsf{Y}^{(\sigma')*}_{i_1\cdots i_s}
  = g^{(s)}_{\sigma\sigma'}.
  \label{eq:12}
\end{equation}
The decomposition of Eq.~(\ref{eq:1}) for the traceless tensor is
unique, because the relation is inverted by the orthonormality
relation of Eq.~(\ref{eq:9}), and we have
\begin{equation}
  T_{s\sigma} =
  \frac{1}{A_s}
  T^{(s)}_{i_1i_2\cdots i_s}
  \mathsf{Y}^{(\sigma)*}_{i_1i_2\cdots i_s}.
  \label{eq:13}
\end{equation}

On one hand, the spherical basis defined above is suitable for working
in Cartesian coordinates system. In the context of observational
cosmology, the statistics of the cosmological field in the
distant-observer approximation are described in Cartesian coordinate
systems, where the lines of sight are approximately considered to have
a fixed direction throughout the space. On the other hand, when the
distant-observer approximation is not sufficiently valid and
wide-angle effects are not negligible, it is natural to work in a
spherical coordinates system in which the coordinates' origin is
located at the position of the observer. The direction of the line of
sight corresponds to the radial direction of the spherical
coordinates, and varies with angular positions.

As explained in Paper~III, the spherical basis of a tensor in a
spherical coordinates system can be constructed by a set of vector
basis,
\begin{equation}
  \tilde{\mathbf{e}}^0 = \hat{\mathbf{e}}_r, \quad
  \tilde{\mathbf{e}}^\pm =
  \mp \frac{\hat{\mathbf{e}}_\theta \mp i \hat{\mathbf{e}}_\phi}{\sqrt{2}},
  \label{eq:14}
\end{equation}
where
$\hat{\mathbf{e}}_r, \hat{\mathbf{e}}_\theta, \hat{\mathbf{e}}_\phi$
are the unit base vectors of the spherical coordinates
$(r,\theta,\phi)$. Contrary to the Cartesian case where the base
vectors have fixed directions throughout the space, the above base
vectors do not have fixed directions and depend on the angular
coordinates $(\theta,\phi)$ of each position. The spherical basis
$\tilde{\mathsf{Y}}^{(\sigma)}_{i_1i_2\cdots i_s}(\theta,\phi)$ of
rank-$s$ tensors in spherical coordinates is defined in the same way
for that in the Cartesian coordinates as given in
Eqs.~(\ref{eq:4})--(\ref{eq:7}), replacing the basis vectors
$\mathbf{e}^\sigma$ with $\tilde{\mathbf{e}}^\sigma$.

Defining a $3\times 3$ orthogonal matrix
\begin{equation}
  S(\theta,\phi) =
  \begin{pmatrix}
    \cos\theta\,\cos\phi & \cos\theta\,\sin\phi & -\sin\theta \\
    -\sin\phi & \cos\phi & 0 \\
    \sin\theta\,\cos\phi & \sin\theta\,\sin\phi & \cos\theta
  \end{pmatrix},
  \label{eq:15}
\end{equation}
the orthonormal bases in the Cartesian and spherical coordinates
systems are related by
\begin{equation}
  \begin{pmatrix}
    \hat{\mathbf{e}}_\theta(\theta,\phi) \\
    \hat{\mathbf{e}}_\phi(\theta,\phi) \\
    \hat{\mathbf{e}}_r(\theta,\phi)
  \end{pmatrix}
  = S(\theta,\phi)
  \begin{pmatrix}
    \hat{\mathbf{e}}_1 \\ \hat{\mathbf{e}}_2 \\ \hat{\mathbf{e}}_3
  \end{pmatrix}.
  \label{eq:16}
\end{equation}
The components of the spherical vectors, Eqs.~(\ref{eq:3}) and
(\ref{eq:14}) are related by
\begin{equation}
  {\tilde{\mathrm{e}}^\sigma\phantom{}}_j(\theta,\phi) =
  {\mathrm{e}^\sigma}_i\, S_{ij}(\theta,\phi),
  \label{eq:17}
\end{equation}
where
$\tilde{\mathrm{e}}^\sigma_{\phantom{\sigma}i} =
[\tilde{\mathbf{e}}^\sigma]_i =
\tilde{\mathbf{e}}^\sigma\cdot\hat{\mathbf{e}}_i$ with $\sigma=0,\pm$
  are Cartesian components of Eq.~(\ref{eq:14}), and
  $S_{ij}(\theta,\phi)$ are matrix elements of Eq.~(\ref{eq:15}).
  Accordingly, the spherical bases of traceless tensors in Cartesian
  and spherical coordinates are related by
\begin{equation}
  \tilde{\mathsf{Y}}^{(\sigma)}_{j_1\cdots j_s}(\theta,\phi)
  = 
  {\mathsf{Y}}^{(\sigma)}_{i_1\cdots i_s} S_{i_1j_1}(\theta,\phi)
  \cdots S_{i_sj_s}(\theta,\phi).
  \label{eq:18}
\end{equation}
As the matrix $S$ of Eq.~(\ref{eq:15}) is an orthogonal matrix,
$S^{-1} = S^\mathrm{T}$, the orthonormality relations of spherical
basis in Cartesian coordinates, Eqs.~(\ref{eq:9}) and (\ref{eq:12})
are translated to those in spherical coordinates,
\begin{align}
  \tilde{\mathsf{Y}}^{(\sigma)*}_{i_1\cdots i_s}(\theta,\phi)
  \tilde{\mathsf{Y}}^{(\sigma')}_{i_1\cdots i_s}(\theta,\phi)
  &= \delta_\sigma^{\sigma'},
  \label{eq:19}\\
  \tilde{\mathsf{Y}}^{(\sigma)}_{i_1\cdots i_s}(\theta,\phi)
  \tilde{\mathsf{Y}}^{(\sigma')}_{i_1\cdots i_s}(\theta,\phi)
  &= g_{(s)}^{\sigma\sigma'},
    \label{eq:20}\\
  \tilde{\mathsf{Y}}^{(\sigma)*}_{i_1\cdots i_s}(\theta,\phi)
  \tilde{\mathsf{Y}}^{(\sigma')*}_{i_1\cdots i_s}(\theta,\phi)
  &= g^{(s)}_{\sigma\sigma'}.
  \label{eq:21}
\end{align}
Similarly to Eqs.~(\ref{eq:1}) and (\ref{eq:13}), traceless tensors
$T^{(s)}_{i_1i_2\cdots i_s}$ of rank-$s$ are uniquely decomposed into
irreducible tensors $\tilde{T}_{s\sigma}$ in spherical coordinates,
depending on angular positions $\theta,\phi$.

The relation of the basis vectors, Eq.~(\ref{eq:16}), is considered as
a local rotation of the coordinate basis at a position of $\bm{r}$.
Applying the rotational transformation of the spherical basis derived
in a previous paper,\footnote{To derive the following equation, we
  substitute
  $R(\phi,\theta,0) = S^\mathrm{T}(\phi,\theta) = S^{-1}(\phi,\theta)$
  in Eqs.~(21), (24) and (25) of Paper~I.} the following identity is
shown to hold:
\begin{equation}
  \tilde{\mathsf{Y}}^{(\sigma)}_{i_1i_2\cdots i_s}(\theta,\phi)
  = \sum_{\sigma'}
  \mathsf{Y}^{(\sigma')}_{i_1i_2\cdots i_s}
  D^{\sigma'*}_{(s)\sigma}(\phi,\theta,0),
  \label{eq:22}
\end{equation}
where $ D^{m'}_{(l)m}(\alpha,\beta,\gamma)$ is the Wigner's rotation
matrix \cite{Edmonds:1955fi,Khersonskii:1988krb} for the rotation with
Euler's angles $(\alpha,\beta,\gamma)$. The Wigner's matrix is
unitary, and the Wigner's matrix of the inverse rotation is given by
\begin{align}
  D^{m'}_{(l)m}(-\gamma,-\beta,-\alpha)
  &= D^{m*}_{(l)m'}(\alpha,\beta,\gamma)
  \label{eq:23} \\
  &= (-1)^{m-m'} D^{-m}_{(l)-m'}(\alpha,\beta,\gamma).
  \label{eq:24}
\end{align}

\subsection{Irreducible tensor fields in Cartesian coordinates}

We generally consider a tensor field $F_{Xi_1\cdots i_s}(\bm{r})$ as
in previous Papers~I--III, where the label $X$ denotes the class of
objects in a given sample, such as a certain type of galaxy, etc. The
traceless part of the tensor, $F^{(s)}_{Xi_1\cdots i_s}(\bm{r})$, can
be decomposed into spherical basis both in Cartesian and spherical
coordinates as
\begin{align}
  F^{(s)}_{Xi_1i_2\cdots i_s}(\bm{r})
  &=
    A_s   
    \sum_{\sigma} F_{Xs\sigma}(\bm{r})
    \mathsf{Y}^{(\sigma)}_{i_1i_2\cdots i_s}
  \label{eq:25}\\
  &=
    A_s   
   \sum_{\sigma} \tilde{F}_{Xs\sigma}(\bm{r})
  \tilde{\mathsf{Y}}^{(\sigma)}_{i_1i_2\cdots i_s}(\theta,\phi),
  \label{eq:26}
\end{align}
where $F_{Xs\sigma}(\bm{r})$ and $\tilde{F}_{Xs\sigma}(\bm{r})$ are
components of the irreducible tensors in Cartesian and spherical
coordinates, respectively, and $(\theta,\phi)$ are angular coordinates
of the position $\bm{r}$. Using the orthonormality of the spherical
bases, Eqs.~(\ref{eq:9}) and (\ref{eq:19}), the irreducible tensors
are given by
\begin{align}
  F_{Xs\sigma}(\bm{r})
  &=
    \frac{1}{A_s} F^{(s)}_{Xi_1i_2\cdots i_s}(\bm{r})
    \mathsf{Y}^{(\sigma)*}_{i_1i_2\cdots i_s},
    \label{eq:27}\\
  \tilde{F}_{Xs\sigma}(\bm{r})
  &=
    \frac{1}{A_s} F^{(s)}_{Xi_1i_2\cdots i_s}(\bm{r})
    \tilde{\mathsf{Y}}^{(\sigma)*}_{i_1i_2\cdots i_s}(\theta,\phi).
    \label{eq:28}
\end{align}
Combining Eqs.~(\ref{eq:22})--(\ref{eq:28}), the two irreducible
tensors are related by
\begin{equation}
  \tilde{F}_{Xs\sigma}(\bm{r}) =
  \sum_{\sigma'} {F}_{Xs\sigma'}(\bm{r})
  D^{\sigma'}_{(s)\sigma}(\phi,\theta,0).
  \label{eq:29}
\end{equation}

We consider a passive rotation of the Cartesian coordinates system
around the origin,
\begin{equation}
  \hat{\mathbf{e}}_i \xrightarrow{\mathbb{R}}
  \hat{\mathbf{e}}_i' = \hat{\mathbf{e}}_j R_{ji},  
  \label{eq:30}
\end{equation}
where $R_{ji} = \hat{\mathbf{e}}_j \cdot \hat{\mathbf{e}}_i'$ are
components of a real orthogonal matrix which satisfies
$R^\mathrm{T}R = I$ and $I$ is the $3\times 3$ unit matrix. Under the
above rotation, the Cartesian tensor field transforms as
\begin{multline}
  F^{(s)}_{Xi_1i_2\cdots i_s}(\bm{r}) 
  \xrightarrow{\mathbb{R}}
  F^{\prime (s)}_{Xi_1i_2\cdots i_s}(\bm{r}')
  \\
  = F^{(s)}_{Xj_1j_2\cdots j_s}(\bm{r})
  R_{j_1i_1} R_{j_2i_2}\cdots R_{j_si_s},
  \label{eq:31}
\end{multline}
and the spherical tensor in Cartesian coordinates transforms as
(Paper~I), 
\begin{equation}
  F_{Xs\sigma}(\bm{r})
  \xrightarrow{\mathbb{R}}
  F'_{Xs\sigma}(\bm{r}') =
  \sum_{\sigma'}
  F_{Xs\sigma'}(\bm{r}) D^{\sigma'}_{(s)\sigma}(R),
  \label{eq:32}
\end{equation}
where $D^{\sigma'}_{(s)\sigma}(R)$ is the Wigner's rotation matrix
corresponding to the rotation $R$.

Because of the property of the rotational transformation above, it is
natural to expand the angular dependence of $F_{Xs\sigma}(\bm{r})$
by the spherical harmonics as
\begin{equation}
  F_{Xs\sigma}(\bm{r}) =
  \sum_{l=0}^\infty \sum_{m=-l}^{l}
  \sqrt{\frac{4\pi}{2l+1}}\,
  F_{Xs\sigma}^{lm}(r)
  Y_{lm}(\theta,\phi),
  \label{eq:33}
\end{equation}
where the functions $\sqrt{4\pi/(2l+1)}\, Y_{lm}(\theta,\phi)$ are
spherical harmonics of Racah's normalization, which are extensively
used in Papers~I--III and are denoted as $C_{lm}(\theta,\phi)$ there.
At a distance $r=|\bm{r}|$ from the observer, the function
$F^{lm}_{Xs\sigma}(r)$ in the above definition corresponds to the
expansion coefficients with the latter normalization of the spherical
harmonics. From the orthonormality of the spherical harmonics, the
coefficients of the expansion are inversely given by
\begin{equation}
  F_{Xs\sigma}^{lm}(r) =
  \sqrt{\frac{2l+1}{4\pi}}
  \int \sin\theta\,d\theta\,d\phi\,
  F_{Xs\sigma}(\bm{r})
  Y_{lm}^*(\theta,\phi).
  \label{eq:34}
\end{equation}

Under the rotation of the coordinates system around the origin,
the spherical harmonics transform as
\begin{equation}
  Y_{lm}(\theta,\phi) \xrightarrow{\mathbb{R}}
  Y_{lm}(\theta',\phi') =
  \sum_{m'} Y_{lm'}(\theta,\phi) D_{(l)m}^{m'}(R),
  \label{eq:35}
\end{equation}
and thus the expansion coefficients of Eq.~(\ref{eq:33}) transform
as
\begin{equation}
  F_{Xs\sigma}^{lm}(r) \xrightarrow{\mathbb{R}}
  F_{Xs\sigma}^{\prime lm}(r) =
  \sum_{\sigma',m'}
  F_{Xs\sigma'}^{lm'}(r) D^{\sigma'}_{(s)\sigma}(R) D^{m'*}_{(l)m}(R).
  \label{eq:36}
\end{equation}
The complex conjugate of the Wigner rotation matrix satisfies
\begin{equation}
  D^{m'*}_{(l)m}(R) = D^m_{(l)m'}(R^{-1})
  = \sum_{m_1,m_1'} g^{(l)}_{m'm_1'} g_{(l)}^{mm_1} D^{m_1'}_{(l)m_1}(R).
  \label{eq:37}
\end{equation}

Below, we consider other symmetries than that of rotation. Assuming the
original tensor field $F^{(s)}_{Xi_1i_2\cdots i_s}(\bm{r})$ is real,
the complex conjugates of Eq.~(\ref{eq:27}) are given by
\begin{equation}
  F_{Xs\sigma}^*(\bm{r})
  = (-1)^\sigma F_{Xs,-\sigma}(\bm{r})
  = \sum_{\sigma'} g_{(s)}^{\sigma\sigma'} F_{Xs\sigma'}(\bm{r}),
  \label{eq:38}
\end{equation}
and the complex conjugate of the multipole coefficients,
Eq.~(\ref{eq:34}), is given by
\begin{equation}
  F^{lm\,*}_{Xs\sigma}(r) =
    (-1)^{m+\sigma} F^{l,-m}_{Xs,-\sigma}(r) =
    \sum_{m',\sigma'} g^{(l)}_{mm'} g_{(s)}^{\sigma\sigma'}
    F^{lm'}_{Xs\sigma'}(r).
  \label{eq:39}
\end{equation}

The parity transformation of the original tensor field is given by
\begin{equation}
  F^{(s)}_{Xi_1\cdots i_s}(\bm{r}) \xrightarrow{\mathbb{P}}
  F^{\prime (s)}_{Xi_1\cdots i_s}(\bm{r}) =
  (-1)^{p_X+s} F^{(s)}_{Xi_1\cdots i_s}(-\bm{r}),
  \label{eq:40}
\end{equation}
where $p_X=0$ for ordinary tensors and $p_X=1$ for pseudotensors.
Correspondingly, the parity transformation of irreducible tensors,
Eq.~(\ref{eq:27}), is given by
\begin{equation}
  F_{Xs\sigma}(\bm{r}) \xrightarrow{\mathbb{P}}
  F_{Xs\sigma}'(\bm{r}) =
  (-1)^{p_X+s} F_{Xs\sigma}(-\bm{r}),
  \label{eq:41}
\end{equation}
and that of multiple coefficients, Eq.~(\ref{eq:34}), is given by
\begin{equation}
  F^{lm}_{Xs\sigma}(r) \xrightarrow{\mathbb{P}}
  F^{\prime lm}_{Xs\sigma}(r) =
  (-1)^{p_X+s+l} F^{lm}_{Xs\sigma}(r).
  \label{eq:42}
\end{equation}

\subsection{Irreducible tensor fields in spherical coordinates}

The radial base vector $\hat{\mathbf{e}}_r$ in spherical coordinates
at a physical position does not change under the rotation, while the
angular base vectors $(\hat{\mathbf{e}}_\theta,\hat{\mathbf{e}}_\phi)$
rotate depending on the rotation $R$. We denote the last rotation
angle by $\gamma$, which is somehow a complicated function of $R_{ij}$
and $(\theta,\phi)$. Thus the base vectors in spherical coordinates at
a given direction transform as
\begin{equation}
  \begin{pmatrix}
    \hat{\mathbf{e}}_\theta(\theta,\phi) \\
    \hat{\mathbf{e}}_\phi(\theta,\phi) \\
    \hat{\mathbf{e}}_r(\theta,\phi)
  \end{pmatrix}
  \xrightarrow{\mathbb{R}}
  \begin{pmatrix}
    \hat{\mathbf{e}}_\theta'(\theta',\phi') \\
    \hat{\mathbf{e}}_\phi'(\theta',\phi') \\
    \hat{\mathbf{e}}_r'(\theta',\phi')
  \end{pmatrix}
  =
  \begin{pmatrix}
    \cos\gamma & \sin\gamma & 0 \\
    -\sin\gamma & \cos\gamma & 0 \\
    0 & 0 & 1
  \end{pmatrix}
  \begin{pmatrix}
    \hat{\mathbf{e}}_\theta(\theta,\phi) \\
    \hat{\mathbf{e}}_\phi(\theta,\phi) \\
    \hat{\mathbf{e}}_r(\theta,\phi)
  \end{pmatrix},
  \label{eq:43}
\end{equation}
where $(\theta',\phi')$ are transformed angular coordinates in the
rotated coordinates system. In terms of the spherical basis of
Eq.~(\ref{eq:14}), the above transformation is equivalent to
\begin{equation}
  \begin{cases}
  \tilde{\mathbf{e}}^0(\theta,\phi)
  \xrightarrow{\mathbb{R}}
  \tilde{\mathbf{e}}'^0(\theta',\phi')
  = \tilde{\mathbf{e}}^0(\theta,\phi),
  \\
  \tilde{\mathbf{e}}^\pm(\theta,\phi)
  \xrightarrow{\mathbb{R}}
  \tilde{\mathbf{e}}'^\pm(\theta',\phi')
  = e^{\pm i\gamma} \tilde{\mathbf{e}}^\pm(\theta,\phi).
  \label{eq:44}
  \end{cases}
\end{equation}
In terms of the Cartesian components of the base
vectors,
\begin{align}
  {\tilde{\mathrm{e}}^\sigma}_{\phantom{\sigma}i}(\theta,\phi)
  &=
  [\tilde{\mathbf{e}}^\sigma(\theta,\phi)]_i =
  \tilde{\mathbf{e}}^\sigma(\theta,\phi)\cdot \hat{\mathbf{e}}_i,
  \label{eq:45}\\
  {\tilde{\mathrm{e}}'^\sigma}_{\phantom{\sigma}i}(\theta',\phi')
  &=
  [\tilde{\mathbf{e}}'^\sigma(\theta',\phi')]_i =
    \tilde{\mathbf{e}}'^\sigma(\theta',\phi')\cdot \hat{\mathbf{e}}'_i
  \label{eq:46}
\end{align}
with $\sigma=0,\pm$, the transformation of Eq.~(\ref{eq:44}) is given
by
\begin{equation}
  {\tilde{\mathrm{e}}^\sigma}_{\phantom{\sigma}i}(\theta,\phi)
  \xrightarrow{\mathbb{R}}
  {\tilde{\mathrm{e}}'^\sigma}_{\phantom{\sigma}i}(\theta',\phi') =
  e^{i\sigma\gamma}
  {\tilde{\mathrm{e}}^\sigma}_{\phantom{\sigma}j}(\theta,\phi)
  R_{ji}
  \label{eq:47}
\end{equation}
According to the transformation rule of Eq.~(\ref{eq:47}),
the spherical tensor basis in spherical coordinates transforms as
\begin{multline}
  \tilde{\mathsf{Y}}^{(\sigma)}_{i_1i_2\cdots i_s}(\theta,\phi)
  \xrightarrow{\mathbb{R}}
  \tilde{\mathsf{Y}}^{\prime (\sigma)}_{i_1i_2\cdots
    i_s}(\theta',\phi')
  \\
  = e^{i\sigma\gamma}
  \tilde{\mathsf{Y}}^{(\sigma)}_{j_1j_2\cdots j_s}(\theta,\phi)
  R_{j_1i_1}R_{j_2i_2}\cdots R_{j_si_s}. 
  \label{eq:48}
\end{multline}
Combining Eqs.~(\ref{eq:28}), (\ref{eq:31}) and (\ref{eq:48}), the
spherical tensor field in spherical coordinates simply transforms as
\begin{equation}
  \tilde{F}_{Xs\sigma}(\bm{r})
  \xrightarrow{\mathbb{R}}
  \tilde{F}'_{Xs\sigma}(\bm{r}') =
  e^{-i\sigma\gamma} \tilde{F}_{Xs\sigma}(\bm{r}).
  \label{eq:49}
\end{equation}

The spherical tensor field is naturally expanded by the spin-weighted
spherical harmonics $\sY{\sigma}{lm}(\theta,\phi)$
\cite{Newman:1966ub,Goldberg:1966uu} as
\begin{equation}
  \tilde{F}_{Xs\sigma}(\bm{r}) =
  \sum_{l=|\sigma|}^\infty \sum_{m=-l}^l
  \sqrt{\frac{4\pi}{2l+1}}\,
  \tilde{F}^{lm}_{Xs\sigma}(r) \,\sY{\sigma}{lm}(\theta,\phi),
  \label{eq:50}
\end{equation}
where $\tilde{F}^{lm}_{Xs\sigma}(r)$ are the expansion coefficients of
the multipoles of angular dependence with a fixed distance $r$ from
the origin of the coordinates. The spin-weighted spherical harmonics are
represented by the Wigner's rotation matrix as
\begin{equation}
  \sY{\sigma}{lm}(\theta,\phi) =
  (-1)^m \sqrt{\frac{2l+1}{4\pi}}
  D^{-m}_{(l)\sigma}(\phi,\theta,0).
  \label{eq:51}
\end{equation}
From the orthonormality of spin-weighted spherical
harmonics,
\begin{equation}
  \int \sin\theta\,d\theta\,d\phi\,
  \sY{\sigma}{lm}^*(\theta,\phi)
  \sY{\sigma}{l'm'}(\theta,\phi) = \delta_{ll'} \delta^m_{m'},
  \label{eq:52}
\end{equation}
the above expansion is inverted as
\begin{equation}
  \tilde{F}^{lm}_{Xs\sigma}(r) =
  \sqrt{\frac{2l+1}{4\pi}}
  \int \sin\theta\,d\theta\,d\phi\,
  \tilde{F}_{Xs\sigma}(\bm{r})\,\sY{\sigma}{lm}^*(\theta,\phi).
  \label{eq:53}
\end{equation}

Under the passive rotation of Eq.~(\ref{eq:30}), the spin-weighted
spherical harmonics transform as \cite{Boyle:2013nka}
\begin{equation}
  \sY{\sigma}{lm}(\theta,\phi) \xrightarrow{\mathbb{R}}
  \sY{\sigma}{lm}(\theta',\phi') =
  e^{-i\sigma\gamma}
  \sum_{m'} \sY{\sigma}{lm'}(\theta,\phi) D_{(l)m}^{m'}(R),
  \label{eq:54}
\end{equation}
where the angle $\gamma$ is exactly the same as the one in
Eq.~(\ref{eq:43}). Therefore, the phase factor $e^{-i\sigma\gamma}$
cancels under the rotational transformations of Eqs.~(\ref{eq:49}) and
(\ref{eq:54}), and the multipole coefficients simply transform as
\begin{equation}
  \tilde{F}^{lm}_{Xs\sigma}(r)
  \xrightarrow{\mathbb{R}}
  \tilde{F}'^{lm}_{Xs\sigma}(r) =
  \sum_{m'}
  \tilde{F}^{lm'}_{Xs\sigma}(r) D^{m'*}_{(l)m}(R),
  \label{eq:55}
\end{equation}
Comparing the above transformation with the one in Cartesian
coordinates, Eq.~(\ref{eq:36}), one notices the absence of the
rotation matrix regarding the index $\sigma$ of irreducible tensor
in spherical coordinates, as the tensor basis rotates together with
the coordinates axis.

Substituting Eq.~(\ref{eq:29}) into Eq.~(\ref{eq:53}), the relation
between the multipoles $F^{lm}_{Xs\sigma}(r)$ and the irreducible
tensor $F_{Xs\sigma}(\bm{r})$ in Cartesian coordinates is obtained.
The spin-weighted spherical harmonics in the integrand of
Eq.~(\ref{eq:53}) is substituted by Eq.~(\ref{eq:51}), and we use a
formula for a product of Wigner's rotation matrices
\cite{Edmonds:1955fi,paperI},
\begin{multline}
  D_{(l_1)m_1}^{m_1'}(R) D_{(l_2)m_2}^{m_2'}(R)
  = (-1)^{l_1+l_2}
  \sum_{l,m,m'} (-1)^l(2l+1)
  \\ \times
  \left(l_1\,l_2\,l\right)_{m_1m_2}^{\phantom{m_1m_2}m}
  \left(l_1\,l_2\,l\right)^{m_1'm_2'}_{\phantom{m_1'm_2'}m'}
  D_{(l)m}^{m'}(R),
  \label{eq:56}
\end{multline}
where the notation introduced in Paper~I for Wigner $3j$-symbols is
employed, i.e., $(l_1\,l_2\,l_3)_{m_1m_2m_3}$ is the usual
$3j$-symbol, and azimuthal indices $m_1, m_2,\ldots$ can be raised by
the spherical metric $g_{(l)}^{mm'}$ (see Paper~I). As a result, we
eventually derive the relation between multipole coefficients in
spherical and Cartesian coordinates:
\begin{multline}
  \tilde{F}^{lm}_{Xs\sigma}(r) =
  (-1)^{\sigma} (2l+1)
  \sum_{l',m',\sigma'}
  \begin{pmatrix}
    s & l' & l \\ \sigma & 0 & -\sigma
  \end{pmatrix}
  \\ \times
  \left(s\,l\,l'\right)^{\sigma'm}_{\phantom{\sigma'm}m'}
  F_{Xs\sigma'}^{l'm'}(r).
  \label{eq:57}
\end{multline}

Below, we consider other symmetries than that of rotation. Assuming the
original tensor field $F^{(s)}_{Xi_1i_2\cdots i_s}(\bm{r})$ is real,
the complex conjugates of Eq.~(\ref{eq:28}) are given by
\begin{equation}
  \tilde{F}_{Xs\sigma}^*(\bm{r})
    = (-1)^\sigma \tilde{F}_{Xs,-\sigma}(\bm{r}) =
    \sum_{\sigma'} g_{(s)}^{\sigma\sigma'}
    \tilde{F}_{Xs\sigma'}(\bm{r}),
  \label{eq:58}
\end{equation}
and the complex conjugate of the multipole coefficients,
Eq.~(\ref{eq:53}), is given by
\begin{equation}
  \tilde{F}^{lm\,*}_{Xs\sigma}(r) =
    (-1)^m \tilde{F}^{l,-m}_{Xs,-\sigma}(r) = \sum_{m'}
    g^{(l)}_{mm'} \tilde{F}^{lm'}_{Xs,-\sigma}(r).
  \label{eq:59}
\end{equation}

The parity transformation of the original tensor field is given by
Eq.~(\ref{eq:40}). Correspondingly, the parity transformation of
irreducible tensors in spherical coordinates, Eq.~(\ref{eq:28}), is
given by
\begin{equation}
  \tilde{F}_{Xs\sigma}(\bm{r}) \xrightarrow{\mathbb{P}}
  \tilde{F}_{Xs\sigma}'(\bm{r}) =
  (-1)^{p_X} \tilde{F}_{Xs,-\sigma}(-\bm{r}),
  \label{eq:60}
\end{equation}
where we use the parity property of the spherical basis,
\begin{equation}
  \tilde{\mathsf{Y}}^{(\sigma)}_{i_1\cdots i_s}(\pi-\theta,\phi+\pi) =
  (-1)^s\,\tilde{\mathsf{Y}}^{(-\sigma)}_{i_1\cdots i_s}(\theta,\phi),
  \label{eq:61}
\end{equation}
which can be shown from the definition. Accordingly, the parity of
multiple coefficients, Eq.~(\ref{eq:53}), is given by
\begin{equation}
 \tilde{F}^{lm}_{Xs\sigma}(r) \xrightarrow{\mathbb{P}}
  \tilde{F}^{\prime lm}_{Xs\sigma}(r) =
  (-1)^{p_X+l} \tilde{F}^{lm}_{Xs,-\sigma}(r),
  \label{eq:62}
\end{equation}
where we use the parity of spin-weighted spherical harmonics,
\begin{equation}
  \sY{\sigma}{lm}(\pi-\theta, \phi+\pi)
  = (-1)^l \sY{-\sigma}{lm}(\theta, \phi).
  \label{eq:63}
\end{equation}

\subsection{%
  The power spectrum
}

The statistics of irreducible tensors can be theoretically predicted
by applying the iPT, as we have shown in previous Papers~I--III in the
limit of flat-sky and distant-observer approximations. In this paper,
we are generalizing the previous results, without assuming the last
approximations. First, we consider statistics of irreducible tensors in
Cartesian coordinates, $F_{Xs\sigma}(\bm{r})$ defined by
Eq.~(\ref{eq:25}), which are naturally treated by the perturbation
theory as we show below. The irreducible tensors in spherical
coordinates, $\tilde{F}_{Xs\sigma}(\bm{r})$ and
$\tilde{F}^{lm}_{Xs\sigma}(r)$ defined by Eqs.~(\ref{eq:26}) and
(\ref{eq:50}), are linearly related to those in Cartesian coordinates
by Eqs.~(\ref{eq:29}) and (\ref{eq:57}), and thus statistics of the
latter variables in spherical coordinates are derived from the former
variables in Cartesian coordinates.

We first define the power spectrum of the multipoles in Cartesian
coordinates by
\begin{equation}
  C^{(s_1s_2)\,l_1l_2;m_1m_2}_{X_1X_2\sigma_1\sigma_2}(r_1,r_2)
  \equiv
  \left\langle
    F^{l_1m_1}_{X_1s_1\sigma_1}(r_1)
    F^{l_2m_2}_{X_2s_2\sigma_2}(r_2)
  \right\rangle.
  \label{eq:64}
\end{equation}
Under the passive rotation of Eq.~(\ref{eq:30}), the multipoles
transform as Eq.~(\ref{eq:36}). Accordingly, the power spectrum above
transforms as
\begin{multline}
  C^{(s_1s_2)\,l_1l_2;m_1m_2}_{X_1X_2\sigma_1\sigma_2}(r_1,r_2)
  \xrightarrow{\mathbb{R}}
  \sum_{\sigma'_1,\sigma'_2,m'_1,m'_2}
  C^{(s_1s_2)\,l_1l_2;m_1'm_2'}_{X_1X_2\sigma'_1\sigma'_2}(r_1,r_2)
  \\ \times
  D^{\sigma'_1}_{(s_1)\sigma_1}(R) D^{\sigma'_2}_{(s_2)\sigma_2}(R)
  D^{m'_1*}_{(l_1)m_1}(R) D^{m'_2*}_{(l_2)m_2}(R).
  \label{eq:65}
\end{multline}

In the statistically isotropic Universe, the correlation function of
the multipoles, Eq.~(\ref{eq:64}), should be rotationally invariant as
the function is given by an ensemble average of the field variables,
and should not depend on any choice of the Cartesian basis
$\hat{\mathbf{e}}_i$. Therefore, the functions are invariant under the
transformation of Eq.~(\ref{eq:65}) for any rotation $R$. Averaging
over the rotation in particular, the functions satisfy an identity,
\begin{multline}
  C^{(s_1s_2)\,l_1l_2;m_1m_2}_{X_1X_2\sigma_1\sigma_2}(r_1,r_2) = 
  \sum_{\sigma'_1,\sigma'_2,m'_1,m'_2}
  C^{(s_1s_2)\,l_1l_2;m_1'm_2'}_{X_1X_2\sigma'_1\sigma'_2}(r_1,r_2)
  \\ \times
  \int \frac{[dR]}{8\pi^2}
  D^{\sigma'_1}_{(s_1)\sigma_1}(R) D^{\sigma'_2}_{(s_2)\sigma_2}(R)
  D^{m'_1*}_{(l_1)m_1}(R) D^{m'_2*}_{(l_2)m_2}(R),
  \label{eq:66}
\end{multline}
where the last integral represents the average over conceivable
rotations $R$, or integrals over the Euler angles of the rotation $R$
divided by the total volume $8\pi^2$ of possible values of the Euler
angles. The products of Wigner's rotation matrices are evaluated by
the method described in Paper~I, and the integral in Eq.~(\ref{eq:66})
reduces to a summation of the products of $3j$-symbols,
\begin{multline}
  \int \frac{[dR]}{8\pi^2}
  D^{\sigma'_1}_{(s_1)\sigma_1}(R) D^{\sigma'_2}_{(s_2)\sigma_2}(R)
  D^{m'_1*}_{(l_1)m_1}(R) D^{m'_2*}_{(l_2)m_2}(R)
  \\
  = (-1)^{s_1+s_2+l_1+l_2} \sum_{l,m,m'} (2l+1)
  \left(s_1\,l_1\,l\right)_{\sigma_1}^{\phantom{\sigma_1}m_1m}
\\ \times  
  \left(s_2\,l_2\,l\right)_{\sigma_2\phantom{m_2}m}^{\phantom{\sigma_2}m_2}
  \left(s_1\,l_1\,l\right)^{\sigma_1'\phantom{m_1'}m'}_{\phantom{\sigma_1'}m_1'}
  \left(s_2\,l_2\,l\right)^{\sigma_2'}_{\phantom{\sigma_2'}m_2'm'}.
  \label{eq:67}
\end{multline}
Substituting the above into Eq.~(\ref{eq:66}), we find the correlation
function is given by
\begin{multline}
  C^{(s_1s_2)\,l_1l_2;m_1m_2}_{X_1X_2\sigma_1\sigma_2}(r_1,r_2) =
  i^{s_1+s_2+l_1+l_2} \sum_{l,m}
  \left(s_1\,l_1\,l\right)_{\sigma_1}^{\phantom{\sigma_1}m_1m}
  \\ \times
  \left(s_2\,l_2\,l\right)_{\sigma_2\phantom{m_2}m}^{\phantom{\sigma_2}m_2}
  C^{s_1s_2;l_1l_2}_{X_1X_2\,l}(r_1,r_2),
  \label{eq:68}
\end{multline}
where
\begin{multline}
  C^{s_1s_2;l_1l_2}_{X_1X_2\,l}(r_1,r_2) \equiv
  i^{s_1+s_2+l_1+l_2}
  (2l+1)
  \sum_{\substack{\sigma_1,\sigma_2\\m_1,m_2,m}}
  \left(s_1\,l_1\,l\right)^{\sigma_1}_{\phantom{\sigma_1}m_1m}
  \\ \times
  \left(s_2\,l_2\,l\right)^{\sigma_2\phantom{m_2}m}_{\phantom{\sigma_2}m_2}
  C^{(s_1s_2)\,l_1l_2;m_1m_2}_{X_1X_2\sigma_1\sigma_2}(r_1,r_2).
  \label{eq:69}
\end{multline}
The last functions $C^{l_1l_2;l}_{X_1X_2;s_1s_2}(r_1,r_2)$ are
invariant under rotations of the coordinates system, and thus
represent physical degrees of freedom of the spectrum, which do not
depend on the choice of the coordinates system. We call the function
the invariant spectrum of the tensor fields in the present paper on
the full-sky analysis. As we show in later sections below, one can
naturally predict the invariant spectrum by the iPT.

The property of complex conjugate, Eq.~(\ref{eq:39}), shows that
\begin{multline}
  C^{(s_1s_2)\,l_1l_2;m_1m_2\,*}_{X_1X_2\sigma_1\sigma_2}(r_1,r_2) =
  \sum_{m_1',m_2',\sigma_1',\sigma_2'}
  g_{(s_1)}^{\sigma_1\sigma_1'} g_{(s_2)}^{\sigma_2\sigma_2'}
  g^{(l_1)}_{m_1m_1'} g^{(l_2)}_{m_2m_2'}
  \\ \times
  C^{(s_1s_2)\,l_1l_2;m_1'm_2'}_{X_1X_2\sigma_1'\sigma_2'}(r_1,r_2).
  \label{eq:70}
\end{multline}
The parity property of Eq.~(\ref{eq:42}) shows that
\begin{multline}
  C^{(s_1s_2)\,l_1l_2;m_1m_2}_{X_1X_2\sigma_1\sigma_2}(r_1,r_2)
  \xrightarrow{\mathbb{P}}
  (-1)^{p_{X_1}+p_{X_2}+s_1+s_2+l_1+l_2}
  \\ \times
  C^{(s_1s_2)\,l_1l_2;m_1m_2}_{X_1X_2\sigma_1\sigma_2}(r_1,r_2).
  \label{eq:71}
\end{multline}
Substituting the above equations into Eq.~(\ref{eq:69}), the invariant
spectrum satisfies the property of complex conjugate,
\begin{equation}
  C^{s_1s_2;l_1l_2\,*}_{X_1X_2\,l}(r_1,r_2) =
  C^{s_1s_2;l_1l_2}_{X_1X_2\,l}(r_1,r_2),
  \label{eq:72}
\end{equation}
i.e., they are real functions, and the parity transformation is given
by
\begin{equation}
  C^{s_1s_2;l_1l_2}_{X_1X_2\,l}(r_1,r_2)
  \xrightarrow{\mathbb{P}}
  (-1)^{p_{X_1}+p_{X_2}+s_1+s_2+l_1+l_2}
  C^{s_1s_2;l_1l_2}_{X_1X_2\,l}(r_1,r_2). 
  \label{eq:73}
\end{equation}

Next, we consider the power spectrum of the irreducible tensor field
$\tilde{F}^{lm}_{Xs\sigma}(r)$ on the basis of spherical coordinates.
Because of the property of rotational transformation of
Eq.~(\ref{eq:55}), and following the similar (and simpler)
considerations as in the above, one finds that the power spectrum of
the fields necessarily have a form,
\begin{equation}
  \left\langle
    \tilde{F}^{l_1m_1}_{X_1s_1\sigma_1}(r_1)
    \tilde{F}^{l_2m_2}_{X_2s_2\sigma_2}(r_2)
  \right\rangle
  = \delta_{l_1l_2}\, g_{(l_1)}^{m_1m_2}\,
  \tilde{C}^{(s_1s_2)\,l_1}_{X_1X_2\sigma_1\sigma_2}(r_1,r_2),
    \label{eq:74}
\end{equation}
and the last factor corresponds to the reduced power spectrum of the
irreducible tensors in the spherical coordinates. Substituting
Eq.~(\ref{eq:57}) into the left-hand side (lhs) of the above equation,
one can confirm the power spectrum has the form of the rhs. Using the
orthogonality relations of the $3j$-symbol and formulas of the
$6j$-symbol quoted in Appendix~C of Paper~I, we find the reduced power
spectrum introduced above is given by
\begin{multline}
  \tilde{C}^{(s_1s_2)\,l}_{X_1X_2\sigma_1\sigma_2}(r_1,r_2)
  =
  i^{s_1+s_2}
  (-1)^{\sigma_1+\sigma_2}
  \sum_{l_1,l_2} i^{l_1+l_2}
  \begin{pmatrix}
    s_1 & l_1 & l \\ \sigma_1 & 0 & -\sigma_1
  \end{pmatrix}
  \\ \times
  \begin{pmatrix}
    s_2 & l_2 & l \\ \sigma_2 & 0 & -\sigma_2
  \end{pmatrix}
  C^{s_1s_2;l_1l_2}_{X_1X_2\,l}(r_1,r_2).
  \label{eq:75}
\end{multline}
Using the orthogonality relations of the $3j$-symbol
\cite{Khersonskii:1988krb}, the above equation can be inverted as
\begin{multline}
  C^{s_1s_2;l_1l_2}_{X_1X_2\,l}(r_1,r_2) =
  (-i)^{s_1+s_2+l_1+l_2}
  (2l_1+1)(2l_2+1)
  \\ \times
  \sum_{\sigma_1,\sigma_2} (-1)^{\sigma_1+\sigma_2}
  \begin{pmatrix}
    s_1 & l_1 & l \\ \sigma_1 & 0 & -\sigma_1
  \end{pmatrix}
  \begin{pmatrix}
    s_2 & l_2 & l \\ \sigma_2 & 0 & -\sigma_2
  \end{pmatrix}
  \\ \times
  \tilde{C}^{(s_1s_2)\,l}_{X_1X_2\sigma_1\sigma_2}(r_1,r_2).
  \label{eq:76}
\end{multline}

The property of complex conjugate, Eq.~(\ref{eq:59}), shows that
\begin{equation}
  \tilde{C}^{(s_1s_2)\,l\,*}_{X_1X_2\sigma_1\sigma_2}(r_1,r_2) =
  \tilde{C}^{(s_1s_2)\,l}_{X_1X_2,-\sigma_1,-\sigma_2}(r_1,r_2),
  \label{eq:77}
\end{equation}
Similarly, the parity property of
Eq.~(\ref{eq:62}) shows that
\begin{equation}
  \tilde{C}^{(s_1s_2)\,l}_{X_1X_2\sigma_1\sigma_2}(r_1,r_2)
  \xrightarrow{\mathbb{P}}
  (-1)^{p_{X_1}+p_{X_2}}
  \tilde{C}^{(s_1s_2)\,l}_{X_1X_2,-\sigma_1,-\sigma_2}(r_1,r_2).
  \label{eq:78}
\end{equation}
Equations~(\ref{eq:77}) and (\ref{eq:78}) are consistent with
Eq.~(\ref{eq:75}) with Eqs.~(\ref{eq:72}) and (\ref{eq:73}).
When the parity symmetry holds and the power spectrum is invariant
under the transformation of Eq.~(\ref{eq:78}), the power spectrum is 
real when $s_1+s_2+p_{X_1}+p_{X_2}=\mathrm{even}$, and pure imaginary
when $s_1+s_2+p_{X_1}+p_{X_2}=\mathrm{odd}$.

\subsection{%
  The correlation function}

\subsubsection{Two-point correlation function in Cartesian coordinates}

The two-point correlation functions of the tensor fields are also
given in terms of the invariant spectrum defined by Eq.~(\ref{eq:69}).
First, we consider the correlation function of the tensor field in
Cartesian coordinates, defined by
\begin{equation}
  \left\langle
    F_{X_1s_1\sigma_1}(\bm{r}_1)
    F_{X_2s_2\sigma_2}(\bm{r}_2)
  \right\rangle =
  \xi^{(s_1s_2)}_{X_1X_2\sigma_1\sigma_2}(\bm{r}_1,\bm{r}_2).
  \label{eq:79}
\end{equation}
Combining Eqs.~(\ref{eq:33}), (\ref{eq:64}) and (\ref{eq:68}),
the above correlation function is given by the invariant spectrum.
A straightforward calculation gives
\begin{multline}
  \xi^{(s_1s_2)}_{X_1X_2\sigma_1\sigma_2}(\bm{r}_1,\bm{r}_2)
  = 4\pi \sum_{\substack{l_1,l_2,l\\m_1,m_2,m}}
  \frac{i^{s_1+s_2+l_1+l_2}}{\sqrt{(2l_1+1)(2l_2+1)}}
  \\ \times
  \left(s_1\,l_1\,l\right)_{\sigma_1}^{\phantom{\sigma_1}m_1m}
  \left(s_2\,l_2\,l\right)_{\sigma_2\phantom{m_2}m}^{\phantom{\sigma_2}m_2}
  Y_{l_1m_1}(\Omega_1) Y_{l_2m_2}(\Omega_2)
  \\ \times
  C^{s_1s_2;l_1l_2}_{X_1X_2\,l}(r_1,r_2),
  \label{eq:80}
\end{multline}
where $\Omega_i=(\theta_i,\phi_i)$ represents the angular coordinates
of $\hat{\bm{r}}_i$ with $i=1,2$. Applying a recoupling formula of
$3j$-symbols using the $6j$-symbol (see Appendix~C of Paper~I in the
present notations), the above equation is equivalent to
\begin{multline}
  \xi^{(s_1s_2)}_{X_1X_2\sigma_1\sigma_2}(\bm{r}_1,\bm{r}_2)
  = i^{s_1-s_2} \sum_{l_1,l_2}
  (-i)^{l_1-l_2}
  \sum_{l,m} (2l+1)
  \\ \times
  \left(s_1\,s_2\,l\right)_{\sigma_1\sigma_2}^{\phantom{\sigma_1\sigma_2}m}
  X^{l_1l_2}_{lm}(\hat{\bm{r}}_1,\hat{\bm{r}}_2)
  \sum_{l'}
  \begin{Bmatrix}
    s_1 & s_2 & l \\ l_2 & l_1 & l'
  \end{Bmatrix}
  \\ \times
  C^{s_1s_2;l_1l_2}_{X_1X_2\,l'}(r_1,r_2),
  \label{eq:81}
\end{multline}
where
\begin{multline}
  X^{l_1l_2}_{lm}(\hat{\bm{r}}_1,\hat{\bm{r}}_2) 
  = \frac{4\pi}{\sqrt{(2l_1+1)(2l_2+1)}}
  \\ \times
  \sum_{m_1,m_2}
  \left(l\,l_1\,l_2\right)_{m}^{\phantom{m}m_1m_2}
  Y_{l_1m_1}(\Omega_1) Y_{l_2m_2}(\Omega_2)
  \label{eq:82}
\end{multline}
are the bipolar spherical harmonics with a normalization introduced in
Paper~I. Using orthogonality relations of the spherical harmonics,
$3j$- and $6j$-symbols, the Eqs.~(\ref{eq:80}) and (\ref{eq:81}) can
be inverted as
\begin{multline}
  C^{s_1s_2;l_1l_2}_{X_1X_2\,l}(r_1,r_2)
  = \frac{i^{s_1+s_2+l_1+l_2}}{4\pi}
    (2l+1)\sqrt{(2l_1+1)(2l_2+1)}
  \\ \times
  \sum_{\sigma_1,\sigma_2,m_1,m_2,m}
  \left(s_1\,l_1\,l\right)^{\sigma_1m_1}_{\phantom{\sigma_1m_1}m}
  \left(s_2\,l_2\,l\right)^{\sigma_2m_2m}
  \\ \times
  \int d\Omega_1 d\Omega_2
  Y_{l_1m_1}(\Omega_1) Y_{l_2m_2}(\Omega_2)
  \xi^{(s_1s_2)}_{X_1X_2\sigma_1\sigma_2}(\bm{r}_1,\bm{r}_2),
  \label{eq:83}
\end{multline}
and
\begin{multline}
  C^{s_1s_2;l_1l_2}_{X_1X_2\,l}(r_1,r_2)
  = i^{s_1-s_2}\, (-i)^{l_1-l_2} (2l+1)(2l_1+1)(2l_2+1)
  \\ \times
  \sum_{l',m',\sigma_1,\sigma_2} (2l'+1)
  \left(s_1\,s_2\,l'\right)^{\sigma_1\sigma_2m'}
  \begin{Bmatrix}
    s_1 & s_2 & l' \\ l_2 & l_1 & l
  \end{Bmatrix}
  \\ \times
  \int \frac{d\Omega_1}{4\pi} \frac{d\Omega_2}{4\pi}
  X^{l_1l_2}_{l'm'}(\hat{\bm{r}}_1,\hat{\bm{r}}_2)
  \xi^{(s_1s_2)}_{X_1X_2\sigma_1\sigma_2}(\bm{r}_1,\bm{r}_2).
  \label{eq:84}
\end{multline}

Substituting Eq.~(\ref{eq:38}) into Eq.~(\ref{eq:79}), the complex
conjugate of the correlation function is given by
\begin{equation}
  \xi^{(s_1s_2)\,*}_{X_1X_2\sigma_1\sigma_2}(\bm{r}_1,\bm{r}_2)
  = (-1)^{\sigma_1+\sigma_2}
  \xi^{(s_1s_2)}_{X_1X_2,-\sigma_1,-\sigma_2}(\bm{r}_1,\bm{r}_2),
  \label{eq:85}
\end{equation}
which is consistent with Eqs.~(\ref{eq:72}) and (\ref{eq:80}).
Similarly, the parity transformation of Eq.~(\ref{eq:41}) shows that
\begin{equation}
  \xi^{(s_1s_2)}_{X_1X_2\sigma_1\sigma_2}(\bm{r}_1,\bm{r}_2)
  \xrightarrow{\mathbb{P}}
  (-1)^{p_{X_1}+p_{X_2}+s_1+s_2}
  \xi^{(s_1s_2)}_{X_1X_2\sigma_1\sigma_2}(-\bm{r}_1,-\bm{r}_2),
  \label{eq:86}
\end{equation}
which is consistent with Eqs.~(\ref{eq:73}) and (\ref{eq:80}).

In the definition of the correlation function, Eq.~(\ref{eq:79}), the
statistical average corresponds to the ensemble average, which cannot
be evaluated for an observer whose position is fixed near the Earth.
In practice, one should average over the sky positions of the
correlation function to obtain statistically meaningful results. In
this regard, one can set a coordinates system for each pair of
astronomical objects to probe the correlation function of tensor
fields. For each pair of objects, 1 and 2, it is always possible to
set the spherical coordinates system $(r,\theta,\phi)$ in which the
first object is located at $\bm{r}_1:(r_1,\theta,0)$ and the second
object is located at $\bm{r}_2:(r_2,0,0)$. Averaging over all the
pairs of objects, one can observationally estimate the correlation
function for these particular values of the argument. We denote the
correlation function with this particular setup by
$\xi^{(s_1s_2)}_{X_1X_2\sigma_1\sigma_2}(r_1,r_2,\theta)$. Just
substituting these particular values, $\Omega_1=(r_1,\theta,0)$ and
$\Omega_2=(r_2,0,0)$, Eq.~(\ref{eq:80}) reduces to
\begin{multline}
  \xi^{(s_1s_2)}_{X_1X_2\sigma_1\sigma_2}(r_1,r_2,\theta)
  = i^{s_1+s_2} (-1)^{\sigma_1}
  \sum_{l_1,l_2,l} i^{l_1+l_2}
  \begin{pmatrix}
    s_1 & l_1 & l \\ \sigma_1 & -\sigma_{12} & \sigma_2
  \end{pmatrix}
  \\ \times
  \begin{pmatrix}
    s_2 & l_2 & l \\ \sigma_2 & 0 & -\sigma_2
  \end{pmatrix}
  \tilde{P}_{l_1}^{\sigma_{12}}(\cos\theta)
  C^{s_1s_2;l_1l_2}_{X_1X_2\,l}(r_1,r_2),
  \label{eq:87}
\end{multline}
where we define normalized Legendre polynomials,
\begin{equation}
  \tilde{P}_l^m(\cos\theta)
  \equiv
  \sqrt{\frac{(l-m)!}{(l+m)!}}\,P_l^m(\cos\theta)
  = \sqrt{\frac{4\pi}{2l+1}}\,Y_l^m(\theta,0).
  \label{eq:88}
\end{equation}

\subsubsection{Two-point correlation function in spherical coordinates}

Next, we consider the correlation function of the irreducible
tensor field $\tilde{F}_{Xs\sigma}(\bm{r})$ on the basis of spherical
coordinates, defined by
\begin{equation}
  \left\langle
    \tilde{F}_{X_1s_1\sigma_1}(\bm{r}_1)
    \tilde{F}_{X_2s_2\sigma_2}(\bm{r}_2)
  \right\rangle =
  \tilde{\xi}^{(s_1s_2)}_{X_1X_2\sigma_1\sigma_2}(\bm{r}_1,\bm{r}_2).
  \label{eq:89}
\end{equation}
Because of the relation of Eq.~(\ref{eq:29}), the correlation function in
spherical coordinates above is related to that in Cartesian
coordinates of Eq.~(\ref{eq:79}) by
\begin{multline}
  \tilde{\xi}^{(s_1s_2)}_{X_1X_2\sigma_1\sigma_2}(\bm{r}_1,\bm{r}_2) =
  \sum_{\sigma_1',\sigma_2'}
  \xi^{(s_1s_2)}_{X_1X_2\sigma_1'\sigma_2'}(\bm{r}_1,\bm{r}_2)
  \\ \times
  D^{\sigma_1'}_{(s_1)\sigma_1}(\phi_1,\theta_1,0)
  D^{\sigma_2'}_{(s_2)\sigma_2}(\phi_2,\theta_2,0).
  \label{eq:90}
\end{multline}
Substituting Eq.~(\ref{eq:80}) into the above equation, using a
formula of Eq.~(\ref{eq:56}) and applying an orthogonality relation of
$3j$-symbols, we derive
\begin{multline}
  \tilde{\xi}^{(s_1s_2)}_{X_1X_2\sigma_1\sigma_2}(\bm{r}_1,\bm{r}_2)
  = i^{s_1+s_2} (-1)^{\sigma_1+\sigma_2}
  \sum_{l_1,l_2,l} i^{l_1+l_2}
  \begin{pmatrix}
    s_1 & l_1 & l \\ \sigma_1 & 0 & -\sigma_1
  \end{pmatrix}
  \\ \times
  \begin{pmatrix}
    s_2 & l_2 & l \\ \sigma_2 & 0 & -\sigma_2
  \end{pmatrix}
  A^{(l)}_{\sigma_1\sigma_2}(\hat{\bm{r}}_1,\hat{\bm{r}}_2)
  C^{s_1s_2;l_1l_2}_{X_1X_2\,l}(r_1,r_2),
  \label{eq:91}
\end{multline}
where
\begin{equation}
  A^{(l)}_{\sigma_1\sigma_2}(\hat{\bm{r}}_1,\hat{\bm{r}}_2)
  \equiv
  \sum_{m_1,m_2} g^{(l)}_{m_1m_2}
  D^{m_1}_{(l)\sigma_1}(\phi_1,\theta_1,0)
  D^{m_2}_{(l)\sigma_2}(\phi_2,\theta_2,0)
  \label{eq:92}
\end{equation}
are functions of the two directions,
$\hat{\bm{r}}_1\equiv \bm{r}_1/r_1$ and
$\hat{\bm{r}}_2\equiv \bm{r}_2/r_2$ with spherical coordinates
$(\theta_1,\phi_1)$ and $(\theta_2,\phi_2)$, respectively. The same
equation can be derived as well by substituting Eqs.~(\ref{eq:50}),
(\ref{eq:74}) and (\ref{eq:75}) into the above Eq.~(\ref{eq:89}).
Because of the properties of Wigner's rotation matrix, the complex
conjugate of the last function is given by
\begin{equation}
  A^{(l)\,*}_{\sigma_1\sigma_2}(\hat{\bm{r}}_1,\hat{\bm{r}}_2) =
  (-1)^{\sigma_1+\sigma_2}
  A^{(l)}_{-\sigma_1,-\sigma_2}(\hat{\bm{r}}_1,\hat{\bm{r}}_2),
  \label{eq:93}
\end{equation}
and the parity property is given by
\begin{equation}
  A^{(l)}_{\sigma_1\sigma_2}(-\hat{\bm{r}}_1,-\hat{\bm{r}}_2) =
  A^{(l)}_{-\sigma_1,-\sigma_2}(\hat{\bm{r}}_1,\hat{\bm{r}}_2).
  \label{eq:94}
\end{equation}
Applying the addition theorem of the Wigner rotation
matrix \cite{Khersonskii:1988krb}, the last functions reduce to an
expression with a single Wigner matrix as
\begin{equation}
  A^{(l)}_{\sigma_1\sigma_2}(\hat{\bm{r}}_1,\hat{\bm{r}}_2)
  = (-1)^{\sigma_2}
    D^{-\sigma_2}_{(l)\;\sigma_1}(\alpha,\beta,\gamma),
 \label{eq:95}
\end{equation}
where $(\alpha,\beta,\gamma)$ are Euler angles of the consecutive
rotations by $(\phi_2,\theta_2,0)$ and $(0,-\theta_1,-\phi_1)$ in this
order. Explicitly, the Euler angles are related to the angular
coordinates of $\Omega_1$ and  $\Omega_2$ by
\begin{align}
  \cot\alpha
  &= \cos\theta_2\cot(\phi_1-\phi_2)
  - \frac{\cot\theta_1\sin\theta_2}{\sin(\phi_1-\phi_2)},
  \label{eq:96}\\
  \cos\beta
  &= \cos\theta_1\cos\theta_2
  + \sin\theta_1\sin\theta_2\cos(\phi_1-\phi_2),
  \label{eq:97}\\
  \cot\gamma
  &= \cos\theta_1\cot(\phi_1-\phi_2)
  - \frac{\sin\theta_1\cot\theta_2}{\sin(\phi_1-\phi_2)}.
  \label{eq:98}
\end{align}

Substituting Eq.~(\ref{eq:58}) into Eq.~(\ref{eq:89}), the complex
conjugate of the correlation function is given by
\begin{equation}
  \tilde{\xi}^{(s_1s_2)\,*}_{X_1X_2\sigma_1\sigma_2}(\bm{r}_1,\bm{r}_2)
  = (-1)^{\sigma_1+\sigma_2}
  \tilde{\xi}^{(s_1s_2)}_{X_1X_2,-\sigma_1,-\sigma_2}(\bm{r}_1,\bm{r}_2),
  \label{eq:99}
\end{equation}
which is consistent with Eqs.~(\ref{eq:72}) and (\ref{eq:91}).
Similarly, the parity transformation of Eq.~(\ref{eq:60}) shows that
\begin{equation}
  \tilde{\xi}^{(s_1s_2)}_{X_1X_2\sigma_1\sigma_2}(\bm{r}_1,\bm{r}_2)
  \xrightarrow{\mathbb{P}}
  (-1)^{p_{X_1}+p_{X_2}}
  \tilde{\xi}^{(s_1s_2)}_{X_1X_2,-\sigma_1,-\sigma_2}(-\bm{r}_1,-\bm{r}_2),
  \label{eq:100}
\end{equation}
which is consistent with Eqs.~(\ref{eq:73}) and (\ref{eq:80}).

As in the previous case, the correlation function of Eq.~(\ref{eq:91})
corresponds to the ensemble average, and we consider the coordinate
systems are set for each pair of objects. Just as in the previous
case, we consider they are chosen as $(\theta_1,\phi_1)=(\theta,0)$
and $(\theta_2,\phi_2)=(0,0)$. Substituting these coordinate values
into Eq.~(\ref{eq:91}), we derive
\begin{multline}
  \tilde{\xi}^{(s_1s_2)}_{X_1X_2\sigma_1\sigma_2}(r_1,r_2,\theta)
  = i^{s_1+s_2} (-1)^{\sigma_1+\sigma_2}
  \sum_{l,l_1,l_2} i^{l_1+l_2}
  u^{(l)}_{\sigma_1\sigma_2}(\theta)
  \\ \times
  \begin{pmatrix}
    s_1 & l_1 & l \\ \sigma_1 & 0 & -\sigma_1
  \end{pmatrix}
  \begin{pmatrix}
    s_2 & l_2 & l \\ \sigma_2 & 0 & -\sigma_2
  \end{pmatrix}
  C^{s_1s_2;l_1l_2}_{X_1X_2\,l}(r_1,r_2),
  \label{eq:101}
\end{multline}
where
\begin{equation}
  u^{(l)}_{m'm}(\theta) \equiv
  (-1)^m D^{-m}_{(l)\,m'}(0,\theta,0)
  = (-1)^m d^{(l)}_{-m,m'}(\theta).
  \label{eq:102}
\end{equation}
The function $d^{(l)}_{m'm}(\theta) \equiv D^{m'}_{(l)m}(0,\theta,0)$
is known to be the Wigner's $d$-function. An explicit formula of
Wigner's rotation matrix \cite{Khersonskii:1988krb} is translated into
a formula,
\begin{multline}
  u^{(l)}_{m'm}(\theta) =
  (-1)^m\sqrt{(l+m)!\,(l-m)!\,(l+m')!\,(l-m')!}
  \\ \times 
  \sum_r
  \frac{(-1)^r
    \left[\cos(\theta/2)\right]^{2l-m-m'-2r}
    \left[\sin(\theta/2)\right]^{m+m'+2r}}
  {r!\,(l-m-r)!\,(l-m'-r)!\,(m+m'+r)!},
  \label{eq:103}
\end{multline}
where $r$ runs over all integers for which the factorial arguments are
non-negative. We have
\begin{equation}
  u^{(l)\,*}_{m'm}(\theta) = u^{(l)}_{m'm}(\theta),
  \label{eq:104}
\end{equation}
i.e., they are real functions, and we also have symmetries,
\begin{equation}
  u^{(l)}_{m'm}(\theta)
  = u^{(l)}_{-m,-m'}(\theta)
  = (-1)^{m'+m} u^{(l)}_{mm'}(\theta),
  \label{eq:105}
\end{equation}
and
\begin{equation}
  u^{(l)}_{m'm}(-\theta) =  u^{(l)}_{mm'}(\theta).
  \label{eq:106}
\end{equation}
From the last symmetry, the correlation function of Eq.~(\ref{eq:101})
satisfies the interchange symmetry:
\begin{equation}
  \tilde{\xi}^{(s_1s_2)}_{X_1X_2\sigma_1\sigma_2}(r_1,r_2,\theta) =
  \tilde{\xi}^{(s_2s_1)}_{X_2X_1\sigma_2\sigma_1}(r_2,r_1,-\theta). 
  \label{eq:107}
\end{equation}

Comparing Eq.~(\ref{eq:101}) with Eq.~(\ref{eq:75}), we have a simple
relation between the correlation function and the power spectrum in
spherical coordinates,
\begin{equation}
  \tilde{\xi}^{(s_1s_2)}_{X_1X_2\sigma_1\sigma_2}(r_1,r_2,\theta)
  =
  \sum_l u^{(l)}_{\sigma_1\sigma_2}(\theta)
  \tilde{C}^{(s_1s_2)\,l}_{X_1X_2\sigma_1\sigma_2}(r_1,r_2).
  \label{eq:108}
\end{equation}
From the orthonormality relation of Wigner's $d$-function
\cite{Khersonskii:1988krb}, we have the orthonormality relation,
\begin{equation}
  \int_0^\pi \sin\theta\, d\theta\,
  u^{(l')}_{m'm}(\theta)\, u^{(l)}_{m'm}(\theta)
  = \frac{2}{2l+1} \delta_{l'l}
  \label{eq:109}
\end{equation}
for arbitrary integers $m$ and $m'$ that satisfy $|m| \leq l$ and
$|m'| \leq l'$. Thus the inverse relation of Eq.~(\ref{eq:108}) is
given by
\begin{multline}
  \tilde{C}^{(s_1s_2)\,l}_{X_1X_2\sigma_1\sigma_2}(r_1,r_2) =
  \frac{2l+1}{2}
  \int_0^\pi \sin\theta\,d\theta\,
  u^{(l)}_{\sigma_1\sigma_2}(\theta)\,
  \\ \times
  \tilde{\xi}^{(s_1s_2)}_{X_1X_2\sigma_1\sigma_2}(r_1,r_2,\theta).
  \label{eq:110}
\end{multline}
The parity transformation of Eq.~(\ref{eq:101}) is given by
\begin{equation}
  \tilde{\xi}^{(s_1s_2)}_{X_1X_2\sigma_1\sigma_2}(r_1,r_2,\theta)
  \xrightarrow{\mathbb{P}}
  (-1)^{p_{X_1}+p_{X_2}}
  \tilde{\xi}^{(s_1s_2)}_{X_1X_2,-\sigma_1,-\sigma_2}(r_1,r_2,\theta).
  \label{eq:111}
\end{equation}

\subsection{%
  Alternative expressions of the correlation functions
}

The evaluations of the correlation functions from the invariant power
spectrum by the formulas of Eqs.~(\ref{eq:87}), (\ref{eq:101}) and
(\ref{eq:108}) are numerically difficult, because we need to evaluate
the power spectrum $C^{s_1s_2;l_1l_2}_{X_1X_2\,l}(r_1,r_2)$
theoretically up to sufficiently large $l$ to ensure the convergence
of the series expansion. In order to see the problem, we consider a
small angle limit $\theta \rightarrow 0$, for example. In this limit,
we have
$u^{(l)}_{\sigma_1\sigma_2}(\theta) \rightarrow
g^{(l)}_{\sigma_1\sigma_2}$, where the rhs is defined by
Eq.~(\ref{eq:11}). Therefore, we have
\begin{equation}
  \lim_{\theta\rightarrow 0}\,
  \tilde{\xi}^{(s_1s_2)}_{X_1X_2\sigma_1\sigma_2}(r_1,r_2,\theta)
  = g^{(l)}_{\sigma_1\sigma_2}
  \sum_l
  \tilde{C}^{(s_1s_2)\,l}_{X_1X_2\sigma_1,-\sigma_1}(r_1,r_2).
  \label{eq:112}
\end{equation}
The rhs of the above equation vanishes when
$\sigma_1 + \sigma_2 \ne 0$, while whole multipoles $l$ of the power
spectrum contribute when $\sigma_1 + \sigma_2 = 0$. Unless the
absolute values of the power spectrum for large $l$ are small enough
and the summation converges fast, the above summation cannot be
evaluated by truncating the infinite sum over $l$.

A useful expression of the correlation function of scalar fields
without assuming the distant-observer approximation is proposed in
Ref.~\cite{Szalay:1997cc}, in which the correlation function is
parametrized by the separation vector,
\begin{equation}
  \bm{x} \equiv \bm{r}_1-\bm{r}_2,
  \label{eq:113}
\end{equation}
and the opening angle $\theta$. This method was recently generalized in
order to describe the correlation function of two-dimensionally
projected shapes of galaxies
\cite{Shiraishi:2020vvj,Shiraishi:2023zda}. Below we consider
essentially the same way of description as our formalism,
generalizing the linear analysis of the previous work to those
applicable in more general situations. For that purpose, we consider a
redundant function
\begin{equation}
  \Xi^{(s_1s_2)}_{X_1X_2\sigma_1\sigma_2}
  (\bm{x};\hat{\bm{z}}_1,\hat{\bm{z}}_2),
  \label{eq:114}
\end{equation}
in which the arguments $\bm{x},\hat{\bm{z}}_1,\hat{\bm{z}}_2$ are
independent vectors, and the last two vectors
$\hat{\bm{z}}_1,\hat{\bm{z}}_2$ are unit vectors. When the arguments
are given by $\bm{x}=\bm{r}_1-\bm{r}_2$,
$\hat{\bm{z}}_1 = \hat{\bm{r}}_1=\bm{r}_1/r_1$,
$\hat{\bm{z}}_2 = \hat{\bm{r}}_2=\bm{r}_2/r_2$, the values of the
above function are defined to be the same as the values of the
correlation function of Eq.~(\ref{eq:79}):
\begin{equation}
  \xi^{(s_1s_2)}_{X_1X_2\sigma_1\sigma_2}(\bm{r}_1,\bm{r}_2) = 
  \Xi^{(s_1s_2)}_{X_1X_2\sigma_1\sigma_2}
  (\bm{r}_1-\bm{r}_2;\hat{\bm{r}}_1,\hat{\bm{r}}_2).
  \label{eq:115}
\end{equation}
The redundant function of Eq.~(\ref{eq:114}) is defined so that the
homogeneity and isotropy are satisfied, i.e., the function is
independent under the translation and rotation of the coordinates
system. As long as the last condition is satisfied, the function by
definition can take arbitrary values when the above relation of
Eq.~(\ref{eq:115}) is not physically satisfied, e.g., when the three
vectors $\bm{x}$, $\bm{z}_1$, $\bm{z}_2$ are not coplanar, etc. As we
will see in the next section, the redundant function with unphysical
arguments is conveniently defined in the iPT so that the lines of
sight $\hat{\bm{z}}_1$, $\hat{\bm{z}}_2$ can be oriented to any
direction even though they do not correspond to physical lines of
sight.

Following the similar derivation with the rotational symmetry given in
Paper~I, the redundant function can be represented by
\begin{multline}
  \Xi^{(s_1s_2)}_{X_1X_2\sigma_1\sigma_2}
  (\bm{x};\hat{\bm{z}}_1,\hat{\bm{z}}_2) =
  i^{s_1+s_2} \hspace{-1pc}
  \sum_{\substack{l_1,l_2,l,l_{z1},l_{z2}\\m_1,m_2,m,m_{z1},m_{z2}}}
  i^l C_{lm}(\hat{\bm{x}})
  \left(l_1\,l_2\,l\right)_{m_1m_2}^{\phantom{m_1m_2}m}
  \\ \times
  \left(s_1\,l_{z1}\,l_1\right)_{\sigma_1}^{\phantom{\sigma_1}m_{z1}m_1}
  \left(s_2\,l_{z2}\,l_2\right)_{\sigma_2}^{\phantom{\sigma_2}m_{z2}m_2}
  C_{l_{z1}m_{z1}}(\hat{\bm{z}}_1)
  C_{l_{z2}m_{z2}}(\hat{\bm{z}}_2)
  \\ \times
  (-1)^{l_2}
  \Xi^{s_1s_2;l_1l_2l}_{X_1X_2l_{z1}l_{z2}}(x),
  \label{eq:116}
\end{multline}
where $\Xi^{s_1s_2;l_1l_2l}_{X_1X_2l_{z1}l_{z2}}(x)$ is a rotationally
invariant function, $x = |\bm{x}|$, and we use a notation
$C_{lm}(\hat{\bm{r}}) = \sqrt{4\pi/(2l+1)}\, Y_{lm}(\theta,\phi)$ for
spherical harmonics with Racah's normalization with spherical
coordinates $(\theta,\phi)$ of the direction $\hat{\bm{r}}$. The
invariant function is evaluated by the iPT as explicitly shown in the
next section.

The flip symmetry in redshift space, which is discussed in Paper~I,
indicates that the correlation function in redshift space is invariant
under the flipping of the direction to the lines of sight. That
requires the function in Eq.~(\ref{eq:116}) is invariant under the
flipping, $\hat{\bm{z}}_1 \rightarrow -\hat{\bm{z}}_1$ and
$\hat{\bm{z}}_2 \rightarrow -\hat{\bm{z}}_2$. In order to satisfy the
flip symmetry, the indices $l_{z1}$ and $l_{z2}$ should be
non-negative even integers. This property is satisfied in predictions
of the iPT, as explicitly shown in the next section. We assume this
property in the following.

Applying a recoupling formula of $3j$-symbols using the $9j$-symbol,
(see Appendix~C of Paper~I in the present notations), the above
Eq.~(\ref{eq:116}) is equivalent to
\begin{multline}
  \Xi^{(s_1s_2)}_{X_1X_2\sigma_1\sigma_2}
  (\bm{x};\hat{\bm{z}}_1,\hat{\bm{z}}_2) =
  i^{s_1+s_2} \sum_{s,\sigma} (2s+1)
  \left(s_1\,s_2\,s\right)_{\sigma_1\sigma_2}^{\phantom{\sigma_1\sigma_2}\sigma}
  \\ \times
  \sum_{l,l_{z1},l_{z2},l_z}
  (-i)^l (-1)^{l_z}  \sqrt{2l_z+1}\,
  X^{l\,l_{z1}l_{z2}}_{l_z;s\sigma}(\hat{\bm{x}},\hat{\bm{z}}_1,\hat{\bm{z}}_2)
  \\ \times
  \sum_{l_1,l_2} (-1)^{l_1}
  \begin{Bmatrix}
    s_1 & s_2 & s \\
    l_{z1} & l_{z2} & l_z \\
    l_1 & l_2 & l
  \end{Bmatrix}
  \Xi^{s_1s_2;l_1l_2l}_{X_1X_2l_{z1}l_{z2}}(x).
  \label{eq:117}
\end{multline}
With the above expressions, the correlation function in Cartesian
coordinates is given by Eq.~(\ref{eq:115}) in terms of the invariant
function.

From the correlation function in Cartesian coordinates given above,
the correlation function in spherical coordinates is given by
Eq.~(\ref{eq:90}). Using the formulas for a product of Wigner's
rotation matrices and an orthogonality relation of $3j$-symbols
\cite{Edmonds:1955fi}, one can show an identity
\begin{multline}
  \sum_{\sigma',m_z}
  \left(s\,l_z\,l\right)_{\sigma'}^{\phantom{\sigma'}m_zm}
  C_{l_zm_z}(\theta,\phi)
  D^{\sigma'}_{(s)\sigma}(\phi,\theta,0)
  \\
  =(-1)^{\sigma}
  \begin{pmatrix}
    s & l_z & l \\ \sigma & 0 & -\sigma
  \end{pmatrix}
  D_{(l)\sigma}^m(\phi,\theta,0).
  \label{eq:118}
\end{multline}
Substituting the form of Eq.~(\ref{eq:116}) into Eqs.~(\ref{eq:90})
and (\ref{eq:115}), and applying the above identity, we derive
a relatively simple result:
\begin{multline}
  \tilde{\xi}^{(s_1s_2)}_{X_1X_2\sigma_1\sigma_2}(\bm{r}_1,\bm{r}_2)
  = i^{s_1+s_2} (-1)^{\sigma_1+\sigma_2}
  \sum_{l,m} i^l C_{lm}(\hat{\bm{x}})
  \\ \times
  \sum_{l_1,l_2,m_1,m_2}
  \left(l_1\,l_2\,l\right)_{m_1m_2}^{\phantom{m_1m_2}m}
  D_{(l_1)\sigma_1}^{m_1}(\phi_1,\theta_1,0)
  D_{(l_2)\sigma_2}^{m_2}(\phi_2,\theta_2,0)
  \\ \times
  (-1)^{l_2}
  \tilde{\Xi}^{s_1s_2;l_1l_2l}_{X_1X_2\sigma_1\sigma_2}(x).
  \label{eq:119}
\end{multline}
where
\begin{equation}
  \tilde{\Xi}^{s_1s_2;l_1l_2l}_{X_1X_2\sigma_1\sigma_2}(x)
  \equiv
  \sum_{l_{z1},l_{z2}}
  \begin{pmatrix}
    s_1 & l_{z1} & l_1 \\ \sigma_1 & 0 & -\sigma_1
  \end{pmatrix}
  \begin{pmatrix}
    s_2 & l_{z2} & l_2 \\ \sigma_2 & 0 & -\sigma_2
  \end{pmatrix}
  \Xi^{s_1s_2;l_1l_2l}_{X_1X_2l_{z1}l_{z2}}(x),
  \label{eq:120}
\end{equation}
and $x = |\bm{r}_1-\bm{r}_2|$. 

As in the previous subsection, one can choose a coordinates system
$(\theta_1,\phi_1)=(\theta,0)$ and $(\theta_2,\phi_2)=(0,0)$ for each
pair of objects. In this case, the above equation reduces to
\begin{multline}
  \tilde{\xi}^{(s_1s_2)}_{X_1X_2\sigma_1\sigma_2}(r_1,r_2,\theta)
  = i^{s_1+s_2} (-1)^{\sigma_1}
  \sum_{l,m} i^l
  \tilde{P}_l^m(\mu_2)
  \\ \times
  \sum_{l_1,l_2}
  u^{(l_1)}_{\sigma_1,\sigma_2-m}(\theta)
  (-1)^{l_2}
  \begin{pmatrix}
    l_1 & l_2 & l \\
    m-\sigma_2 & \sigma_2 & -m
  \end{pmatrix}
  \tilde{\Xi}^{s_1s_2;l_1l_2l}_{X_1X_2\sigma_1\sigma_2}(x),
  \label{eq:121}
\end{multline}
where
\begin{equation}
  x = \sqrt{{r_1}^2 + {r_2}^2 - 2 r_1r_2\cos\theta}, \quad
  \mu_2 = \frac{r_1\cos\theta - r_2}{x}.
  \label{eq:122}
\end{equation}
Geometrically, the variable $\mu_2$ corresponds to the cosine of the
angle between $\bm{r}_2$ and $\bm{x}$. The expression of
Eq.~(\ref{eq:121}) is useful to numerically evaluate the correlation
function from the perturbation theory, as demonstrated in the next
section.

\subsection{%
  Correspondences to the distant-observer limit
}

In Papers~I--III, the statistics of tensor fields are considered in
the flat-sky and distant-observer limits. These previous results
should be recovered by taking the corresponding limits of the full-sky
formulation defined above. In Paper~III, the correspondence is already
shown for projected tensor fields. One can generalize the derivation
to the general case without projection onto the sky, as outlined
below.

\subsubsection{The power spectrum}

First, we consider correspondences between the power spectra of
distant-observer limit and those of full-sky formulation. It is
convenient to define a weighted sum of the multipole expansion in
spherical coordinates by
\begin{equation}
  \tilde{g}_{Xs\sigma}(\bar{\bm{l}},r)
  \equiv
  \frac{2\pi}{\bar{l}}
  \sum_m i^m e^{im\phi_l} \tilde{F}^{lm}_{Xs\sigma}(r),
  \label{eq:123}
\end{equation}
where $\phi_l$ is the azimuthal angle  of the two-dimensional vector
$\bar{\bm{l}}$, and
\begin{equation}
  \bar{l} \equiv l+\frac{1}{2}, \quad
  \bar{\bm{l}} \equiv
  \left(\bar{l} \cos\phi_l, \bar{l} \sin\phi_l,0\right).
  \label{eq:124}
\end{equation}
The inverse relation of Eq.~(\ref{eq:123}) is given by
\begin{equation}
  \tilde{F}^{lm}_{Xs\sigma}(r) =
  (-i)^m \frac{\bar{l}}{2\pi}
  \int \frac{d\phi_l}{2\pi} e^{-im\phi_l}
  \tilde{g}_{Xs\sigma}(\bar{\bm{l}},r).
  \label{eq:125}
\end{equation}

In the flat-sky and distant-observer limits, we consider the vicinity of
$\bm{r} \approx (0,0,r_0)$, where $r_0$ is much larger than the
clustering scales we are interested in. Substituting the above
equation into Eq.~(\ref{eq:50}) in the limits, and following a
derivation similar to the one in Paper~III, we derive
\begin{equation}
  \tilde{F}_{Xs\sigma}(\bm{r})
  \approx
  (-i)^\sigma  e^{-i\sigma\phi}
  \int \frac{d^2\bar{l}}{(2\pi)^2}
  e^{i\bar{\bm{l}}\cdot\bm{\theta}} e^{i\sigma\phi_l}
  \tilde{g}_{Xs\sigma}(\bar{\bm{l}},r),
  \label{eq:126}
\end{equation}
where
\begin{equation}
  \bm{r} = \left(r_0\theta\cos\phi,r_0\theta\sin\phi,r\right),\quad
  \bm{\theta} = \left(\theta\cos\phi,\theta\sin\phi,0\right).
  \label{eq:127}
\end{equation}
The variables $\theta$ and $\phi$ correspond to the spherical
coordinates of position vector $\bm{r}$ in the approximation, in which
we have
\begin{equation}
  \theta \ll 1, \quad l \gg 1.
  \label{eq:128}
\end{equation}
In the same limit, the two-point correlation of
$\tilde{g}_{Xs\sigma}(\bar{\bm{l}},r)$ is derived by following exactly
the same manner as Paper~III, resulting in
\begin{equation}
  \left\langle
    \tilde{g}_{X_1s_1\sigma_1}(\bar{\bm{l}},r)
    \tilde{g}_{X_2s_2\sigma_2}(\bar{\bm{l}}',r')
  \right\rangle
  \approx
  \frac{2\pi}{\bar{l}}
  (2\pi)^2 \delta_\mathrm{D}^2(\bar{\bm{l}} + \bar{\bm{l}}')
  \tilde{C}^{(s_1s_2)\,l}_{X_1X_2\sigma_1\sigma_2}(r,r').
  \label{eq:129}
\end{equation}

In the limit of $\theta \ll 1$, the rotation matrix of
Eq.~(\ref{eq:29}) is given by $D^{\sigma'}_{(s)\sigma}(\phi,0,0) =
\delta^{\sigma'}_\sigma e^{-i\sigma\phi}$, and thus we have
\begin{equation}
  \tilde{F}_{Xs\sigma}(\bm{r}) \approx
  e^{-i\sigma\phi} F_{Xs\sigma}(\bm{r}).
  \label{eq:130}
\end{equation}
We define the tensor field with a displaced origin in the
distant-observer approximation,
\begin{equation}
  \bar{F}_{Xs\sigma}(\bm{x}) \equiv
  F_{Xs\sigma}(\bm{x} + r_0\hat{\mathbf{e}}_3).
  \label{eq:131}
\end{equation}
Combining Eqs.~(\ref{eq:126}) and (\ref{eq:130}),
the Fourier transform of Eq.~(\ref{eq:131}) is given by
(note that we use the same symbols of functions for configuration
space and Fourier space)
\begin{align}
  \bar{F}_{Xs\sigma}(\bm{k})
  &= \int d^3x\,e^{-i\bm{k}\cdot\bm{x}}  \bar{F}_{Xs\sigma}(\bm{x})
  \nonumber\\
  &\approx (-i)^\sigma e^{i\sigma\phi_k}
    {r_0}^2
    \int dx\, e^{-ik_\parallel x}
    \tilde{g}_{Xs\sigma}(\bar{\bm{l}},x+r_0),
    \label{eq:132}
\end{align}
where the wave vector $\bm{k}$ in the last expression is parametrized
by
\begin{equation}
  \bm{k} =
  \left(
    k_\perp\cos\phi_k,k_\perp\sin\phi_k,k_\parallel
  \right),
  \label{eq:133}
\end{equation}
with
\begin{equation}
  k_\perp \equiv \frac{\bar{l}}{r_0}, \quad
  \phi_k \equiv \phi_l.
  \label{eq:134}
\end{equation}

From Eqs.~(\ref{eq:129}) and (\ref{eq:132}), we
derive
\begin{equation}
  \left\langle
    \bar{F}_{X_1s_1\sigma_1}(\bm{k})
    \bar{F}_{X_2s_2\sigma_2}(\bm{k}')
  \right\rangle
  \approx (2\pi)^3 \delta_\mathrm{D}^3(\bm{k}+\bm{k}')
  P^{(s_1s_2)}_{X_1X_2\sigma_1\sigma_2}(\bm{k}),
  \label{eq:135}
\end{equation}
where
\begin{multline}
  P^{(s_1s_2)}_{X_1X_2\sigma_1\sigma_2}(\bm{k})
  =
  (-i)^{\sigma_1-\sigma_2}
  e^{i(\sigma_1+\sigma_2)\phi_k}
  \\ \times 
 \frac{2\pi r_0}{k_\perp} \int dx\,e^{-ik_\parallel x}
  \tilde{C}^{(s_1s_2)\,l}_{X_1X_2\sigma_1\sigma_2}(x+r_0,r_0),
  \label{eq:136}
\end{multline}
and we have applied a relation
$\tilde{C}^{(s_1s_2)\,l}_{X_1X_2\sigma_1\sigma_2}(x+r_0,x'+r_0)
\approx
\tilde{C}^{(s_1s_2)\,l}_{X_1X_2\sigma_1\sigma_2}(x-x'+r_0,r_0)$ in the
distant-observer approximation. The appearance of the delta function
in Eq.~(\ref{eq:135}) shows that the translational invariance of the
two-point statistics is recovered in the Cartesian coordinates of the
distant-observer limit. Equation~(\ref{eq:136}) is the correspondence of
the full-sky power spectrum
$\tilde{C}^{(s_1s_2)\,l}_{X_1X_2\sigma_1\sigma_2}(r,r')$ with the
flat-sky, distant-observer power spectrum
$P^{(s_1s_2)}_{X_1X_2\sigma_1\sigma_2}(\bm{k})$ in Cartesian
coordinates. Inverting Eq.~(\ref{eq:136}), the full-sky power spectrum
is approximately given by an integral of flat-sky, distant-observer
power spectrum as
\begin{multline}
  \tilde{C}^{(s_1s_2)\,l}_{X_1X_2\sigma_1\sigma_2}(r_1,r_2)
  \approx
  i^{\sigma_1-\sigma_2}
  e^{-i(\sigma_1+\sigma_2)\phi_k}
  \\ \times 
  \frac{k_\perp}{2\pi r_0}
  \int \frac{dk_\parallel}{2\pi}\,e^{ik_\parallel (r_1-r_2)}
  P^{(s_1s_2)}_{X_1X_2\sigma_1\sigma_2}(\bm{k}),
  \label{eq:137}
\end{multline}
where $|r_1-r_2| \ll r_0 \equiv (r_1+r_2)/2$ and
$l = k_\perp r_0 - 1/2 \gg 1$.

Substituting Eq.~(\ref{eq:77}) into the complex
conjugate of Eq.~(\ref{eq:136}), we derive
\begin{equation}
  P^{(s_1s_2)\,*}_{X_1X_2\sigma_1\sigma_2}(\bm{k})
  = (-1)^{\sigma_1+\sigma_2}
  P^{(s_1s_2)}_{X_1X_2,-\sigma_1,-\sigma_2}(-\bm{k}).
  \label{eq:138}
\end{equation}
This property of the complex conjugate is consistent with the power
spectrum in the distant-observer approximation described in Paper~I.
Similarly, the parity transformation of Eq.~(\ref{eq:136}) is given by
\begin{equation}
  P^{(s_1s_2)}_{X_1X_2\sigma_1\sigma_2}(\bm{k})
  \xrightarrow{\mathbb{P}}
  (-1)^{p_{X_1}+p_{X_2}+\sigma_1+\sigma_2}
  e^{2i(\sigma_1+\sigma_2)\phi_k}
  P^{(s_1s_2)}_{X_1X_2,-\sigma_1,-\sigma_2}(\bm{k}).
  \label{eq:139}
\end{equation}
On one hand, the last transformation corresponds to the parity around
the origin of the spherical coordinates. In Paper~I, on the other
hand, the parity transformation in the distant-observer approximation
is differently defined and corresponds to the parity around the center
of the sample, say, $\bm{r} = r_0 \hat{\mathbf{e}}_3$. As discussed in
Paper~III, the parity transformation around the origin of spherical
coordinates corresponds to the combined transformation of the parity
and flipping around the center of the sample, instead of the origin of
the spherical coordinates.

\subsubsection{The correlation function}

Next, we consider correspondences between the correlation functions of
distant-observer limit and those of full-sky formulation. In the
distant-observer approximation of Papers~I--III, the correlation
function of an irreducible tensor is defined by the field
$F_{Xs\sigma}(\bm{r})$ in Cartesian coordinates. Thus the correlation
function in the distant-observer limit corresponds to
$\xi^{(s_1s_2)}_{X_1X_2\sigma_1\sigma_2}(r_1,r_2)$ of
Eq.~(\ref{eq:79}). The correlation function in spherical coordinates,
$\tilde{\xi}^{(s_1s_2)}_{X_1X_2\sigma_1\sigma_2}(r_1,r_2)$ of
Eq.~(\ref{eq:89}) is related to that in Cartesian coordinates by
Eq.~(\ref{eq:90}). One can always choose the coordinates with
$(\theta_1,\phi_1) = (\theta,0)$ and $(\theta_2,\phi_2) = (0,0)$ as
done in Eq.~(\ref{eq:101}), and in this case we have
\begin{equation}
  \tilde{\xi}^{(s_1s_2)}_{X_1X_2\sigma_1\sigma_2}(r_1,r_2,\theta) =
  \sum_{m'} d^{(s_1)}_{m'\sigma_1}(\theta)
  \xi^{(s_1s_2)}_{X_1X_2m'\sigma_2}(r_1,r_2,\theta).
  \label{eq:140}
\end{equation}
In the distant-observer approximation with $\theta \ll 1$, the Wigner's
$d$-function is approximated by $d^{(l)}_{m'm}(\theta) \approx
\delta_{m'm}$, and we have
\begin{equation}
  \tilde{\xi}^{(s_1s_2)}_{X_1X_2\sigma_1\sigma_2}(r_1,r_2,\theta)
  \approx
  \xi^{(s_1s_2)}_{X_1X_2\sigma_1\sigma_2}(r_1,r_2,\theta).
  \label{eq:141}
\end{equation}
The above property can also be understood by noting that the spherical
basis $\tilde{\mathbf{e}}^m$ of Eq.~(\ref{eq:14}) in spherical
coordinates and the basis $\mathbf{e}^m$ of Eq.~(\ref{eq:3}) in
Cartesian coordinates are approximately the same on the $x$-$z$ plane
or the plane of $\phi=0$.

The correlation function in the distant-observer limit,
$\bar{\xi}^{(s_1s_2)}_{X_1X_2\sigma_1\sigma_2}(\bm{x})$ is the
three-dimensional Fourier transform of the power spectrum in the same
limit,
\begin{equation}
  \bar{\xi}^{(s_1s_2)}_{X_1X_2\sigma_1\sigma_2}(\bm{x}) =
  \int \frac{d^3k}{(2\pi)^3}
  P^{(s_1s_2)}_{X_1X_2\sigma_1\sigma_2}(\bm{k}).
  \label{eq:142}
\end{equation}
In Paper~I, we have derived an equation to describe the correlation
function in distant-observer approximation when the line of sight is
directed to the $z$ axis, and the result is given by
\begin{multline}
  \bar{\xi}^{(s_1s_2)}_{X_1X_2\sigma_1\sigma_2}(\bm{x}) =
  i^{s_1+s_2}
  \sum_{l,l_z,s}  i^l (-1)^s 
  \sqrt{2s+1}
  \begin{pmatrix}
    s_1 & s_2 & s \\ \sigma_1 & \sigma_2 & -\sigma_{12}
  \end{pmatrix}
  \\ \times
  \begin{pmatrix}
    s & l_z & l \\ \sigma_{12} & 0 & -\sigma_{12}
  \end{pmatrix}
  C_{l\sigma_{12}}(\hat{\bm{x}})
  \xi^{s_1s_2;l\,l_z;s}_{X_1X_2}(x),
  \label{eq:143}
\end{multline}
where $\sigma_{12}=\sigma_1+\sigma_2$. Corresponding to the full-sky
correlation function of
$\tilde{\xi}^{(s_1s_2)}_{X_1X_2\sigma_1\sigma_2}(r_1,r_2,\theta)$ in
the naturally chosen coordinates system, Eq.~(\ref{eq:101}), one
considers the coordinates system with the azimuthal angle of $\bm{x}$
is zero, $\phi_x=0$, and in this case we have
\begin{multline}
  \bar{\xi}^{(s_1s_2)}_{X_1X_2\sigma_1\sigma_2}(x,\mu_x) =
  i^{s_1+s_2} \sum_{l,l_z,s}  i^l (-1)^s 
  \sqrt{2s+1}
  \begin{pmatrix}
    s_1 & s_2 & s \\ \sigma_1 & \sigma_2 & -\sigma_{12}
  \end{pmatrix}
  \\ \times
  \begin{pmatrix}
    s & l_z & l \\ \sigma_{12} & 0 & -\sigma_{12}
  \end{pmatrix}
  \tilde{P}_l^{\sigma_{12}}(\mu_x)
  \xi^{s_1s_2;l\,l_z;s}_{X_1X_2}(x),
  \label{eq:144}
\end{multline}
where $\mu_x = \cos\theta_x$ is the direction cosine of the vector
$\bm{x}$ relative to the line of sight.

In the distant-observer limit of $\theta\rightarrow 0$, we have
$u^{(s_1)}_{\sigma_1,\sigma_2-m}(\theta) \rightarrow
(-1)^{\sigma_1}\delta_{\sigma_1+\sigma_2,m}$ in the expression of the
correlation function with wide-angle effects, Eq.~(\ref{eq:121}).
Comparing the result of Eq.~(\ref{eq:121}) in this limit with the
expression of Eq.~(\ref{eq:144}), the following correspondence between
the invariant functions is derived:
\begin{multline}
  \xi^{s_1s_2;l\,l_z;s}_{X_1X_2}(x)
  =
  (-1)^{l_z}(2l_z+1) \sqrt{2s+1}
  \sum_{l_1,l_2,l_{z1},l_{z2}}
  (-1)^{l_1}
  \\ \times
  \begin{pmatrix}
    l_{z1} & l_{z2} & l_z \\ 0 & 0 & 0
  \end{pmatrix}
  \begin{Bmatrix}
    s_1 & s_2 & s \\ l_{z1} & l_{z2} & l_z \\ l_1 & l_2 & l
  \end{Bmatrix}
  \Xi^{s_1s_2;l_1l_2l}_{X_1X_2l_{z1}l_{z2}}(x)
  \label{eq:145}
\end{multline}
in the distant-observer limit. Otherwise the expression of
Eq.~(\ref{eq:121}) is inconsistent with the expression of
Eq.~(\ref{eq:144}) in the distant-observer limit. In the next section,
the above correspondence is explicitly shown to hold in linear theory.

\section{\label{sec:iPTFullSky}%
  Evaluation from the integrated perturbation theory
}

\subsection{%
  Applying the perturbation theory to full-sky statistics}

A plausible property of the tensor fields in Cartesian coordinates is
the statistical homogeneity of the field in real space, because the
directions of the spherical basis in Cartesian coordinates,
$\mathbf{e}^m$, are fixed throughout the whole space, while those in
spherical coordinates, $\tilde{\mathbf{e}}^m$, vary and depend on the
angular position $(\theta,\phi)$ in space. However, in redshift space,
the tensor fields are not statistically homogeneous, because the
statistical properties of spatial clustering are distorted along the
lines of sight, which directions depend on the angular positions of
objects. In Papers~I--III, where the distant-observer approximation is
applied, the direction to the line of sight is fixed in Cartesian
coordinates, and therefore the tensor fields are still homogeneous in
space, even though they are anisotropic due to the redshift-space
distortions.

To reconcile the redshift-space distortions even in the full-sky case,
we consider virtually homogeneous tensor fields
$F_{Xs\sigma}(\bm{r};\hat{\bm{z}})$ with a fictitious line-of-sight
direction, $\hat{\bm{z}}$, which is fixed throughout the space. In
reality, the line-of-sight direction $\hat{\bm{z}}$ must be the same
as the radial direction $\hat{\bm{r}}$ from the observer, and
observable tensor fields correspond to a special case of the virtual
field with a constraint $\hat{\bm{z}} = \hat{\bm{r}}$, i.e.,
\begin{equation}
  F_{Xs\sigma}(\bm{r}) = F_{Xs\sigma}(\bm{r};\hat{\bm{r}}).
  \label{eq:146}
\end{equation}
As we explicitly show below, the introduction of the virtually
homogeneous field with a fixed line-of-sight direction turns out to be
convenient in applying the perturbation theory to derive the
theoretical predictions of the clustering in redshift space. In real
space, such introduction of the virtually homogeneous field is not
necessary.

We consider the Fourier transform of the virtually homogeneous field in
redshift space,
\begin{align}
  F_{Xs\sigma}(\bm{r};\hat{\bm{z}})
  &=
  \int \frac{d^3k}{(2\pi)^3} e^{i\bm{k}\cdot\bm{r}}
  F_{Xs\sigma}(\bm{k};\hat{\bm{z}}),
  \label{eq:147}\\
  F_{Xs\sigma}(\bm{k};\hat{\bm{z}})
  &=
  \int d^3r\, e^{-i\bm{k}\cdot\bm{r}}
  F_{Xs\sigma}(\bm{r};\hat{\bm{z}}).
  \label{eq:148}
\end{align}
The field $F_{Xs\sigma}(\bm{k};\hat{\bm{z}})$ is the Fourier transform
and not the same as the function $F_{Xs\sigma}(\bm{r};\hat{\bm{z}})$
in configuration space, as we use the same symbols in both
configuration and Fourier spaces for simplicity. In the above
Eqs.~(\ref{eq:147}) and (\ref{eq:148}), the line-of-sight direction
$\hat{\bm{z}}$ is temporarily independent to the position $\bm{r}$.

The two-point correlation function of the virtual fields with
different lines of sight in configuration space is given by
\begin{equation}
  \left\langle
    F_{X_1s_1\sigma_1}(\bm{r}_1;\hat{\bm{z}}_1)
    F_{X_2s_2\sigma_2}(\bm{r}_2;\hat{\bm{z}}_2)
  \right\rangle =
  \Xi^{(s_1s_2)}_{X_1X_2\sigma_1\sigma_2}
  (\bm{r}_1-\bm{r}_2;\hat{\bm{z}}_1,\hat{\bm{z}}_2).
  \label{eq:149}
\end{equation}
That is, the correlation function of virtual fields with fixed lines
of sight only depends on the relative position $\bm{r}_1-\bm{r}_2$ due
to the homogeneity by construction, while the function explicitly
depends on the two different lines of sight, $\hat{\bm{z}}_1$ and
$\hat{\bm{z}}_2$. The power spectrum of the virtual fields in the
Fourier space is given by
\begin{multline}
  \left\langle
    F_{X_1s_1\sigma_1}(\bm{k}_1;\hat{\bm{z}}_1)
    F_{X_2s_2\sigma_2}(\bm{k}_2;\hat{\bm{z}}_2)
  \right\rangle
  \\ =
  (2\pi)^3 \delta_\mathrm{D}^3(\bm{k}_1 + \bm{k}_2)
  P^{(s_1s_2)}_{X_1X_2\sigma_1\sigma_2}
  (\bm{k}_1;\hat{\bm{z}}_1,\hat{\bm{z}}_2),
  \label{eq:150}
\end{multline}
where the Dirac's delta function appears due to the homogeneity of the
virtual fields mentioned above. The power spectrum and the correlation
function with virtually fixed two lines of sight are related by the
Fourier transform,
\begin{equation}
  \Xi^{(s_1s_2)}_{X_1X_2\sigma_1\sigma_2}(\bm{r};\hat{\bm{z}}_1,\hat{\bm{z}}_2)
  = \int \frac{d^3k}{(2\pi)^3} e^{i\bm{k}\cdot\bm{r}}
  P^{(s_1s_2)}_{X_1X_2\sigma_1\sigma_2}(\bm{k};\hat{\bm{z}}_1,\hat{\bm{z}}_2).
  \label{eq:151}
\end{equation}
As we show below, the power spectrum
$P^{(s_1s_2)}_{X_1X_2\sigma_1\sigma_2}
(\bm{k};\hat{\bm{z}}_1,\hat{\bm{z}}_2)$ is a natural quantity to be
theoretically predicted by the perturbation theory.

Once the last power spectrum is analytically evaluated by the
perturbation theory, the correlation function
$\Xi^{(s_1s_2)}_{X_1X_2\sigma_1\sigma_2}(\bm{r};\hat{\bm{z}}_1,\hat{\bm{z}}_2)$
of Eq.~(\ref{eq:151}) is obtained, and from that, the correlation
function of Eq.~(\ref{eq:79}) is given by Eq.~(\ref{eq:115}).
Substituting the result obtained in the above way into
Eqs.~(\ref{eq:83}) or (\ref{eq:84}), the expression of the invariant
power spectrum $C^{s_1s_2;l_1l_2}_{X_1X_2\,l}(r_1,r_2)$ is
analytically evaluated by the perturbation theory. The power spectrum
$\tilde{C}^{(s_1s_2)\,l}_{X_1X_2\sigma_1\sigma_2}(r_1,r_2)$ and the
correlation function
$\tilde{\xi}^{(s_1s_2)}_{X_1X_2\sigma_1\sigma_2}(r_1,r_2,\theta)$ in
spherical coordinates are finally evaluated by Eqs.~(\ref{eq:75}) or
(\ref{eq:101}), respectively. The starting point of the above
procedure is to evaluate the power spectrum
$P^{(s_1s_2)}_{X_1X_2\sigma_1\sigma_2}(\bm{k};\hat{\bm{z}}_1,\hat{\bm{z}}_2)$,
to which we apply the iPT in the following.

\subsection{
  Invariant power spectrum in Fourier space
}

In Paper~I--III, the line of sight is fixed to a single direction
$\hat{\bm{z}}$ according to the distant-observer approximation. While
we do not assume the distant-observer approximation in this paper, the
methods in previous papers can be generalized to include the
possibility of having multiple numbers of line-of-sight directions. In
terms of the iPT, we just use the propagators in redshift space
(Paper~I),
$\Gamma^{(n)}_{Xs\sigma}(\bm{k}_1,\ldots,\bm{k}_n;\hat{\bm{z}}_i)$,
with $i=1,2$ in evaluating the power spectrum of Eq.~(\ref{eq:150}).
In the course of applying the iPT in this generalized situation, it is
convenient to decompose the direction dependencies of the arguments in
the power spectrum of Eq.~(\ref{eq:150}) by polypolar spherical
harmonics, which are extensively used in Paper~I.

Following the same kind of calculations illustrated in Paper~I, the
decomposition is given by
\begin{multline}
  P^{(s_1s_2)}_{X_1X_2\sigma_1\sigma_2}
  (\bm{k};\hat{\bm{z}}_1,\hat{\bm{z}}_2) =
  i^{s_1+s_2} \sum_{s,\sigma}
  (-1)^s \sqrt{2s+1}
  \left(s_1\,s_2\,s\right)_{\sigma_1\sigma_2}^{\phantom{\sigma_1\sigma_2}\sigma}
  \\ \times
  \sum_{l,l_z,l_{z1},l_{z2}}
  X^{l\,l_{z1}l_{z2}}_{l_z;s\sigma}(\hat{\bm{k}},\hat{\bm{z}}_1,\hat{\bm{z}}_2)
  P^{s_1s_2;s\,l_z}_{X_1X_2;l\,l_{z1}l_{z2}}(k),
  \label{eq:152}
\end{multline}
where $X^{l_1l_2l_3}_{L;lm}(\bm{n}_1,\bm{n}_2,\bm{n}_3)$ is the
tripolar spherical harmonics and the explicit definition is given in
Paper~I. The last factor corresponds to the invariant spectrum. From
the orthonormality relation of the tripolar spherical harmonics and
the orthogonality relation of the $3j$-symbols (Paper~I), the above
expansion is unique, i.e., the invariant spectrum is uniquely
determined by the power spectrum on the lhs.

Substituting the expansion of Eq.~(\ref{eq:152}) into
Eq.~(\ref{eq:151}), using the Rayleigh expansion formula for the
factor $e^{i\bm{k}\cdot\bm{r}}$, and various formulas regarding
$3j$-symbols and polypolar spherical harmonics (listed in Appendixes
of Paper~I), Eq.~(\ref{eq:115}) eventually reduces to
\begin{multline}
  \xi^{(s_1s_2)}_{X_1X_2\sigma_1\sigma_2}(\bm{r}_1,\bm{r}_2)
  = i^{s_1+s_2}
  \sum_{s,\sigma} (-1)^{s} \sqrt{\{s\}}
  \left(s_1\,s_2\,s\right)_{\sigma_1\sigma_2}^{\phantom{\sigma_1\sigma_2}\sigma}
  \\ \times
  \sum_{\substack{l,l_z,l_{z1},l_{z2}\\l_1,l_2,l_1',l_2'}}
  i^{l_1'-l_2'} (-1)^{l_1+l_2+l} \{l_1\}\{l_2\} \{l_1'\}\{l_2'\}
  \sqrt{\{l_z\}}
  \begin{pmatrix}
    l_1' & l_2' & l \\ 0 & 0 & 0
  \end{pmatrix}
  \\ \times
  \begin{pmatrix}
    l_1 & l_{z1} & l_1' \\ 0 & 0 & 0
  \end{pmatrix}
  \begin{pmatrix}
    l_2 & l_{z2} & l_2' \\ 0 & 0 & 0
  \end{pmatrix}
  \begin{Bmatrix}
    l_1 & l_2 & s \\
    l_1' & l_2' & l \\
    l_{z1} & l_{z2} & l_z
  \end{Bmatrix}
  X^{l_1l_2}_{s\sigma}(\hat{\bm{r}}_1,\hat{\bm{r}}_2)
  \\ \times
  \int \frac{k^2dk}{2\pi^2}
  j_{l_1'}(kr_1) j_{l_2'}(kr_2)
  P^{s_1s_2;s\,l_z}_{X_1X_2;l\,l_{z1}l_{z2}}(k),
  \label{eq:153}
\end{multline}
where the $3\times 3$ matrix with curly brackets is the $9j$-symbol,
and we use simple notations introduced in Paper~I,
\begin{equation}
  \left\{s\right\} \equiv 2s+1, \quad
  \left\{l_1\right\} \equiv 2l_1+1,
  \label{eq:154}
\end{equation}
and so forth. Substituting the above result into Eq.~(\ref{eq:84}), we
have
\begin{multline}
  C^{s_1s_2;l_1l_2}_{X_1X_2\,l}(r_1,r_2)
  = (-1)^{s_2} (-i)^{l_1-l_2}
  \{l\} \{l_1\} \{l_2\}
  \sum_{s} (-1)^s \sqrt{\{s\}}
  \\ \times
  \begin{Bmatrix}
    s_1 & s_2 & s \\
    l_2 & l_1 & l
  \end{Bmatrix}
  \sum_{\substack{l',l_1',l_2'\\l_z,l_{z1},l_{z2}}}
  (-i)^{l_1'-l_2'} \{l_1'\}\{l_2'\} \sqrt{\{l_z\}}
  \begin{pmatrix}
    l_1' & l_2' & l' \\
    0 & 0 & 0
  \end{pmatrix}
  \\ \times
  \begin{pmatrix}
    l_1 & l_{z1} & l_1' \\
    0 & 0 & 0
  \end{pmatrix}
  \begin{pmatrix}
    l_2 & l_{z2} & l_2' \\
    0 & 0 & 0
  \end{pmatrix}
  \begin{Bmatrix}
    l_1 & l_2 & s \\
    l_1' & l_2' & l' \\
    l_{z1} & l_{z2} & l_z
  \end{Bmatrix}
  \\ \times
  \int \frac{k^2dk}{2\pi^2}
  j_{l_1'}(kr_1) j_{l_2'}(kr_2)
  P^{s_1s_2;s\,l_z}_{X_1X_2;l'l_{z1}l_{z2}}(k).
  \label{eq:155}
\end{multline}
Once the lhs of Eq.~(\ref{eq:152}) is evaluated by the perturbation
theory, the invariant function
$P^{s_1s_2;s\,l_z}_{X_1X_2;l\,l_{z1}l_{z2}}(k)$ is uniquely
determined, and finally the invariant spectrum
$C^{s_1s_2;l_1l_2}_{X_1X_2l}(r_1,r_2)$ in the perturbation theory is
evaluated by the above equation.

\subsection{%
  The linear power spectra
}

In the lowest-order approximation, or in linear theory of
gravitational evolution, the power spectrum of the iPT is given by
\cite{Matsubara:2011ck}
\begin{equation}
  P^{(s_1s_2)}_{X_1X_2\sigma_1\sigma_2}(\bm{k};\hat{\bm{z}}_1,\hat{\bm{z}}_2)
  = i^{s_1+s_2}
  \Gamma^{(1)}_{X_1s_1\sigma_1}(\bm{k};\hat{\bm{z}}_1)
  \Gamma^{(1)}_{X_2s_2\sigma_2}(-\bm{k};\hat{\bm{z}}_2)
  P_\mathrm{L}(k),
  \label{eq:156}
\end{equation}
where $P_\mathrm{L}(k)$ is the linear matter power spectrum, and
$\Gamma^{(1)}_{Xs\sigma}(\bm{k},\hat{\bm{z}})$ is the first-order
propagator of the irreducible tensor field in redshift space with a
line of sight $\hat{\bm{z}}$ \cite{paperI}. Explicitly we have
\begin{equation}
  \label{eq:157}
  \Gamma^{(1)}_{Xs\sigma}(\bm{k},\hat{\bm{z}})
  = \sum_{l_z,l}
  \Gamma^{(1)s\,l_z}_{Xl}(k)
  X^{l_zl}_{s\sigma}(\hat{\bm{z}},\hat{\bm{k}})
\end{equation}
in a form of expansion by bipolar spherical harmonics, where
\begin{multline}
  \Gamma^{(1)s\,l_z}_{Xl}(k) =
  \delta_{l_z0}\delta_{ls}\, c^{(1)}_{Xs}(k)
  \\
  + \delta_{s0} c^{(0)}_X
  \left[
    \left(1 + \frac{f}{3}\right)
    \delta_{l_z0}\delta_{l0} +
    \frac{2\sqrt{5}f}{3} \delta_{l_z2}\delta_{l2}
  \right]
  \label{eq:158}
\end{multline}
is the invariant propagator in redshift space,
$f = d\ln D(t)/d\ln a(t)$ is the linear growth rate, $a(t)$, $D(t)$,
respectively, are the scale factor and the linear growth factor, and
$c^{(0)}_X$, $c^{(1)}_{Xs}(k)$ are invariant renormalized bias
functions of zeroth and first orders, which are defined in Paper~I. In
Paper~I, we consider the general case that the propagators can
explicitly depend on the direction cosine
$\mu=\hat{\bm{k}}\cdot\hat{\bm{z}}$. In the above, we expand all the
direction dependence in the propagator and the propagator does not
depend on $\mu$. We consider literally the lowest-order perturbations,
and do not include the resummation factor $\Pi(k,\mu)$ explained in
Paper~I.

The Kronecker's delta symbols in Eq.~(\ref{eq:158}) ensure that the
indices always satisfy $l=s+l_z$. Defining
$G_{Xs\,l_z}(k) \equiv \Gamma^{(1)s\,l_z}_{X\,s+l_z}(k)$,
Eq.~(\ref{eq:158}) reduces to
\begin{equation}
  \Gamma^{(1)s\,l_z}_{Xl}(k) = \delta_{l,s+l_z} G_{Xs\,l_z}(k),
  \label{eq:159}
\end{equation}
where
\begin{equation}
  G_{Xs\,l_z}(k) =
  \delta_{l_z0}\, c^{(1)}_{Xs}(k)
  + \delta_{s0} c^{(0)}_X
  \left[
    \left(1 + \frac{f}{3}\right) \delta_{l_z0} +
    \frac{2\sqrt{5}f}{3} \delta_{l_z2}
  \right].
  \label{eq:160}
\end{equation}
The index $l_z$ above takes only $0$ and $2$, and is an even number.

Substituting Eq.~(\ref{eq:157}) into Eq.~(\ref{eq:156}), and
applying formulas of $3j$- and $9j$-symbols, products spherical
harmonics, etc.~(see, e.g., Appendixes of Paper~I), the direction
dependencies of the power spectrum of Eq.~(\ref{eq:156}) are expanded
in a form of Eq.~(\ref{eq:152}), from which the invariant power
spectrum can be read off. The result is given by
\begin{multline}
  P^{s_1s_1;s\,l_z}_{X_1X_2;l\,l_{z1}l_{z2}}(k)
  =
  (-1)^{s+l_z+s_2} \{l\} \sqrt{\{s\}\{l_z\}}
  \\ \times
  \begin{pmatrix}
    s_1+l_{z1} & s_2+l_{z2} & l \\ 0 & 0 & 0
  \end{pmatrix}
  \begin{Bmatrix}
    s_1 & s_2 & s \\
    l_{z1} & l_{z2} & l_z \\
    s_1+l_{z1} & s_2+l_{z2} & l
  \end{Bmatrix}
  \\ \times
  G_{X_1s_1l_{z1}}(k) G_{X_2s_2l_{z2}}(k)
  P_\mathrm{L}(k).
  \label{eq:161}
\end{multline}
Substituting the above equation into Eq.~(\ref{eq:155}), and using
various kinds of sum rules involving $3j$-, $6j$-, and $9j$-symbols
\cite{Khersonskii:1988krb}, one can show that the invariant spectrum
of linear theory finally reduces to
\begin{multline}
  C^{s_1s_2;l_1l_2}_{X_1X_2\,l}(r_1,r_2) =
  (-i)^{l_1+l_2} (-1)^{s_1+s_2+l}\{l\}\{l_1\}\{l_2\}
  \\ \times
  \sum_{l_1',l_2',l_{z1},l_{z2}}
  (-i)^{l_1'+l_2'}\{l_1'\}\{l_2'\}
  \begin{pmatrix}
    l_1 & l_{z1} & l_1' \\ 0 & 0 & 0
  \end{pmatrix}
  \begin{pmatrix}
    l_2 & l_{z2} & l_2' \\ 0 & 0 & 0
  \end{pmatrix}
  \\ \times
  \begin{pmatrix}
    l & l_1' & s_1+l_{z1} \\ 0 & 0 & 0
  \end{pmatrix}
  \begin{pmatrix}
    l & l_2' & s_2+l_{z2} \\ 0 & 0 & 0
  \end{pmatrix}
  \\ \times
  \begin{Bmatrix}
    s_1 & l_{z1} & s_1+l_{z1} \\ l_1' & l & l_1
  \end{Bmatrix}
  \begin{Bmatrix}
    s_2 & l_{z2} & s_2+l_{z2} \\ l_2' & l & l_2
  \end{Bmatrix}
  \\ \times
  \int \frac{k^2dk}{2\pi^2}
  j_{l_1'}(kr_1) j_{l_2'}(kr_2)
  G_{X_1s_1l_{z1}}(k) G_{X_2s_2l_{z2}}(k)
  P_\mathrm{L}(k).
  \label{eq:162}
\end{multline}

With the last equation, the power spectra
$C^{(s_1s_2)l_1l_2;m_1m_2}_{X_1X_2\sigma_1\sigma_2}(r_1,r_2)$,
$\tilde{C}^{(s_1s_2)\,l}_{X_1X_2\sigma_1\sigma_2}(r_1,r_2)$, and
correlation functions
$\xi^{(s_1s_2)}_{X_1X_2\sigma_1\sigma_2}(r_1,r_2,\theta)$,
$\tilde{\xi}^{(s_1s_2)}_{X_1X_2\sigma_1\sigma_2}(r_1,r_2,\theta)$ are
predicted by Eqs.~(\ref{eq:68}), (\ref{eq:75}), (\ref{eq:87}) and
(\ref{eq:101}). Among them, the multipole power spectrum in spherical
coordinates,
$\tilde{C}^{(s_1s_2)\,l}_{X_1X_2\sigma_1\sigma_2}(r_1,r_2)$, reduces
to a form without $6j$-symbols. Applying the sum rules involving $3j$-
and $6j$-symbols \cite{Khersonskii:1988krb}, the result is given by
\begin{multline}
  \tilde{C}^{(s_1s_2)\,l}_{X_1X_2\sigma_1\sigma_2}(r_1,r_2)
  = (-i)^{s_1+s_2} (-1)^l
  \{l\}
  \sum_{l_1,l_2,l_{z1},l_{z2}}
  (-i)^{l_1+l_2}\{l_1\}\{l_2\}
  \\ \times
  \begin{bmatrix}
    l & l_1 & s_1+l_{z1} \\ \sigma_1 & 0 & -\sigma_1
  \end{bmatrix}
  \begin{bmatrix}
    l & l_2 & s_2+l_{z2} \\ \sigma_2 & 0 & -\sigma_2
  \end{bmatrix}
  \\ \times
  \begin{pmatrix}
    s_1 & l_{z1} & s_1+l_{z1} \\ \sigma_1 & 0 & -\sigma_1
  \end{pmatrix}
  \begin{pmatrix}
    s_2 & l_{z2} & s_2+l_{z2} \\ \sigma_2 & 0 & -\sigma_2
  \end{pmatrix}
  \\ \times
  \int \frac{k^2dk}{2\pi^2}
  j_{l_1}(kr_1) j_{l_2}(kr_2)
  G_{X_1s_1l_{z1}}(k) G_{X_2s_2l_{z2}}(k) P_\mathrm{L}(k).
  \label{eq:163}
\end{multline}
In the above equation, we employ a notation,
\begin{equation}
  \begin{bmatrix}
    l_1 & l_2 & l_3 \\ m_1 & m_2 & m_3
  \end{bmatrix}
  \equiv
  \begin{pmatrix}
    l_1 & l_2 & l_3 \\ 0 & 0 & 0
  \end{pmatrix}
  \begin{pmatrix}
    l_1 & l_2 & l_3 \\ m_1 & m_2 & m_3
  \end{pmatrix},
  \label{eq:164}
\end{equation}
which is totally symmetric under permutations of the three columns,
and nonzero only when $l_1+l_2+l_3=\mathrm{even}$. Substituting
Eq.~(\ref{eq:160}), the summations over $l_{z1}$ and $l_{z2}$ are
trivially taken. The above results are derived in redshift space from
the beginning of the derivation. The results in real space are
trivially derived by substituting $l_z=0$ and $f=0$ in
Eq.~(\ref{eq:160}).
  
For the reader's convenience, we note below more explicit
forms of the results. In the form of the linear propagator,
Eq.~(\ref{eq:160}), it is convenient to separately consider the two
cases, $s=0$ and $s\ge 1$, as the last term on the rhs of
Eq.~(\ref{eq:160}) vanishes when $s \ge 1$. Accordingly, it is also
convenient to separately write down the formula of the power
spectrum in three cases, $\tilde{C}^{(00)\,l}_{X_1X_200}$,
$\tilde{C}^{(s0)\,l}_{X_1X_2\sigma 0}$ and
$\tilde{C}^{(s_1s_2)\,l}_{X_1X_2\sigma_1\sigma_2}$ with
$s,s_1,s_2 \geq 1$.

When both $X_1$ and $X_2$ are scalar fields, we have $s_1=s_2=0$ in
Eq.~(\ref{eq:163}). The result of substituting Eq.~(\ref{eq:160}) into
Eq.~(\ref{eq:163}) in this case is given by
\begin{multline}
  \frac{\tilde{C}^{(00)\,l}_{X_1X_200}(r_1,r_2)}{2l+1}
  = I^{(00)}_{X_1X_2ll}(r_1,r_2)
  + \left(1 + \frac{f}{3}\right)^2 c^{(0)}_{X_1} c^{(0)}_{X_2}
  I_{ll}(r_1,r_2)
  \\
  + \left(1 + \frac{f}{3}\right) c^{(0)}_{X_1}
  I^{(0)}_{X_2ll}(r_1,r_2) + (1\leftrightarrow 2)
  \\
  + \frac{2f}{3}  c^{(0)}_{X_1}
  \sum_{l_1} (-1)^{(l_1-l)/2}(2l_1+1)
  \begin{pmatrix} l & l_1 & 2 \\ 0 & 0 & 0 \end{pmatrix}^2
  \\ \times
  \left[
    I^{(0)}_{X_2l_1l}(r_1,r_2) +
    \left(1 + \frac{f}{3}\right) c^{(0)}_{X_2} I_{l_1l}(r_1,r_2)
  \right] + (1\leftrightarrow 2)
  \\
  + \frac{4f^2}{9} c^{(0)}_{X_1} c^{(0)}_{X_2}
  \sum_{l_1,l_2} (-1)^{(l_1+l_2-2l)/2} (2l_1+1) (2l_2+1)
  \\ \times
  \begin{pmatrix} l & l_1 & 2 \\ 0 & 0 & 0 \end{pmatrix}^2
  \begin{pmatrix} l & l_2 & 2 \\ 0 & 0 & 0 \end{pmatrix}^2
  I_{l_1l_2}(r_1,r_2),
  \label{eq:165}
\end{multline}
where $+\,(1\leftrightarrow 2)$ indicates adding a previous term with
replacements $X_1 \leftrightarrow X_2$, $l_1 \leftrightarrow l_2$ and
$r_1 \leftrightarrow r_2$, and we define integrals,
\begin{align}
  I_{l_1l_2}(r_1,r_2)
  &\equiv \int \frac{k^2dk}{2\pi^2} j_{l_1}(kr_1) j_{l_2}(kr_2)
    P_\mathrm{L}(k),
  \label{eq:166}\\
  I_{Xl_1l_2}^{(s)}(r_1,r_2)
  &\equiv \int \frac{k^2dk}{2\pi^2} j_{l_1}(kr_1) j_{l_2}(kr_2)
   c^{(1)}_{Xs}(k) P_\mathrm{L}(k),
  \label{eq:167}\\
  I_{X_1X_2l_1l_2}^{(s_1s_2)}(r_1,r_2)
  &\equiv \int \frac{k^2dk}{2\pi^2} j_{l_1}(kr_1) j_{l_2}(kr_2)
    \nonumber \\
  & \hspace{4.8pc} \times
    c^{(1)}_{X_1s_1}(k) c^{(1)}_{X_2s_2}(k) P_\mathrm{L}(k).
  \label{eq:168}
\end{align}

When $X_1$ is a scalar field and $X_2$ is a nonscalar field, we have
$s_1=0$ and $s_2=s$ where $s \geq 1$ in Eq.~(\ref{eq:163}). The
explicit form in this case is given by
\begin{multline}
  \frac{\tilde{C}^{(s0)\,l}_{X_1X_2\sigma 0}(r_1,r_2)}{2l+1}
  = \frac{(-1)^\sigma i^s}{\sqrt{2s+1}}
  \sum_{l_1} (-i)^{l_1}(2l_1+1)
  \begin{bmatrix}
    s & l_1 & l \\ \sigma & 0 & -\sigma
  \end{bmatrix}
  \\ \times
  \left[
    i^l I^{(s0)}_{X_1X_2l_1l}(r_1,r_2)
    + i^l c^{(0)}_{X_2} \left(1+\frac{f}{3}\right)
    I^{(s)}_{X_1l_1l}(r_1,r_2)
  \right.
  \\
  \left.
    + \frac{2f}{3} c^{(0)}_{X_2}
    \sum_{l_2} i^{l_2} (2l_2+1)
    \begin{pmatrix}
      l & l_2 & 2 \\ 0 & 0 & 0
    \end{pmatrix}^2
    I^{(s)}_{X_1l_1l_2}(r_1,r_2)
  \right].
  \label{eq:169}
\end{multline}
When both $X_1$ and $X_2$ are nonscalar fields, we have
$s_1, s_2 \geq 1$ in Eq.~(\ref{eq:163}). The explicit form in this
case is given by
\begin{multline}
  \frac{\tilde{C}^{(s_1s_2)\,l}_{X_1X_2\sigma_1\sigma_2}(r_1,r_2)}{2l+1}
  = \frac{(-1)^{l+\sigma_1+\sigma_2} i^{s_1+s_2}}{\sqrt{(2s_1+1)(2s_2+1)}}
  \sum_{l_1,l_2} (-i)^{l_1+l_2}
  \\ \times
  (2l_1+1)(2l_2+1)
  \begin{bmatrix}
    s_1 & l_1 & l \\ \sigma_1 & 0 & -\sigma_1
  \end{bmatrix}
  \begin{bmatrix}
    s_2 & l_2 & l \\ \sigma_2 & 0 & -\sigma_2
  \end{bmatrix}
  \\ \times
  I^{(s_1s_2)l_1l_2}_{X_1X_2}(r_1,r_2).
  \label{eq:170}
\end{multline}

\subsection{The linear correlation function}

The numerical evaluations of the linear correlation function are
conveniently evaluated by applying the expression of
Eq.~(\ref{eq:121}), rather than using a formal relation of
Eq.~(\ref{eq:108}) between the correlation function and power spectrum
in spherical coordinates.

Identifying the virtual correlation function in Eq.~(\ref{eq:149})
with that of the redundant function in Eq.~(\ref{eq:114}), the
decomposition of the power spectrum is given by the inverse Fourier
transform of Eq.~(\ref{eq:151}) with Eq.~(\ref{eq:117}). Applying the
Rayleigh expansion formula, the following result is derived:
\begin{multline}
  P^{(s_1s_2)}_{X_1X_2\sigma_1\sigma_2}
  (\bm{k};\hat{\bm{z}}_1,\hat{\bm{z}}_2)
  = i^{s_1+s_2} \sum_{s,\sigma} (2s+1)
  \left(s_1\,s_2\,s\right)_{\sigma_1\sigma_2}^{\phantom{\sigma_1\sigma_2}\sigma}
  \\ \times
  \sum_{l,l_{z1},l_{z2},l_z}
  (-1)^{l+
    l_z}  \sqrt{2l_z+1}\,
  X^{l\,l_{z1}l_{z2}}_{l_z;s\sigma}(\hat{\bm{k}},\hat{\bm{z}}_1,\hat{\bm{z}}_2)
  \\ \times
  \sum_{l_1,l_2} (-1)^{l_1}
  \begin{Bmatrix}
    s_1 & s_2 & s \\
    l_{z1} & l_{z2} & l_z \\
    l_1 & l_2 & l
  \end{Bmatrix}
  \mathcal{P}^{s_1s_2;l_1l_2l}_{X_1X_2l_{z1}l_{z2}}(k),
  \label{eq:171}
\end{multline}
where
\begin{equation}
  \mathcal{P}^{s_1s_2;l_1l_2l}_{X_1X_2l_{z1}l_{z2}}(k) =
  4\pi \int x^2dx\, j_l(kx)\, \Xi^{s_1s_2;l_1l_2l}_{X_1X_2l_{z1}l_{z2}}(x).
  \label{eq:172}
\end{equation}
Substituting Eq.~(\ref{eq:161}) into Eq.~(\ref{eq:152}), the resulting
expression for the power spectrum really has the form of
Eq.~(\ref{eq:171}) with the invariant spectrum there given by
\begin{multline}
  \mathcal{P}^{s_1s_2;l_1l_2l}_{X_1X_2l_{z1}l_{z2}}(k)
  = \delta_{l_1,s_1+l_{z1}} \delta_{l_2,s_2+l_{z2}}
  (2l+1)
  \begin{pmatrix}
    l_1 & l_2 & l \\ 0 & 0 & 0
  \end{pmatrix}
  \\ \times
  G_{X_1s_1l_{z1}}(k) G_{X_2s_2l_{z2}}(k) P_\mathrm{L}(k).
  \label{eq:173}
\end{multline}
Applying the inverse Hankel transform of Eq.~(\ref{eq:172}), we have
\begin{multline}
  \Xi^{s_1s_2;l_1l_2l}_{X_1X_2l_{z1}l_{z2}}(x)
  = \delta_{l_1,s_1+l_{z1}} \delta_{l_2,s_2+l_{z2}}
  (2l+1)
  \begin{pmatrix}
    l_1 & l_2 & l \\ 0 & 0 & 0
  \end{pmatrix}
  \\ \times
  \int \frac{k^2dk}{2\pi} j_l(kx)
  G_{X_1s_1l_{z1}}(k) G_{X_2s_2l_{z2}}(k) P_\mathrm{L}(k),
  \label{eq:174}
\end{multline}
and Eq.~(\ref{eq:120}) in this case reduces to
\begin{multline}
  \tilde{\Xi}^{s_1s_2;l_1l_2l}_{X_1X_2\sigma_1\sigma_2}(x)
  = (2l+1)
  \begin{pmatrix}
    l_1 & l_2 & l \\ 0 & 0 & 0
  \end{pmatrix}
  \\ \times
  \begin{pmatrix}
    s_1 & l_1-s_1 & l_1 \\ \sigma_1 & 0 & -\sigma_1
  \end{pmatrix}
  \begin{pmatrix}
    s_2 & l_2-s_2 & l_2 \\ \sigma_2 & 0 & -\sigma_2
  \end{pmatrix}
  \\ \times
  \int \frac{k^2dk}{2\pi} j_l(kx)
  \tilde{G}_{X_1s_1l_1}(k) \tilde{G}_{X_2s_2l_2}(k) P_\mathrm{L}(k),
  \label{eq:175}
\end{multline}
where
\begin{align}
  \tilde{G}_{Xsl}(k)
  &\equiv G_{Xs,l-s}(k)
    \nonumber \\
  &=
  \delta_{ls}\, c^{(1)}_{Xs}(k)
  + \delta_{s0} c^{(0)}_X
  \left[
    \left(1 + \frac{f}{3}\right) \delta_{l0} +
    \frac{2\sqrt{5}f}{3} \delta_{l2}
  \right],
  \label{eq:176}
\end{align}
or we have
\begin{equation}
  \Gamma^{(1)s\,l_z}_{Xl}(k) = \delta_{l_z,l-s} \tilde{G}_{Xsl}(k).
  \label{eq:177}
\end{equation}
Substituting the above equation into Eqs.~(\ref{eq:119}) and
(\ref{eq:121}), the predictions for the correlation functions in
linear theory are given.

\section{\label{sec:ProjectedTensor}%
Projected tensor fields}

\subsection{The two-dimensionally projected irreducible tensors}

In many practical observations, three-dimensional tensors are
projected onto the sky, and in these cases, we only observe projected
two-dimensional tensors of astronomical objects such as galaxy shapes,
galaxy angular momenta, etc. In Paper~III, the relation between the
irreducible decompositions of three-dimensional tensors and those of
projected two-dimensional tensors are derived in Cartesian
coordinates, assuming the distant-observer limit. The same relation
holds also in spherical coordinates, and we briefly summarize the
result below.

Any symmetric two-dimensional, rank-$s$ tensor
$f_{Xi_1\cdots i_s}$ can be decomposed into traceless parts
$f^{(s)}_{Xi_1\cdots i_s}$, which are decomposed by spherical basis on
the tangential plane of the sphere as
\begin{equation}
  f^{(s)}_{Xi_1\cdots i_s} =
    \tilde{f}^{(+s)}_X \tilde{\mathrm{m}}^+_{i_1} \cdots
    \tilde{\mathrm{m}}^+_{i_s} + 
    \tilde{f}^{(-s)}_X \tilde{\mathrm{m}}^-_{i_1} \cdots
    \tilde{\mathrm{m}}^-_{i_s},
  \label{eq:178}
\end{equation}
where $\tilde{\mathrm{m}}^\pm_i$ are angular components of spherical
basis $\tilde{\mathbf{e}}^\pm$ of Eq.~(\ref{eq:14}), i.e.,
$\tilde{\mathrm{m}}^\pm_i =\tilde{\mathrm{e}}^\pm_i$, and
$\tilde{f}^{(\pm s)}_X$ are the decomposed irreducible tensors. The
relations between irreducible components of the three-dimensional
tensor fields and those of projected two-dimensional tensor fields are
given by (Paper~III)
\begin{equation}
  \tilde{f}^{(\pm s)}_X(\bm{r}) =
  A_s \tilde{F}_{Xs,\pm s}(\bm{r}),
  \label{eq:179}
\end{equation}
where $\tilde{F}_{Xs,\pm s}$ on the rhs corresponds to
$\tilde{F}_{Xs\sigma}$ of Eq.~(\ref{eq:28}) with a substitution
$\sigma=\pm s$. As noted in Paper~III, the above formula is only valid
if the rank $s$ of the traceless part $f^{(s)}_{Xi_1\cdots i_s}$ is
the same as the rank of the original tensor $f_{Xi_1\cdots i_s}$, and
does not hold for trace parts. Because of Eq.~(\ref{eq:58}), the
complex conjugate Eq.~(\ref{eq:179}) is given by
\begin{equation}
  \tilde{f}^{(\pm s) *}_X(\bm{r})
  =  (-1)^s \tilde{f}^{(\mp s)}_X(\bm{r}).
  \label{eq:180}
\end{equation}
Because of Eq.~(\ref{eq:60}), the parity transformation of
Eq.~(\ref{eq:179}) is given by
\begin{equation}
  \tilde{f}^{(\pm s)}_X(\bm{r}) \xrightarrow{\mathbb{P}}
  \tilde{f}^{(\pm s)\prime}_X(\bm{r}) =
  (-1)^{p_X} \tilde{f}^{(\mp s)}_X(-\bm{r}).
  \label{eq:181}
\end{equation}

\subsection{The power spectrum of projected tensors}

\subsubsection{General formulas}

The irreducible tensors are expanded by the
spin-weighted spherical harmonics by
\begin{equation}
  \tilde{f}^{(\pm s)}_X(\bm{r}) =
  \sum_{l,m} \sqrt{\frac{4\pi}{2l+1}}\,
  \tilde{f}^{(\pm s)m}_{Xl}(r) \,\sY{\pm s}{lm}(\hat{\bm{r}}).
  \label{eq:182}
\end{equation}
Comparing the expansion with Eq.~(\ref{eq:50}), we have
\begin{equation}
  \tilde{f}^{(\pm s)m}_{Xl}(r) =
  A_s \tilde{F}^{lm}_{Xs,\pm s}(\bm{r}).
  \label{eq:183}
\end{equation}
It is convenient to define expansion coefficients with lowered
azimuthal indices as
\begin{equation}
  \tilde{f}^{(\pm s)}_{Xlm}(r) \equiv
  \sum_{m'} g^{(s)}_{mm'} \tilde{f}^{(\pm s)m'}_{Xl}(r) =
  (-1)^m \tilde{f}^{(\pm s)\,-m}_{Xl}(r).
  \label{eq:184}
\end{equation}
The symmetric properties of Eq.~(\ref{eq:184}), such as complex
conjugate and parity, are explicitly given in Paper~III.

The two-dimensional irreducible tensors are decomposed into the
so-called E and B modes, which are defined by
\begin{equation}
  \tilde{f}^{(\pm s)}_{Xlm}(r) 
  = \tilde{f}^{\mathrm{E}(s)}_{Xlm}(r) \pm i \tilde{f}^{\mathrm{B}(s)}_{Xlm}(r).
  \label{eq:185}
\end{equation}
or equivalently 
\begin{align}
  \tilde{f}^{\mathrm{E}(s)}_{Xlm}(r)
  &= \frac{1}{2}
    \left[
    \tilde{f}^{(+s)}_{Xlm}(r) + \tilde{f}^{(-s)}_{Xlm}(r)
    \right],
  \label{eq:186}\\
  \tilde{f}^{\mathrm{B}(s)}_{Xlm}(r)
  &= \frac{1}{2i}
    \left[
    \tilde{f}^{(+s)}_{Xlm}(r) - \tilde{f}^{(-s)}_{Xlm}(r)
    \right].
  \label{eq:187}
\end{align}
The two-point statistics of E/B modes are given by (Paper~III)
\begin{align}
  \left\langle
    \tilde{f}^{\mathrm{E}(s_1)}_{X_1lm}(r_1)
    \tilde{f}^{\mathrm{E}(s_2)}_{X_2l'm'}(r_2)
  \right\rangle
  &= \delta_{ll'} g^{(l)}_{mm'} C^{\mathrm{EE}(s_1s_2)}_{X_1X_2l}(r_1,r_2),
  \label{eq:188}\\
  \left\langle
    \tilde{f}^{\mathrm{B}(s_1)}_{X_1lm}(r_1)
    \tilde{f}^{\mathrm{B}(s_2)}_{X_2l'm'}(r_2)
  \right\rangle
  &= \delta_{ll'} g^{(l)}_{mm'} C^{\mathrm{BB}(s_1s_2)}_{X_1X_2l}(r_1,r_2),
  \label{eq:189}\\
  \left\langle
    \tilde{f}^{\mathrm{E}(s_1)}_{X_1lm}(r_1)
    \tilde{f}^{\mathrm{B}(s_2)}_{X_2l'm'}(r_2)
  \right\rangle
  &= \delta_{ll'} g^{(l)}_{mm'} C^{\mathrm{EB}(s_1s_2)}_{X_1X_2l}(r_1,r_2),
  \label{eq:190}\\
  \left\langle
    \tilde{f}^{\mathrm{B}(s_1)}_{X_1lm}(r_1)
    \tilde{f}^{\mathrm{E}(s_2)}_{X_2l'm'}(r_2)
  \right\rangle
  &= \delta_{ll'} g^{(l)}_{mm'} C^{\mathrm{BE}(s_1s_2)}_{X_1X_2l}(r_1,r_2),
  \label{eq:191}
\end{align}
where $C^{\mathrm{EE}(s_1s_2)}_{X_1X_2l}$,
$C^{\mathrm{BB}(s_1s_2)}_{X_1X_2l}$,
$C^{\mathrm{EB}(s_1s_2)}_{X_1X_2l}$, and
$C^{\mathrm{BE}(s_1s_2)}_{X_1X_2l}$ are E/B power spectra. The last
spectra are rotationally invariant and real functions. Symmetric
properties of the E/B power spectra are explicitly given in Paper~III.

Using the correspondences of Eqs.~(\ref{eq:183}), (\ref{eq:186}) and
(\ref{eq:187}), the power spectra of E/B modes of
Eqs.~(\ref{eq:188})--(\ref{eq:191}) are linearly related to the power
spectra of three-dimensional irreducible tensor fields
$\tilde{C}^{(s_1s_2)l}_{X_1X_2\sigma_1\sigma_2}(r_1,r_2)$, defined by
Eq.~(\ref{eq:74}). The relations are explicitly given by
\begin{align}
  C^{\mathrm{EE}}_{l}
  &= \frac{1}{4}
    \left[
    \left(C^{++}_{l} + C^{--}_{l}\right) + 
    \left(C^{+-}_{l} + C^{-+}_{l}\right)
    \right],
  \label{eq:192}\\
  C^{\mathrm{BB}}_{l}
  &= -\frac{1}{4}
    \left[
    \left(C^{++}_{l} + C^{--}_{l}\right) - 
    \left(C^{+-}_{l} + C^{-+}_{l}\right)
    \right],
  \label{eq:193}\\
  C^{\mathrm{EB}}_{l}
  &= \frac{1}{4i}
    \left[
    \left(C^{++}_{l} - C^{--}_{l}\right) - 
    \left(C^{+-}_{l} - C^{-+}_{l}\right)
    \right],
  \label{eq:194}\\
  C^{\mathrm{BE}}_{l}
  &= \frac{1}{4i}
    \left[
    \left(C^{++}_{l} - C^{--}_{l}\right) + 
    \left(C^{+-}_{l} - C^{-+}_{l}\right)
    \right],
  \label{eq:195}
\end{align}
where abbreviations,
\begin{align}
  &
    C^{\pm \pm}_{l}
    = A_{s_1} A_{s_2}
    \tilde{C}^{(s_1s_2)l}_{X_1X_2,\pm s_1,\pm s_2}(r_1,r_2),
  \label{eq:196}\\
  &
    C^{\pm \mp}_{l}
    = A_{s_1} A_{s_2}
    \tilde{C}^{(s_1,s_2)l}_{X_1X_2,\pm s_1,\mp s_2}(r_1,r_2),
  \label{eq:197}\\
  &
    C^{\mathrm{EE}}_{l} = C^{\mathrm{EE}(s_1s_2)}_{X_1X_2l}(r_1,r_2), \quad
    C^{\mathrm{BB}}_{l} = C^{\mathrm{BB}(s_1s_2)}_{X_1X_2l}(r_1,r_2),
  \label{eq:198}\\
  &
    C^{\mathrm{EB}}_{l} = C^{\mathrm{EB}(s_1s_2)}_{X_1X_2l}(r_1,r_2), \quad
    C^{\mathrm{BE}}_{l} = C^{\mathrm{BE}(s_1s_2)}_{X_1X_2l}(r_1,r_2), 
  \label{eq:199}
\end{align}
are adopted just for simplicity of presentation. Combining
Eqs.~(\ref{eq:77}), (\ref{eq:78}) and (\ref{eq:192})--(\ref{eq:195}),
the symmetries of complex conjugate and parity are given by
\begin{equation}
  C^{\mathrm{EE}*}_l = C^{\mathrm{EE}}_l, \ 
  C^{\mathrm{BB}*}_l = C^{\mathrm{BB}}_l, \ 
  C^{\mathrm{EB}*}_l = C^{\mathrm{EB}}_l, \ 
  C^{\mathrm{BE}*}_l = C^{\mathrm{BE}}_l,
  \label{eq:200}
\end{equation}
i.e., they are all real functions, and
\begin{align}
  C^{\mathrm{EE}}_l
  &\xrightarrow{\mathbb{P}}
  (-1)^{p_{X_1}+p_{X_2}} C^{\mathrm{EE}}_l,
  \label{eq:201}\\
  C^{\mathrm{BB}}_l
  &\xrightarrow{\mathbb{P}}
  (-1)^{p_{X_1}+p_{X_2}} C^{\mathrm{BB}}_l.
  \label{eq:202}\\
  C^{\mathrm{EB}}_l
  &\xrightarrow{\mathbb{P}}
  (-1)^{p_{X_1}+p_{X_2}+1} C^{\mathrm{EB}}_l,
  \label{eq:203}\\
  C^{\mathrm{BE}}_l
  &\xrightarrow{\mathbb{P}}
  (-1)^{p_{X_1}+p_{X_2}+1} C^{\mathrm{BE}}_l.
  \label{eq:204}
\end{align}

Substituting Eq.~(\ref{eq:75}) into
Eqs.~(\ref{eq:192})--(\ref{eq:195}), we derive
\begin{align}
  C^{\mathrm{EE}(s_1s_2)}_{X_1X_2l}(r_1,r_2)
  &= \sum_{\substack{l_1,l_2\\l_1-s_1+l=\mathrm{even}\\l_2-s_2+l=\mathrm{even}}}
  \hspace{-1pc}
  (-1)^{(l_1+l_2-s_1-s_2)/2} \hat{C}^{s_1s_2;l_1l_2}_{X_1X_2l}(r_1,r_2),
  \label{eq:205}\\
  C^{\mathrm{BB}(s_1s_2)}_{X_1X_2l}(r_1,r_2)
  &=
    \sum_{\substack{l_1,l_2\\l_1-s_1+l=\mathrm{odd}\\l_2-s_2+l=\mathrm{odd}}}
  \hspace{-1pc}
  (-1)^{(l_1+l_2-s_1-s_2-2)/2} \hat{C}^{s_1s_2;l_1l_2}_{X_1X_2l}(r_1,r_2),
  \label{eq:206}\\
  C^{\mathrm{EB}(s_1s_2)}_{X_1X_2l}(r_1,r_2)
  &= \sum_{\substack{l_1,l_2\\l_1-s_1+l=\mathrm{even}\\l_2-s_2+l=\mathrm{odd}}}
  \hspace{-1pc}
  (-1)^{(l_1+l_2-s_1-s_2-1)/2} \hat{C}^{s_1s_2;l_1l_2}_{X_1X_2l}(r_1,r_2),
  \label{eq:207}\\
  C^{\mathrm{BE}(s_1s_2)}_{X_1X_2l}(r_1,r_2)
  &= \sum_{\substack{l_1,l_2\\l_1-s_1+l=\mathrm{odd}\\l_2-s_2+l=\mathrm{even}}}
  \hspace{-1pc}
  (-1)^{(l_1+l_2-s_1-s_2-1)/2} \hat{C}^{s_1s_2;l_1l_2}_{X_1X_2l}(r_1,r_2),
  \label{eq:208}
\end{align}
where
\begin{multline}
  \hat{C}^{s_1s_2;l_1l_2}_{X_1X_2l}(r_1,r_2) \equiv
  A_{s_1}A_{s_2}
  \begin{pmatrix}
    s_1 & l_1 & l \\ s_1 & 0 & -s_1
  \end{pmatrix}
  \begin{pmatrix}
    s_2 & l_2 & l \\ s_2 & 0 & -s_2
  \end{pmatrix}
  \\ \times
  C^{s_1s_2;l_1l_2}_{X_1X_2l}(r_1,r_2).
  \label{eq:209}
\end{multline}
The above results explicitly give the formula for the power spectra of
projected E/B modes represented in terms of the invariant spectrum of
three-dimensional tensor fields.

\subsubsection{Relation to the power spectrum in the distant-observer
  limit}

The power spectrum $P^{(s_1s_2)}_{X_1X_2\sigma_1\sigma_2}(\bm{k})$ of
the tensor fields in the distant-observer limit, defined by
Eq.~(\ref{eq:135}), is related to the full-sky power spectrum
$\tilde{C}^{(s_1s_2)l}_{X_1X_2\sigma_1\sigma_2}(r_1,r_2)$ by
Eqs.~(\ref{eq:136}) and (\ref{eq:137}). These relations are
transformed to the relations between the power spectra of E/B modes,
where the power spectra of E/B modes are defined in Paper~III. The
calculations are straightforward, and the results are naturally given
by
\begin{multline}
  P^{\mathrm{EE/BB/EB/BE}(s_1s_2)}_{X_1X_2}(\bm{k})
  \\
  \approx
  \frac{2\pi r_0}{k_\perp} \int dx\,e^{-ik_\parallel x}
  \tilde{C}^{\mathrm{EE/BB/EB/BE}(s_1s_2)}_{X_1X_2l}(x+r_0,r_0),
  \label{eq:210}
\end{multline}
and
\begin{multline}
  C^{\mathrm{EE/BB/EB/BE}(s_1s_2)}_{X_1X_2l}(r_1,r_2)
  \\
  \approx \frac{k_\perp}{2\pi r_0}
  \int \frac{dk_\parallel}{2\pi}\,e^{ik_\parallel (r_1-r_2)}
  P^{\mathrm{EE/BB/EB/BE}(s_1s_2)}_{X_1X_2}(\bm{k}),
  \label{eq:211}
\end{multline}
where $|r_1-r_2| \ll r_0 \equiv (r_1+r_2)/2$ and
$l = k_\perp r_0 - 1/2 \gg 1$. In a later section, we use the above
correspondences to check the validity of the distant-observer
approximation in the power spectrum. While the power spectra
$C^{\mathrm{EE/BB/EB/BE}}$ in spherical coordinates are all real
functions as noted above, $P^\mathrm{EE/BB}$ are real and
$P^\mathrm{EB/BE}$ are pure imaginary as shown in Paper~III.
Therefore, the exponential factors in the integrals of
Eqs.~(\ref{eq:210}) and (\ref{eq:211}) can be replaced by
$e^{-ik_\parallel x} \rightarrow \cos(k_\parallel x)$ and
$e^{ik_\parallel (r_1-r_2)} \rightarrow \cos[k_\parallel(r_1-r_2)]$,
respectively for EE/BB power spectra, while they can be replaced by
$e^{-ik_\parallel x} \rightarrow -i\sin(k_\parallel x)$ and
$e^{ik_\parallel (r_1-r_2)} \rightarrow i\sin[k_\parallel (r_1-r_2)]$
for EB/BE power spectra.

\subsubsection{The linear power spectrum}

In the lowest-order approximation with iPT, the power spectra appearing
in Eqs.~(\ref{eq:196}) and (\ref{eq:197}) are given by
Eq.~(\ref{eq:163}) with $(\sigma_1,\sigma_2) = (\pm s_1, \pm s_2)$ and
$(\pm s_1, \mp s_2)$, respectively. Substituting the expression of the
linear theory into Eqs.~(\ref{eq:192}) and (\ref{eq:193}), we have
\begin{align}
  C^{\mathrm{EE}(s_1s_2)}_{X_1X_2l}(r_1,r_2)
  &= \sum_{\substack{l_1,l_2\\l_1-s_1+l=\mathrm{even}\\l_2-s_2+l=\mathrm{even}}}
  \hspace{-1pc}
  (-1)^{(l_1+l_2-s_1-s_2)/2} E^{s_1s_2;l_1l_2}_{X_1X_2l}(r_1,r_2),
  \label{eq:212}\\
  C^{\mathrm{BB}(s_1s_2)}_{X_1X_2l}(r_1,r_2)
  &=
    \sum_{\substack{l_1,l_2\\l_1-s_1+l=\mathrm{odd}\\l_2-s_2+l=\mathrm{odd}}}
  \hspace{-1pc}
  (-1)^{(l_1+l_2-s_1-s_2-2)/2} E^{s_1s_2;l_1l_2}_{X_1X_2l}(r_1,r_2),
  \label{eq:213}\\
  C^{\mathrm{EB}(s_1s_2)}_{X_1X_2l}(r_1,r_2)
  &= \sum_{\substack{l_1,l_2\\l_1-s_1+l=\mathrm{even}\\l_2-s_2+l=\mathrm{odd}}}
  \hspace{-1pc}
  (-1)^{(l_1+l_2-s_1-s_2-1)/2} E^{s_1s_2;l_1l_2}_{X_1X_2l}(r_1,r_2),
  \label{eq:214}\\
  C^{\mathrm{BE}(s_1s_2)}_{X_1X_2l}(r_1,r_2)
  &= \sum_{\substack{l_1,l_2\\l_1-s_1+l=\mathrm{odd}\\l_2-s_2+l=\mathrm{even}}}
  \hspace{-1pc}
  (-1)^{(l_1+l_2-s_1-s_2-1)/2} E^{s_1s_2;l_1l_2}_{X_1X_2l}(r_1,r_2),
  \label{eq:215}
\end{align}
where
\begin{multline}
  E^{s_1s_2;l_1l_2}_{X_1X_2l}(r_1,r_2)
  \equiv
  A_{s_1} A_{s_2} (-1)^{s_1+s_2+l} (2l+1)(2l_1+1)(2l_2+1)
  \\ \times
  \sum_{l_{z1},l_{z2}}
  \begin{bmatrix}
    l & l_1 & l_{z1}+s_1 \\ s_1 & 0 & -s_1
  \end{bmatrix}
  \begin{bmatrix}
    l & l_2 & l_{z2}+s_2 \\ s_2 & 0 & -s_2
  \end{bmatrix}
  \\ \times
  \begin{pmatrix}
    s_1 & l_{z1} & l_{z1}+s_1 \\ s_1 & 0 & -s_1
  \end{pmatrix}
  \begin{pmatrix}
    s_2 & l_{z2} & l_{z2}+s_2 \\ s_2 & 0 & -s_2
  \end{pmatrix}
  \\ \times
  \int \frac{k^2dk}{2\pi^2}
  j_{l_1}(kr_1) j_{l_2}(kr_2)
  G_{X_1s_1l_{z1}}(k) G_{X_2s_2l_{z2}}(k)
  P_\mathrm{L}(k),
  \label{eq:216}
\end{multline}
and $G_{Xsl_z}(k)$ is given by Eq.~(\ref{eq:160}). 
In deriving the above results, we use the properties that $l_{z1}$
and $l_{z2}$ are even numbers in Eq.~(\ref{eq:160}). While the
expressions of Eqs.~(\ref{eq:212})--(\ref{eq:215}) have the same forms
with Eqs.~(\ref{eq:205})--(\ref{eq:208}), the summands
$E^{s_1s_2;l_1l_2}_{X_1X_2l}$ are not the counterparts of
$\hat{C}^{s_1s_2;l_1l_2}_{X_1X_2l}$ in linear theory. They give the
same results only after the summations over $l_1$ and $l_2$ are taken.
The above results can also be derived by
Eqs.~(\ref{eq:205})--(\ref{eq:209}), where the invariant power
spectrum $C^{s_1s_2;l_1l_2}_{X_1X_2l}(r_1,r_2)$ is given by
Eq.~(\ref{eq:162}) in linear theory, applying appropriate sum rules
involving $3j$- and $6j$-symbols \cite{Khersonskii:1988krb}.

The summand in the expression of Eq.~(\ref{eq:216}) survives only when
$l_1-s_1+l=\mathrm{even}$ and $l_2-s_2+l=\mathrm{even}$, because
$G_{Xsl_z}(k)$ is nonzero only when $l_z=\mathrm{even}$ and the
symmetry of $3j$-symbols ensures that $l+l_1+l_{z1}+s_1$ and
$l+l_2+l_{z2}+s_2$ are even numbers in order to have nonzero values.
Therefore, BB/EB/BE power spectra vanish in linear theory:
\begin{equation}
  C^{\mathrm{BB}(s_1s_2)}_{X_1X_2l}(r_1,r_2) =
  C^{\mathrm{EB}(s_1s_2)}_{X_1X_2l}(r_1,r_2) =
  C^{\mathrm{BE}(s_1s_2)}_{X_1X_2l}(r_1,r_2) = 0.
  \label{eq:217}
\end{equation}
This property is the general prediction of the iPT in the lowest-order
approximation with renormalized bias functions, and is consistent with
the results derived in Paper~III with the distant-observer
approximation, in which the BB/EB/BE power spectra vanish:
$P^{\mathrm{BB}(s_1s_2)}_{X_1X_2}(\bm{k}) =
P^{\mathrm{EB}(s_1s_2)}_{X_1X_2}(\bm{k}) =
P^{\mathrm{BE}(s_1s_2)}_{X_1X_2}(\bm{k}) = 0$, and only EE power
spectrum $P^{\mathrm{BB}(s_1s_2)}_{X_1X_2}(\bm{k})$ survives in the
lowest-order approximation.\footnote{When the parity symmetry holds
  both in initial conditions and dynamical evolutions, the EB power
  spectra always vanish. The results here state more than that
  without assuming the parity symmetry, and hold only in linear
  theory.}

Substituting Eqs.~(\ref{eq:160}) and (\ref{eq:216}) into
Eq.~(\ref{eq:212}), the EE power spectrum in linear theory is
explicitly given. When both $X_1$ and $X_2$ are scalar fields,
$s_1=s_2=0$, we have
\begin{multline}
  (2l+1)^{-1} C^{\mathrm{EE}(00)}_{X_1X_2l}(r_1,r_2)
  = I^{(00)}_{X_1X_2ll}(r_1,r_2)
  \\
  + \left(1 + \frac{f}{3}\right)
  \left[
    c^{(0)}_{X_1} I^{(0)}_{X_2ll}(r_1,r_2) 
    + c^{(0)}_{X_2} I^{(0)}_{X_1ll}(r_1,r_2)
  \right]
    \\
  + \left(1 + \frac{f}{3}\right)^2 c^{(0)}_{X_1} c^{(0)}_{X_2}
  I_{ll}(r_1,r_2)
  \\
  +  \frac{2f}{3}
  \sum_{l'} (-1)^{(l'-l)/2} (2l'+1)
  \begin{pmatrix}
    l & l' & 2 \\ 0 & 0 & 0
  \end{pmatrix}^2
  \\ \times
  \Biggl\{
    c^{(0)}_{X_1} I^{(0)}_{X_2l'l}(r_1,r_2)
    + c^{(0)}_{X_2} I^{(0)}_{X_1ll'}(r_1,r_2)
    \\
    \hspace{5pc}
    + \left(1 + \frac{f}{3}\right) c^{(0)}_{X_1} c^{(0)}_{X_2}
    \left[
      I_{l'l}(r_1,r_2) + I_{ll'}(r_1,r_2)
      \right]
  \Biggr\}
  \\
  + \frac{4f^2}{9} c^{(0)}_{X_1} c^{(0)}_{X_2}
  \sum_{l_1,l_2} (-1)^{(l_1+l_2-2l)/2} (2l_1+1) (2l_2+1)
  \\ \times
  \begin{pmatrix} l & l_1 & 2 \\ 0 & 0 & 0 \end{pmatrix}^2
  \begin{pmatrix} l & l_2 & 2 \\ 0 & 0 & 0 \end{pmatrix}^2
  I_{l_1l_2}(r_1,r_2),
  \label{eq:218}
\end{multline}
where the integrals defined in Eqs.~(\ref{eq:166})--(\ref{eq:168}) are
employed. When $X_1$ is a nonscalar field and $X_2$ is a scalar
field, we have $s_1=s$ and $s_2=0$ where $s \geq 1$ in
Eq.~(\ref{eq:216}). The explicit form in this case is given by
\begin{multline}
  C^{\mathrm{EE}(s0)}_{X_1X_2l}(r_1,r_2)
  = \frac{(2l+1)A_s}{\sqrt{2s+1}}
  \sum_{l_1} (-1)^{(l_1+s-l)/2} (2l_1+1)
  \\ \times
  \begin{bmatrix}
    s & l_1 & l \\ s & 0 & -s
  \end{bmatrix}
  \left[
    I^{(s0)}_{X_1X_2l_1l}(r_1,r_2)
    + \left(1+\frac{f}{3}\right)
    c^{(0)}_{X_2} I^{(s)}_{X_1l_1l}(r_1,r_2)
  \right.
  \\
  \left.
    + \frac{2f}{3} c^{(0)}_{X_2}
    \sum_{l_2} (-1)^{(l_2-l)/2} (2l_2+1)
    \begin{pmatrix}
      l & l_2 & 2 \\ 0 & 0 & 0
    \end{pmatrix}^2
    I^{(s)}_{X_1l_1l_2}(r_1,r_2)
  \right].
  \label{eq:219}
\end{multline}
When both $X_1$ and $X_2$ are nonscalar fields, we have
$s_1, s_2 \geq 1$ in Eq.~(\ref{eq:216}). The explicit form in this
case is given by
\begin{multline}
  C^{\mathrm{EE}(s_1s_2)}_{X_1X_2l}(r_1,r_2)
  = \frac{(2l+1)A_{s_1}A_{s_2}}{\sqrt{(2s_1+1)(2s_2+1)}}
  \\ \times
  \sum_{l_1,l_2} (-1)^{(l_1+l_2+s_1+s_2-2l)/2} (2l_1+1)(2l_2+1)
  \\ \times
  \begin{bmatrix}
    s_1 & l_1 & l \\ s_1 & 0 & -s_1
  \end{bmatrix}
  \begin{bmatrix}
    s_2 & l_2 & l \\ s_2 & 0 & -s_2
  \end{bmatrix}
  I^{(s_1s_2)}_{X_1X_2l_1l_2}(r_1,r_2).
  \label{eq:220}
\end{multline}
For given values of $l$, $s_1$ and $s_2$, the number of terms in the
summations over $l_1,l_2,l'$ appearing in
Eqs.~(\ref{eq:218})--(\ref{eq:220}) are finite because of the
triangular inequality in $3j$-symbols. Therefore numerical evaluations
of the above expressions are straightforward with
one-dimensional integrations of finite numbers.

\subsection{The correlation function of projected tensors}

\subsubsection{General formulas}

We define the tangential/cross ($+/\times$) modes of the projected
tensors in spherical coordinates by
\begin{equation}
  \tilde{f}^{(\pm s)}_X(\bm{r})
  = (\pm 1)^s
  \left[
    \tilde{f}^{+(s)}_X(\bm{r}) \pm i \tilde{f}^{\times(s)}_X(\bm{r})
  \right]
  \label{eq:221}
\end{equation}
for non-negative integers $s$. Equivalently, we have
\begin{align}
  \tilde{f}^{+(s)}_X(\bm{r})
  &=
    \frac{1}{2}
    \left[
    \tilde{f}^{(+s)}_X(\bm{r}) + (-1)^s \tilde{f}^{(-s)}_X(\bm{r})
    \right],
  \label{eq:222}\\
  \tilde{f}^{\times(s)}_X(\bm{r})
  &=
    \frac{1}{2i}
    \left[
    \tilde{f}^{(+s)}_X(\bm{r}) -
    (-1)^s \tilde{f}^{(-s)}_X(\bm{r})
    \right].
  \label{eq:223}
\end{align}
These definitions for $+/\times$ modes are consistent with those
defined in Paper~III in the distant-observer limit.\footnote{As
  explicitly shown in Paper~III, the $+/\times$ modes in Cartesian
  coordinates are given by
  $f^{(\pm s)}_X(\bm{x}) = e^{\pm is\phi} \tilde{f}^{(\pm
    s)}_X(\bm{r})$.} Because of Eq.~(\ref{eq:180}), complex conjugates
of Eqs.~(\ref{eq:222}) and (\ref{eq:223}) are given by
\begin{equation}
  \tilde{f}^{+(s)*}_X(\bm{r}) =
  \tilde{f}^{+(s)}_X(\bm{r}), \quad
  \tilde{f}^{\times (s)*}_X(\bm{r}) =
  \tilde{f}^{\times (s)}_X(\bm{r}),
  \label{eq:224}
\end{equation}
i.e., they are real functions. Because of Eq.~(\ref{eq:181}), parity
transformations of Eqs.~(\ref{eq:222}) and (\ref{eq:223}) are given by
\begin{align}
  \tilde{f}^{+(s)}_X(\bm{r})
  &\xrightarrow{\mathbb{P}}
  \tilde{f}^{+(s)\prime}_X(\bm{r}) =
  (-1)^{p_X+s} \tilde{f}^{+(s)}_X(-\bm{r}),
  \label{eq:225}\\
  \tilde{f}^{\times(s)}_X(\bm{r})
  &\xrightarrow{\mathbb{P}}
  \tilde{f}^{\times(s)\prime}_X(\bm{r}) =
  (-1)^{p_X+s+1} \tilde{f}^{\times(s)}_X(-\bm{r}).
  \label{eq:226}
\end{align}

The two-point correlation functions of $+/\times$ modes in spherical
coordinates can be defined by
\begin{align}
  \xi^{++(s_1s_2)}_{X_1X_2}(\bm{r}_1,\bm{r}_2)
  &\equiv
    \left\langle
    \tilde{f}^{+(s_1)}_{X_1}(\bm{r}_1)
    \tilde{f}^{+(s_2)}_{X_2}(\bm{r}_2)
    \right\rangle,
  \label{eq:227}\\
  \xi^{\times\times(s_1s_2)}_{X_1X_2}(\bm{r}_1,\bm{r}_2)
  &\equiv
    \left\langle
    \tilde{f}^{\times(s_1)}_{X_1}(\bm{r}_1)
    \tilde{f}^{\times(s_2)}_{X_2}(\bm{r}_2)
    \right\rangle,
  \label{eq:228}\\
  \xi^{+\times(s_1s_2)}_{X_1X_2}(\bm{r}_1,\bm{r}_2)
  &\equiv
    \left\langle
    \tilde{f}^{+(s_1)}_{X_1}(\bm{r}_1)
    \tilde{f}^{\times(s_2)}_{X_2}(\bm{r}_2)
    \right\rangle,
  \label{eq:229}\\
  \xi^{\times+(s_1s_2)}_{X_1X_2}(\bm{r}_1,\bm{r}_2)
  &\equiv
    \left\langle
    \tilde{f}^{\times(s_1)}_{X_1}(\bm{r}_1)
    \tilde{f}^{+(s_2)}_{X_2}(\bm{r}_2)
    \right\rangle.
  \label{eq:230}
\end{align}
Because of the properties of Eq.~(\ref{eq:224}), the correlation
functions of $+/\times$ modes defined above are all real functions.
These functions are linearly related to the correlation functions of
three-dimensional tensors of Eq.~(\ref{eq:89}) through
Eqs.~(\ref{eq:179}), (\ref{eq:221})--(\ref{eq:223}). Explicit
relations are straightforwardly derived, and the results are given by
\begin{align}
  \xi^{++}
  &= \frac{1}{4}
    \left[
    \left(\tilde{X}^{++} + \tilde{X}^{--}\right) + 
    \left(\tilde{X}^{+-} + \tilde{X}^{-+}\right)
    \right],
  \label{eq:231}\\
  \xi^{\times\times}
  &= -\frac{1}{4}
    \left[
    \left(\tilde{X}^{++} + \tilde{X}^{--}\right) - 
    \left(\tilde{X}^{+-} + \tilde{X}^{-+}\right)
    \right],
  \label{eq:232}\\
  \xi^{+\times}
  &= \frac{1}{4i}
    \left[
    \left(\tilde{X}^{++} - \tilde{X}^{--}\right) - 
    \left(\tilde{X}^{+-} - \tilde{X}^{-+}\right)
    \right],
  \label{eq:233}\\
  \xi^{\times+}
  &= \frac{1}{4i}
    \left[
    \left(\tilde{X}^{++} - \tilde{X}^{--}\right) + 
    \left(\tilde{X}^{+-} - \tilde{X}^{-+}\right)
    \right],
  \label{eq:234}
\end{align}
where the abbreviations,
\begin{align}
  &
    \tilde{X}^{\pm \pm} =
    (\pm i)^{s_1+s_2} A_{s_1} A_{s_2}
    \tilde{\xi}^{(s_1s_2)}_{X_1X_2,\pm s_1,\pm s_2}(\bm{r}_1,\bm{r}_2),
  \label{eq:235}\\
  &
    \tilde{X}^{\pm \mp} =
    (\pm i)^{s_1-s_2} A_{s_1} A_{s_2}
    \tilde{\xi}^{(s_1s_2)}_{X_1X_2,\pm s_1,\mp s_2}(\bm{r}_1,\bm{r}_2),
  \label{eq:236}\\
  &
    \xi^{++} =
    \xi^{++(s_1s_2)}_{X_1X_2}(\bm{r}_1,\bm{r}_2), \quad 
    \xi^{+\times} =
    \xi^{+\times(s_1s_2)}_{X_1X_2}(\bm{r}_1,\bm{r}_2), 
  \label{eq:237}\\
  &
    \xi^{\times+}
    = \xi^{\times+(s_1s_2)}_{X_1X_2}(\bm{r}_1,\bm{r}_2), \quad
    \xi^{\times\times}
    = \xi^{\times\times(s_1s_2)}_{X_1X_2}(\bm{r}_1,\bm{r}_2),
  \label{eq:238}
\end{align}
are adopted just for simplicity of representation. The correlation
functions in Eqs.~(\ref{eq:235}) and (\ref{eq:236}) are given by
$\tilde{\xi}^{(s_1s_2)}_{X_1X_2\sigma_1\sigma_2}$ of Eq.~(\ref{eq:89})
with $(\sigma_1,\sigma_2) = (\pm s_1, \pm s_2)$ and
$(\pm s_1, \mp s_2)$, respectively. Substituting Eq.~(\ref{eq:91})
into Eqs.~(\ref{eq:235}) and (\ref{eq:236}),
Eqs.~(\ref{eq:231})--(\ref{eq:234}) give the correlation functions of
$+/\times$ modes of the projected tensors. Because of
Eq.~(\ref{eq:99}), the properties of complex conjugates of
Eqs.~(\ref{eq:235}) and (\ref{eq:236}) are simply given by
$\tilde{X}^{\pm\pm*} = \tilde{X}^{\mp\mp}$ and
$\tilde{X}^{\pm\mp*} = \tilde{X}^{\mp\pm}$. Therefore, the correlation
functions of $+/\times$ modes, Eqs.~(\ref{eq:231})--(\ref{eq:234}) are
all real functions, as noted above.

Instead of the expression of Eq.~(\ref{eq:91}), however, it is
practically convenient to substitute the alternative expression of
Eq.~(\ref{eq:121}) for numerical evaluations of the correlation
function of projected tensors with a special choice of the coordinates
system for each pair of tensors, $(\theta_1,\phi_1) = (\theta,0)$ and
$(\theta_2,\phi_2) = (0,0)$. For convenience, we give explicit
expressions of the ingredients on the rhs of
Eqs.~(\ref{eq:231})--(\ref{eq:234}):
\begin{multline}
  \tilde{X}^{++} \pm \tilde{X}^{--}
  = (-i)^{s_1-s_2} A_{s_1} A_{s_2}
  \sum_{l,m}
  \left[
    1 \pm (-1)^{l-s_1-s_2}
  \right]
  i^l \tilde{P}_l^m(\mu_2)
  \\ \times
  u^{(l_1)}_{s_1,s_2-m}(\theta)
  \sum_{l_1,l_2} (-1)^{l_2}
  \begin{pmatrix}
    l_1 & l_2 & l \\ m-s_2 & s_2 & -m
  \end{pmatrix}
  \tilde{\xi}^{s_1s_2;l_1l_2}_{X_1X_2l}(x),
  \label{eq:239}
\end{multline}
and
\begin{multline}
  \tilde{X}^{+-} \pm \tilde{X}^{-+}
  = (-i)^{s_1-s_2} A_{s_1} A_{s_2}
  \sum_{l,m}
  \left[
    1 \pm (-1)^{l-s_1-s_2}
  \right]
  i^l \tilde{P}_l^m(\mu_2)
  \\ \times
  u^{(l_1)}_{s_1,-s_2-m}(\theta)
  \sum_{l_1,l_2}
  \begin{pmatrix}
    l_1 & l_2 & l \\ m+s_2 & -s_2 & -m
  \end{pmatrix}
  \tilde{\xi}^{s_1s_2;l_1l_2}_{X_1X_2l}(x),
  \label{eq:240}
\end{multline}
where $\mu_2$ is given by Eq.~(\ref{eq:122}), and
$\tilde{\xi}^{s_1s_2;l_1l_2}_{X_1X_2l}(x)$ is the function
$\tilde{\Xi}^{s_1s_2;l_1l_2l}_{X_1X_2\sigma_1\sigma_2}(x)$ of
Eq.~(\ref{eq:120}) substituted by $\sigma_1=s_1$ and $\sigma_2=s_2$,
i.e.,
\begin{equation}
  \tilde{\xi}^{s_1s_2;l_1l_2}_{X_1X_2l}(x) \equiv
  \sum_{l_{z1},l_{z2}}
  \begin{pmatrix}
    s_1 & l_{z1} & l_1 \\ s_1 & 0 & -s_1
  \end{pmatrix}
  \begin{pmatrix}
    s_2 & l_{z2} & l_2 \\ s_2 & 0 & -s_2
  \end{pmatrix}
  \Xi^{s_1s_2;l_1l_2l}_{X_1X_2l_{z1}l_{z2}}(x).
  \label{eq:241}
\end{equation}
The flip symmetry in redshift space as discussed in the sentences
below Eq.~(\ref{eq:116}) indicates that the summations of $l_{z1}$ and
$l_{z2}$ are taken only over even integers. In the case of the plus
sign of the double sign, i.e., $\tilde{X}^{+\pm} + \tilde{X}^{-\mp}$,
the values are real numbers, because the summations are nonzero only
when $l-s_1-s_2=\mathrm{even}$. Similarly, in the cases of the minus
sign, i.e., $\tilde{X}^{+\pm} - \tilde{X}^{-\mp}$, the values are pure
imaginary numbers, because the summations are nonzero only when
$l-s_1-s_2=\mathrm{odd}$. These properties are consistent with the
property that the correlation functions given in
Eqs.~(\ref{eq:231})--(\ref{eq:234}) are all real functions.

Substituting Eqs.~(\ref{eq:239}) and (\ref{eq:240}) into
Eqs.~(\ref{eq:231})--(\ref{eq:234}), we have expressions of the
correlation functions of $+/\times$ modes of the specific coordinates
system for each pair of tensors. The results are given by
\begin{align}
  \xi^{++(s_1s_2)}_{X_1X_2}(r_1,r_2,\theta)
  &=
  \sum_{\substack{l\\l+s_1-s_2=\mathrm{even}}}
  (-1)^{(l+s_1-s_2)/2} Z^{s_1s_2;l}_{X_1X_2+}(r_1,r_2,\theta),
  \label{eq:242}\\
  \xi^{\times\times(s_1s_2)}_{X_1X_2}(r_1,r_2,\theta)
  &=
  \sum_{\substack{l\\l+s_1-s_2=\mathrm{even}}}
  (-1)^{(l+s_1-s_2+2)/2} Z^{s_1s_2;l}_{X_1X_2-}(r_1,r_2,\theta),,
  \label{eq:243}\\
  \xi^{+\times(s_1s_2)}_{X_1X_2}(r_1,r_2,\theta)
  &=
  \sum_{\substack{l\\l+s_1-s_2=\mathrm{odd}}}
  (-1)^{(l+s_1-s_2+1)/2} Z^{s_1s_2;l}_{X_1X_2-}(r_1,r_2,\theta),
  \label{eq:244}\\
  \xi^{\times+(s_1s_2)}_{X_1X_2}(r_1,r_2,\theta)
  &=
  \sum_{\substack{l\\l+s_1-s_2=\mathrm{odd}}}
  (-1)^{(l+s_1-s_2+1)/2} Z^{s_1s_2;l}_{X_1X_2+}(r_1,r_2,\theta),
  \label{eq:245}
\end{align}
where
\begin{multline}
 Z^{s_1s_2;l}_{X_1X_2\pm}(r_1,r_2,\theta) =
 \frac{1}{2} A_{s_1} A_{s_2}
  \sum_{l_1,l_2,m}
  \tilde{P}_l^m(\mu_2)
  \begin{pmatrix}
    l_1 & l_2 & l \\ s_2-m & -s_2 & m
  \end{pmatrix}
  \\ \times
  \left[
    (-1)^{l_1} u^{(l_1)}_{s_1,s_2-m}(\theta) \pm
    u^{(l_1)}_{-s_1,s_2-m}(\theta)
  \right]
  \tilde{\xi}^{s_1s_2;l_1l_2}_{X_1X_2l}(x),
  \label{eq:246}
\end{multline}
and the variables $(x, \mu_2)$ in the last expression are related to
$(r_1,r_2,\theta)$ by Eq.~(\ref{eq:122}).

If we take the distant-observer limit in the above expressions,
Eqs.~(\ref{eq:241})--(\ref{eq:246}), one can confirm that the
corresponding results derived in Paper~III in the same limit from the
first place are reproduced. The outline of the proof is given as
follows: first, we note that
$u^{(l)}_{m'm}(\theta) \rightarrow g^{(l)}_{m'm} = (-1)^m
\delta_{m',-m}$ in the distant-observer limit with
$\theta \rightarrow 0$. In this limit, the summation over $m$ in
Eq.~(\ref{eq:246}) is trivially taken. Substituting Eq.~(\ref{eq:241})
into the result, and applying the recoupling formula with the $9j$-symbol
(given in Appendix~C of Paper~I), they are represented by a function
$\xi^{s_1s_2;l\,l_z;s}_{X_1X_2}(x)$ given in Eq.~(\ref{eq:145}). The
relation between the last function and a normalized function
$\hat{\xi}^{s_1s_2;l\,l_z;s}_{X_1X_2}(x)$ introduced in Paper~III is
given by
$\hat{\xi}^{s_1s_2;l\,l_z;s}_{X_1X_2}(x) =
A_{s_1}A_{s_2}\sqrt{4\pi/(2l+1)}\, \xi^{s_1s_2;l\,l_z;s}_{X_1X_2}(x)$.
After some calculations described above, the corresponding formulas
derived in Paper~III in the distant-observer limit are exactly
reproduced.

\subsubsection{The linear correlation function}

As derived above, the correlation functions of $+/\times$ modes are
given by the invariant function
$\tilde{\xi}^{s_1s_2;l_1l_2}_{X_1X_2l}(x)$. In the linear theory,
substituting Eq.~(\ref{eq:174}) into Eq.~(\ref{eq:241}), we have
\begin{multline}
  \tilde{\xi}^{s_1s_2;l_1l_2}_{X_1X_2l}(x) =
  (2l+1)
  \begin{pmatrix}
    l_1 & l_2 & l \\ 0 & 0 & 0
  \end{pmatrix}
  \\ \times
  \begin{pmatrix}
    s_1 & l_1-s_1 & l_1 \\ s_1 & 0 & -s_1
  \end{pmatrix}
  \begin{pmatrix}
    s_2 & l_2-s_2 & l_2 \\ s_2 & 0 & -s_2
  \end{pmatrix}
  \\ \times
  \int \frac{k^2dk}{2\pi} j_l(kx)
  \tilde{G}_{X_1s_1l_1}(k)
  \tilde{G}_{X_2s_2l_2}(k)
  P_\mathrm{L}(k),
  \label{eq:247}
\end{multline}
where the function $\tilde{G}_{Xsl}(k)$ is given by
Eq.~(\ref{eq:176}).

As in the previous subsections, we give explicit expressions depending
on whether the spins are zero or not. In order to represent the
results, we define the following integrals of the linear power
spectrum:
\begin{align}
  \xi_l(x)
  &\equiv
    \int \frac{k^2dk}{2\pi^2} j_l(kx) P_\mathrm{L}(k),
  \label{eq:248}
  \\
  \xi^{(s)}_{Xl}(x)
  &\equiv
    \int \frac{k^2dk}{2\pi^2} j_l(kx)
    c^{(1)}_{Xs}(k) P_\mathrm{L}(k),
  \label{eq:249}
  \\
  \xi^{(s_1s_2)}_{X_1X_2l}(x)
  &\equiv
    \int \frac{k^2dk}{2\pi^2} j_l(kx)
    c^{(1)}_{X_1s_1}(k) c^{(1)}_{X_2s_2}(k) P_\mathrm{L}(k).
  \label{eq:250}
\end{align}
When both $X_1$ and $X_2$ are scalar fields, $s_1=s_2=0$, we have
\begin{multline}
  \tilde{\xi}^{00;l_1l_2}_{X_1X_2l}(x)
  =
  \delta_{l_10}\delta_{l_20}\delta_{l0}
  \left[
    \xi^{(00)}_{X_1X_20}(x)
    + \left(1 + \frac{f}{3}\right)
    c^{(0)}_{X_1} \xi^{(0)}_{X_20}(x)
    \right.
    \\
    \left.
    + (1 \leftrightarrow 2)
    + \left(1 + \frac{f}{3}\right)^2 
    c^{(0)}_{X_1} c^{(0)}_{X_2} \xi_{0}(x)
  \right]
  \\
  + \delta_{l_12}\delta_{l_20}\delta_{l2}
  \frac{2\sqrt{5}}{3}f
  c^{(0)}_{X_1}
  \left[
    \xi^{(0)}_{X_22} +
    \left(1 + \frac{f}{3}\right)
    c^{(0)}_{X_2} \xi_2(x)
  \right]
  + (1 \leftrightarrow 2)
  \\
  + \delta_{l_12}\delta_{l_22}
  \frac{4f^2}{9\sqrt{5}} c^{(0)}_{X_1} c^{(0)}_{X_2} 
  \left[
    \delta_{l0} \xi_0(x)
    - \delta_{l2}\,\frac{5\sqrt{2}}{\sqrt{7}} \xi_2(x)
    \right.
    \\
    \left.
    + \delta_{l4}\, \frac{9\sqrt{2}}{\sqrt{7}} \xi_4(x)
  \right].
  \label{eq:251}
\end{multline}
When $X_1$ is a nonscalar field and $X_2$ is a scalar
field, we have $s_1=s$ and $s_2=0$ where $s\geq 1$ in
Eq.~(\ref{eq:247}). The explicit form in this case is given by
\begin{multline}
  \tilde{\xi}^{s0;l_1l_2}_{X_1X_2l}(x)
  = (-1)^s
  \delta_{l_1s}\delta_{l_20}\delta_{ls}
  \left[
    \xi^{(s0)}_{X_1X_2s}(x)
    + \left(1 + \frac{f}{3}\right)
    \xi^{(s)}_{X_1s}(x) c^{(0)}_{X_2}
  \right]
  \\
  + \delta_{l_1s}\delta_{l_22}
  \frac{2l+1}{\sqrt{2s+1}}
  \begin{pmatrix}
    s & 2 & l \\ 0 & 0 & 0
  \end{pmatrix}
  \frac{2f}{3}
  \xi^{(s)}_{X_1l}(x) c^{(0)}_{X_2}.
  \label{eq:252}
\end{multline}
When both $X_1$ and $X_2$ are nonscalar fields, we have $s_1,s_2\geq
1$ in Eq.~(\ref{eq:247}). The explicit form in this case is given by
\begin{multline}
  \tilde{\xi}^{s_1s_2;l_1l_2}_{X_1X_2l}(x) =
  \delta_{l_1s_1} \delta_{l_2s_2}
  \frac{2l+1}{\sqrt{(2s_1+1)(2s_2+1)}}
  \\ \times
  \begin{pmatrix}
    s_1 & s_2 & l \\ 0 & 0 & 0
  \end{pmatrix}
  \xi^{(s_1s_2)}_{X_1X_2l}(x).
  \label{eq:253}
\end{multline}
Substituting the above equations into Eq.~(\ref{eq:246}) and
Eqs.~(\ref{eq:242})--(\ref{eq:245}), the predictions of the linear
theory for correlation functions are explicitly given.

On one hand, inspecting the rhs of
Eqs.~(\ref{eq:251})--(\ref{eq:253}), it is straightforward to see that
$l+s_1-s_2=\mathrm{even}$ in all cases. Therefore, Eqs.~(\ref{eq:244})
and (\ref{eq:245}) are zero, i.e.,
\begin{equation}
  \xi^{+\times(s_1s_2)}_{X_1X_2}(r_1,r_2,\theta) =
  \xi^{\times+(s_1s_2)}_{X_1X_2}(r_1,r_2,\theta) = 0
  \label{eq:254}
\end{equation}
in linear theory. In Paper~III, exactly the same property is shown in
the distant-observer limit. The above result shows that the last
property generally holds without assuming the distant-observer
limit.\footnote{When the parity symmetry holds both in the initial
  conditions and dynamical evolutions, the $+/\times$ correlations
  always vanish. The results here state more than that, and hold only
  in linear theory.}

On the other hand, Eqs.~(\ref{eq:242}) and (\ref{eq:243}) survive in
linear theory. As the integers $l_1$ and $l_2$ satisfy
$0 \leq l_1 \leq \mathrm{max}(s_1,2)$ and
$0 \leq l_2 \leq \mathrm{max}(s_2,2)$ in Eq.~(\ref{eq:247}) because of
the specific form of Eq.~(\ref{eq:176}) in linear theory, the
non-negative integer $l$ in Eq.~(\ref{eq:246}) satisfies
$0 \leq l \leq \mathrm{max}(s_1+s_2,4)$ because of the triangular
inequality among $(l,l_1,l_2)$ of the $3j$-symbol. Therefore, the
summation over $l$ in Eqs.~(\ref{eq:242}) and (\ref{eq:243}) is finite
for a given set of spins $s_1$ and $s_2$.

Using the explicit expressions of the rotation matrix with
trigonometric functions in Eq.~(\ref{eq:103}), and substituting
Eqs.~(\ref{eq:251})--(\ref{eq:253}) into Eq.~(\ref{eq:246}), and then
into Eqs.~(\ref{eq:242}) and (\ref{eq:243}), explicit results of the
linear theory are derived for a given set of spins, $s_1$ and $s_2$.
For the reader's convenience, the explicit results with
$s_1, s_2 \leq 2$ are given below.

\begin{figure}
\centering
\includegraphics[width=19pc]{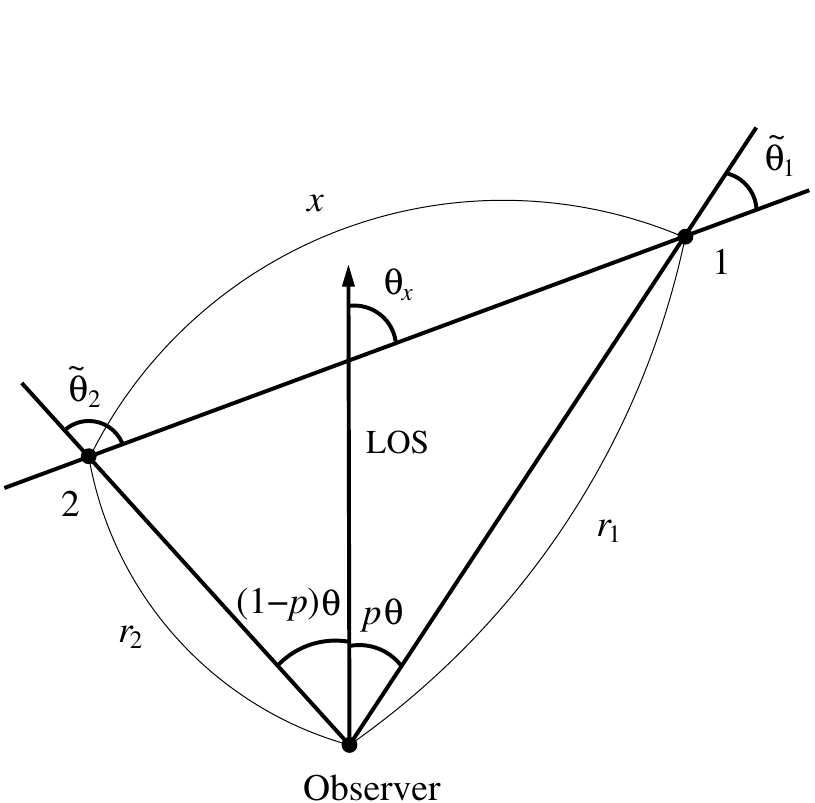}
\caption{\label{fig:1} The variables to characterize geometries of
  correlation function in the full-sky formulation. For a fixed value
  of $p$, the correlation functions are regarded as functions of
  either $(r_1,r_2,\theta)$ or $(x,\theta_x,\theta)$, etc.
  Practically, the choice of $p=1/2$ is convenient and symmetric under
  the interchange of the two points, $1 \leftrightarrow 2$. We use the
  last choice in the numerical demonstrations in the main text. }
\end{figure}

When both $X_1$ and $X_2$ are scalar fields, $s_1=s_2=0$, we have
\begin{multline}
  \xi^{++(00)}_{X_1X_2}(r_1,r_2,\theta)
  = \xi^{(00)}_{X_1X_20}(x)
  + \left(1 + \frac{f}{3}\right)
  c^{(0)}_{X_1} \xi^{(0)}_{X_20}(x) + (1\leftrightarrow 2)
  \\
  +
  \left[
    1 + \frac{2f}{3} + \frac{f^2}{5}
    \left(
      1 - \frac{2}{3} \sin^2\theta
    \right)
  \right]
  c^{(0)}_{X_1} c^{(0)}_{X_2}
  \xi_0(x)
  \\
  - f
  \left(
    \cos^2\tilde{\theta}_1 - \frac{1}{3}
  \right)
  c^{(0)}_{X_1} \xi^{(0)}_{X_22}(x)
  + (1 \leftrightarrow 2)
  \\
  - f
  \left[
    \left(1 + \frac{3f}{7}\right)
    \left(
      \cos^2\tilde{\theta}_1 + \cos^2\tilde{\theta}_2 - \frac{2}{3}
    \right)
  \right.
  \\
  \left.
    - \frac{2f}{21} \sin^2\theta
  \right]
  c^{(0)}_{X_1} c^{(0)}_{X_2}
  \xi_2(x)
  \\
  + \frac{f^2}{4}
  \left[
    \frac{12}{35}
    + 2 \cos\tilde{\theta}_1 \cos\tilde{\theta}_2
    \cos\left(\tilde{\theta}_1+\tilde{\theta}_2\right)
    \phantom{\frac{10}{10}}
  \right.
  \\
  \left.
    - \frac{5}{7}
    \left(
      \cos^2\tilde{\theta}_1 + \cos^2\tilde{\theta}_2
    \right)
    - \frac{3}{35} \sin^2\theta
  \right]
  c^{(0)}_{X_1} c^{(0)}_{X_2}
  \xi_4(x),
  \label{eq:255}
\end{multline}
where $x$, $\tilde{\theta}_1$ and $\tilde{\theta}_2$ are defined by
\begin{align}
  x
  &= \sqrt{{r_1}^2 + {r_2}^2 - 2r_1 r_2 \cos\theta}\,,
    \label{eq:256}\\
  \tilde{\theta}_1
  &= \arccos\left(\frac{r_1 - r_2\cos\theta}{x}\right),
    \label{eq:257}\\
  \tilde{\theta}_2
  &= \arccos\left(\frac{r_1 \cos\theta - r_2}{x}\right).
    \label{eq:258}
\end{align}
Geometric meanings of the above variables are illustrated in
Fig.~\ref{fig:1}.
The cross-correlation function for scalar fields vanishes,
$\xi^{\times\times(00)}_{X_1X_2} = 0$, because the cross component of
the scalar field identically vanishes, as obviously seen from
Eq.~(\ref{eq:223}). The above result, Eq.~(\ref{eq:255}), is
consistent with the known expressions for the linear correlation
function of galaxies in redshift space with wide-angle effects
\cite{Szalay:1997cc,Matsubara:1999du}.

When $X_1$ is a spin-1 field and $X_2$ is a scalar field, $s_1=1$ and
$s_2=0$, we have
\begin{multline}
  \xi^{++(10)}_{X_1X_2}
  = - \frac{1}{\sqrt{6}}
  \Biggl\{
  \sin\tilde{\theta}_1 \xi^{(10)}_{X_1X_21}(x)
  \\
  +
  \left[
    \left(1 + \frac{2f}{5} \right)
    \sin\tilde{\theta}_1
    + \frac{f}{5}
    \sin\left(\tilde{\theta}_1 - 2\tilde{\theta}_2\right)
    \right]
    \xi^{(1)}_{X_11}(x) c^{(0)}_{X_2}
  \\
    - \frac{f}{20}
    \left[
      2 \sin\tilde{\theta}_1
      + 5 \sin\left(\tilde{\theta}_1 + 2\tilde{\theta}_2\right)
    \right.
    \\
    \left.
      + \sin\left(\tilde{\theta}_1 - 2\tilde{\theta}_2\right)
    \right]
    \xi^{(1)}_{X_13}(x) c^{(0)}_{X_2}
    \Biggr\}
  \label{eq:259}
\end{multline}
and $\xi^{\times\times(10)}_{X_1X_2} = 0$. The arguments
$(r_1,r_2,\theta)$ of the correlation function on the lhs are omitted
for simplicity. When $X_1$ is a spin-2 field and $X_2$ is a scalar
field, $s_1=2$ and $s_2=0$, we have
\begin{multline}
  \xi^{++(20)}_{X_1X_2}
  = \frac{1}{2\sqrt{5}}
  \Biggl\{
  \sin^2\tilde{\theta}_1\, \xi^{(20)}_{X_1X_22}(x)
  - \frac{2f}{15} \sin^2\theta\, \xi^{(2)}_{X_10}(x) c^{(0)}_{X_2}
  \\
  +
  \left[
   \left(1+\frac{3f}{7}\right) \sin^2\tilde{\theta}_1
    - \frac{2f}{7} \sin^2\tilde{\theta}_2
    + \frac{2f}{21} \sin^2\theta
  \right]
  \xi^{(2)}_{X_12}(x) c^{(0)}_{X_2}
  \\
  - \frac{f}{28}
  \left[
    \frac{3}{2}
    - \frac{7}{2}
    \cos\left(2\tilde{\theta}_1 + 2\tilde{\theta}_2\right)
    - \cos 2\tilde{\theta}_1
  \right.
  \\
  \left.
    + 3 \cos 2\tilde{\theta}_2
    + \frac{3}{5} \sin^2\theta
  \right]
  \xi^{(2)}_{X_14}(x) c^{(0)}_{X_2}
  \Biggr\},
  \label{eq:260}
\end{multline}
and $\xi^{\times\times(20)}_{X_1X_2} = 0$.

When both $X_1$ and $X_2$ are both spin-1 fields, $s_1=s_2=1$, we have
\begin{align}
  \xi^{++(11)}_{X_1X_2}
  &= \frac{1}{36}
  \left\{
    2 \cos\theta\, \xi^{(11)}_{X_1X_20}(x)
    \right.
    \nonumber\\
  & \hspace{3pc}
    \left.
    +
    \left[
    3\cos\left(\tilde{\theta}_1+\tilde{\theta}_2\right)
    - \cos\theta
    \right]
    \xi^{(11)}_{X_1X_22}(x)
    \right\},
  \label{eq:261}\\
  \xi^{\times\times(11)}_{X_1X_2}
  &= \frac{1}{18}
  \left[
    \xi^{(11)}_{X_1X_20}(x) + \xi^{(11)}_{X_1X_22}(x)
    \right].
  \label{eq:262}
\end{align}
When $s_1=1$ and $s_2=2$, we have
\begin{align}
  \xi^{++(12)}_{X_1X_2}
  &= \frac{\sin\tilde{\theta}_1}{20\sqrt{30}}
    \left\{
    4 \cos\theta\, \xi^{(12)}_{X_1X_21}(x)
    \right.
    \nonumber\\
  & \hspace{2pc}
    \left.
    +
    \left[
    5\cos\left(\tilde{\theta}_1+ \tilde{\theta}_2\right)
    - \cos\theta
    \right]
    \xi^{(12)}_{X_1X_23}(x)
    \right\},
  \label{eq:263}\\
  \xi^{\times\times(12)}_{X_1X_2}
  &= \frac{\sin\tilde{\theta}_1}{5\sqrt{30}}
  \left[
    \xi^{(12)}_{X_1X_21}(x) + \xi^{(12)}_{X_1X_23}(x)
    \right].
  \label{eq:264}
\end{align}
Lastly, when $s_1=s_2=2$, we have
\begin{multline}
  \xi^{++(22)}_{X_1X_2}
  = \frac{1}{525}
  \Biggl\{
  7\left(1 - \frac{1}{2}\sin^2\theta\right)
  \xi^{(22)}_{X_1X_20}(x)
  \\
  + \frac{5}{4}
  \left[
    2 + 3\cos 2\tilde{\theta}_1 + 3\cos 2\tilde{\theta}_2
    + \sin^2\theta
  \right]
  \xi^{(22)}_{X_1X_22}(x)
  \\
  + \frac{3}{32}
  \Biggl[
    57 - 6 \sin^2\theta
    - 30\cos 2\tilde{\theta}_1 - 30\cos 2\tilde{\theta}_2
    \\
      + 35
      \cos\left(2\tilde{\theta}_1+2\tilde{\theta}_2\right)
    \Biggr] 
    \xi^{(22)}_{X_1X_24}(x)
  \Biggr\},
  \label{eq:265}
\end{multline}
and
\begin{multline}
  \xi^{\times\times(22)}_{X_1X_2}
  = \frac{1}{525}
  \Biggl\{
  7 \xi^{(22)}_{X_1X_20}(x)
  \\
  + \frac{5}{2}
  \left[
    3 \cos\left(\tilde{\theta}_1+\tilde{\theta}_2\right)
    + \cos\theta
  \right]
  \xi^{(22)}_{X_1X_22}(x)
  \\
  + \frac{3}{2}
  \left[
    5 \cos\left(\tilde{\theta}_1+\tilde{\theta}_2\right)
    - 3 \cos\theta
  \right]
  \xi^{(22)}_{X_1X_24}(x)
  \Biggr\}.
  \label{eq:266}
\end{multline}

The correlation functions of $+/\times$ modes in the distant-observer
limit with spins less than or equal to two in linear theory is given
by taking the limit of $\theta \rightarrow 0$ in
Eqs.~(\ref{eq:255})--(\ref{eq:266}), where the separation $x$ is fixed
on the rhs in the expressions. The parameters $\tilde{\theta}_1$ and
$\tilde{\theta}_2$ are the same in this limit, and they are given by
$\cos\tilde{\theta}_1 = \cos\tilde{\theta}_2 = (r_1-r_2)/x$. One can
verify that the resulting expressions agree with those derived in
Paper~III. The expressions of the linear correlation functions of
$+/\times$ modes in the large-scale limit, where the scale dependencies
of the renormalized bias functions $c^{(1)}_{Xs\sigma}(k)$ are
neglected and replaced by constants, are explicitly given in
the Appendix for spins less than or equal to 2.

\subsection{\label{subsec:LinearNumerical}
  Comparisons with distant-observer approximation in linear
  theory}

We have derived explicit formulas of power spectra and correlation
functions of projected tensor fields. The results of the lowest-order
approximation of iPT are explicitly exemplified for spins less than or
equal to 2 in the above. In this subsection, we numerically
demonstrate the behavior of these functions and compare them with the
corresponding expressions in the distant-observer approximation
derived in Paper~III.

\subsubsection{The power spectrum}

\begin{figure*}
\centering
\includegraphics[width=40pc]{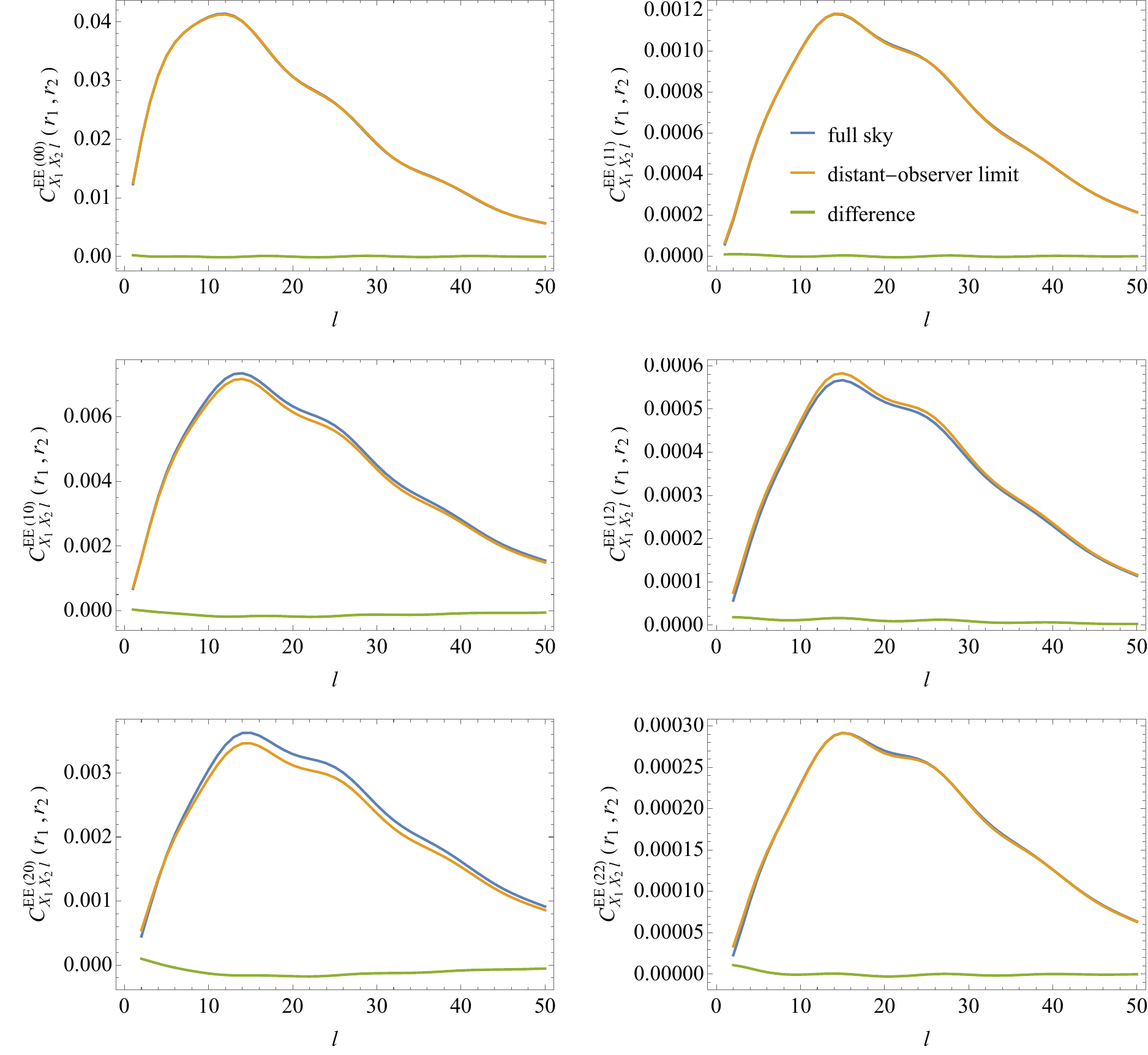}
\caption{\label{fig:2} The EE power spectra
  $C^{\mathrm{EE}(s_1s_2)}_{X_1X_2l}(r_1,r_2)$ for spins less than or
  equal to 2 for simple models of tensor fields (see text) in linear
  theory. The distances are fixed to $r_1=200\,h^{-1}\mathrm{Mpc}$ and
  $r_2=210\,h^{-1}\mathrm{Mpc}$. Blue lines: the full-sky results.
  Orange lines: results with applying the distant-observer limit.
  Green lines: differences between the two results.}
\end{figure*}

The power spectra of E/B modes in the distant-observer approximation
are given in Paper~III. The BB/EB/BE power spectra vanish and only the
EE power spectrum survives in linear theory. The results can be read
off from equations in Paper~III and are given by\footnote{These
  results are straightforwardly derived from Eqs.~(303) and (307) of
Paper~III.}
\begin{align}
  P^{\mathrm{EE}(00)}_{X_1X_2}(\bm{k})
  &=
    b_{X_1}(k) b_{X_2}(k)
    \left[ 1 + \beta_{X_1}(k) \mu^2 \right]
    \nonumber \\
  & \hspace{6pc}
  \times  \left[ 1 + \beta_{X_2}(k) \mu^2 \right]
    P_\mathrm{L}(k),
  \label{eq:267}\\
  P^{\mathrm{EE}(s0)}_{X_1X_2}(\bm{k})
  &=
    \frac{(-1)^s (1-\mu^2)^{s/2}}{2^{s/2}\sqrt{2s+1}}
    c^{(1)}_{X_1s}(k) b_{X_2}(k)
    \nonumber \\
  & \hspace{6pc}
    \times \left[ 1 + \beta_{X_2}(k) \mu^2 \right]
    P_\mathrm{L}(k),
  \label{eq:268}\\
  P^{\mathrm{EE}(s_1s_2)}_{X_1X_2}(\bm{k})
  &=
    \frac{(-1)^{s_1+s_2} (1-\mu^2)^{(s_1+s_2)/2}}
    {2^{(s_1+s_2)/2}\sqrt{(2s_1+1)(2s_2+1)}}
    \nonumber \\
  & \hspace{6pc}
  \times c^{(1)}_{X_1s_1}(k) c^{(1)}_{X_1s_2}(k)
    P_\mathrm{L}(k),
  \label{eq:269}
\end{align}
where
\begin{equation}
  b_X(k) \equiv c^{(0)}_X + c^{(1)}_{X0}(k), \quad
  \beta_X(k) \equiv \frac{c^{(0)}_X f}{b_X(k)},
  \label{eq:270}
\end{equation}
and $s,s_1,s_2\geq 1$ in the last two equations. Substituting the
above functions into Eq.~(\ref{eq:211}), the distant-observer
approximation for the full-sky EE power spectra
$C^{\mathrm{EE}(s_1s_2)}_{X_1X_2l}(r_1,r_2)$ are given. In this way,
we compare the EE power spectra in linear theory with and without
distant-observer approximation.

In Fig.~\ref{fig:2}, the power spectra
$C^{\mathrm{EE}(s_1s_2)}_{X_1X_2l}(r_1,r_2)$ in linear theory with
$(s_1,s_2)=(0,0),(1,0),(2,0),(1,1),(1,2),(2,2)$ are compared between
the full-sky formula of Eqs.~(\ref{eq:218})--(\ref{eq:220}) and those
calculated from distant-observer approximation, Eqs.~(\ref{eq:211})
and (\ref{eq:267})--(\ref{eq:269}). The linear power spectrum
$P_\mathrm{L}(k)$ of the mass density is calculated by a Boltzmann
code \textsc{CLASS} \cite{Lesgourgues:2011re,Blas:2011rf} with a flat
$\Lambda$CDM model and cosmological parameters $h=0.6732$,
$\Omega_{\mathrm{b}0}h^2=0.02238$, $\Omega_{\mathrm{cdm}}h^2=0.1201$,
$n_\mathrm{s}=0.9660$, $\sigma_8=0.8120$ (Planck 2018
\cite{Planck:2018vyg}). Just for simplicity, the renormalized bias
functions are simply modeled by\footnote{A reason why we assume the
  negative values for the renormalized bias functions of odd spins is
  that otherwise the power spectra $C^{\mathrm{EE}(s_1s_2)}_{X_1X_2l}$
  become negative when $s_1+s_2$ is an odd number.} $c^{(0)}_X = 1$
and $c^{(n)}_{Xs}(k) = (-1)^s W(kR)$, where the Gaussian window
function $W(kR) = \exp (-k^2R^2/2)$ is adopted with a smoothing radius
of $R=5\,h^{-1}\mathrm{Mpc}$. The distances are fixed to
$r_1=200\,h^{-1}\mathrm{Mpc}$ and $r_2=210\,h^{-1}\mathrm{Mpc}$ in
this figure.

As obviously seen from the figure, the power spectra
$C^{\mathrm{EE}(s_1s_2)}_{X_1X_2l}$ derived from the distant-observer
limit of the power spectra $P^{\mathrm{EE}(s_1s_2)}_{X_1X_2}(\bm{k})$
using Eq.~(\ref{eq:211}) are not so different from those of full-sky
formulas, and thus consistencies of the two formalisms are seen.
Looking into more detail, the differences are roughly within 1\% for
power spectra with the same spins, i.e.,
$(s_1,s_2)=(0,0),(1,1),(2,2)$, while the differences can be relatively
larger for power spectra with different spins, i.e.,
$(s_1,s_2)=(1,0),(2,0),(1,2)$, and can be roughly 5\% in the case of
$(2,0)$. When the multipole $l$ is large, the component $k_\perp$ of
the wave vector is as large as the order of $1/r_0$ as seen in
Eq.~(\ref{eq:134}), while the line-of-sight component $k_\parallel$ of
much larger order $\gg 1/r_0$ also contributes in the integral of
Eq.~(\ref{eq:211}), because the relative distance $|r_1-r_2|$ can be
much smaller than $r_0$. Therefore, the distant-observer approximation
for the Cartesian power spectrum $P^{(s_1s_2)}_{X_1X_2}(\bm{k})$ needs
not to be accurate enough for mildly large $l$. However, one does not
have to calculate the power spectrum by applying the distant-observer
approximation, as we now have full formulas. The purpose of
comparisons here is just to numerically check the consistency between
two approaches.

\subsubsection{The correlation functions}

\begin{figure*}
\centering
\includegraphics[width=40pc]{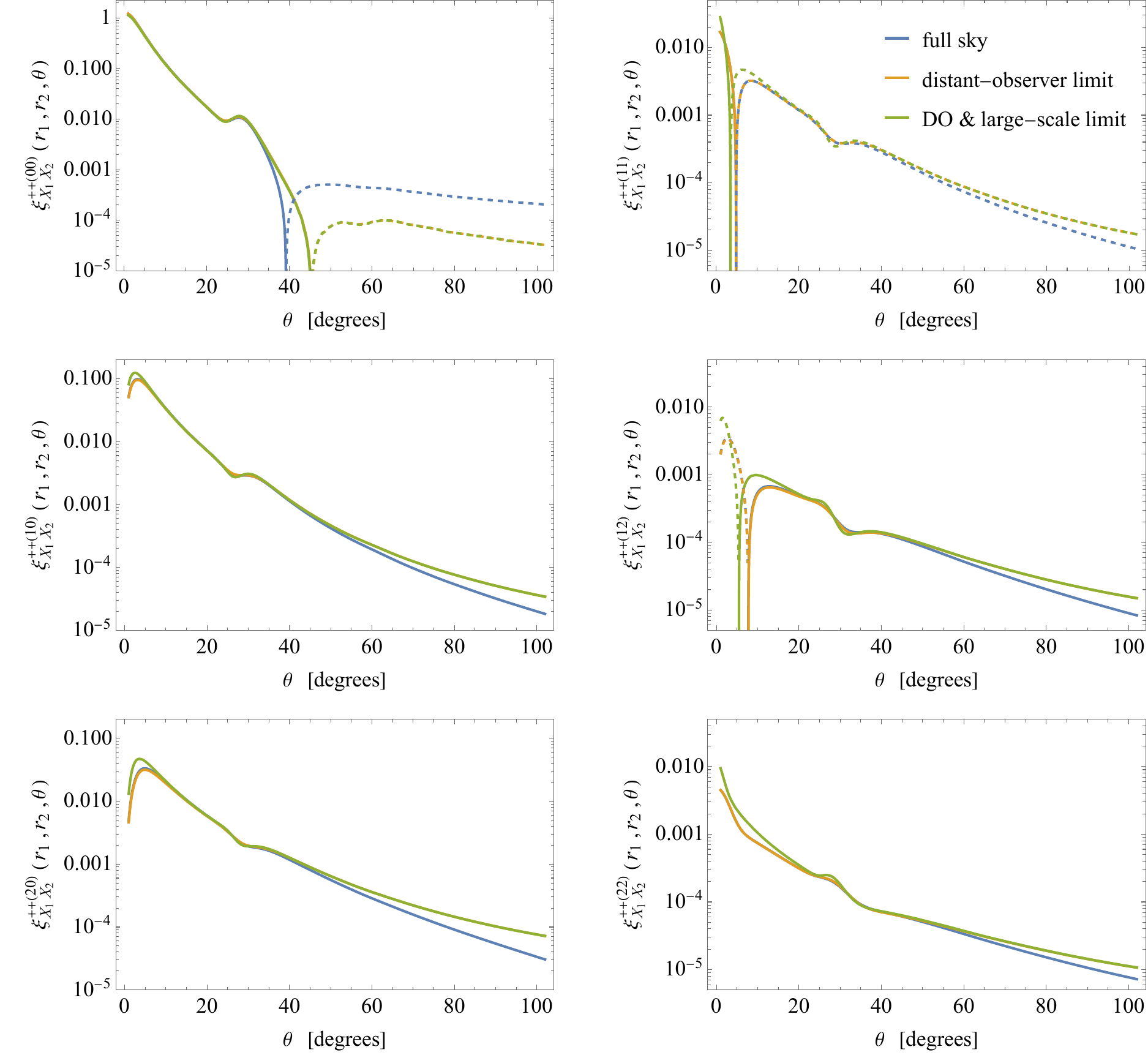}
\caption{\label{fig:3} The $++$ correlation functions
  $\xi^{++(s_1s_2)}_{X_1X_2}(r_1,r_2,\theta)$ for spins less
  than or equal to two for simple models of tensor fields (see text)
  in linear theory. The distances are fixed to
  $r_1=200\,h^{-1}\mathrm{Mpc}$ and $r_2=210\,h^{-1}\mathrm{Mpc}$.
  Blue lines: the full-sky results. Orange lines: results with
  applying the distant-observer limit. Green lines: results with
  applying both the distant-observer and large-scale limits. The solid
  lines correspond to positive values, while the dashed lines
  correspond to negative values.}
\end{figure*}
\begin{figure*}
\centering
\includegraphics[width=40pc]{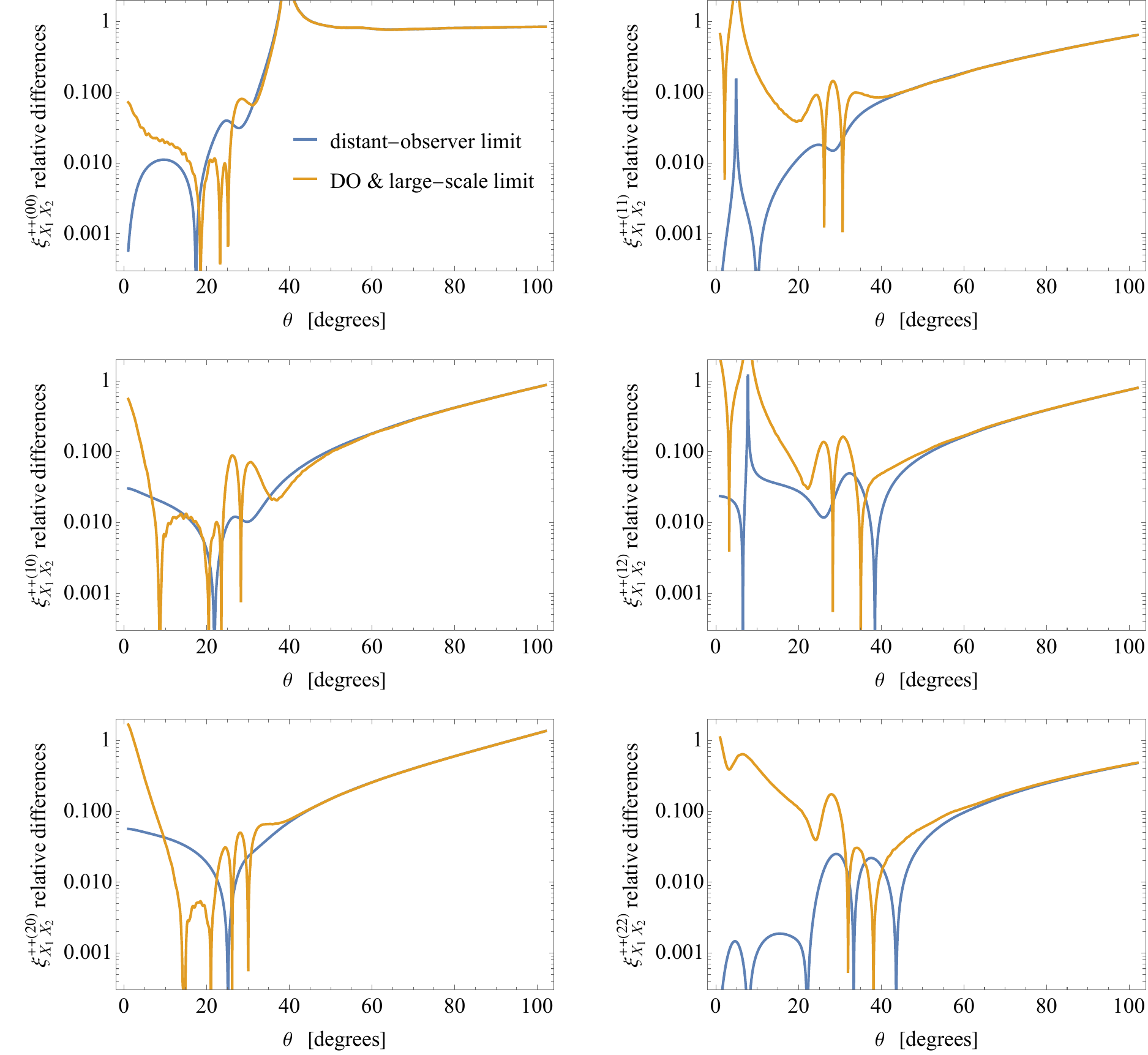}
\caption{\label{fig:4} Relative differences of the distant-observer
  and large-scale limits from the full-sky correlation functions shown
  in Fig.~\ref{fig:3}. The relative differences are defined by
  absolute values as
  $|\xi(\mathrm{limits})/\xi(\mathrm{full\,sky})-1|$. Blue lines: the
  distant-observer limits. Orange lines: the distant-observer and
  large-scale limits.}
\end{figure*}
\begin{figure}
\centering
\includegraphics[width=21pc]{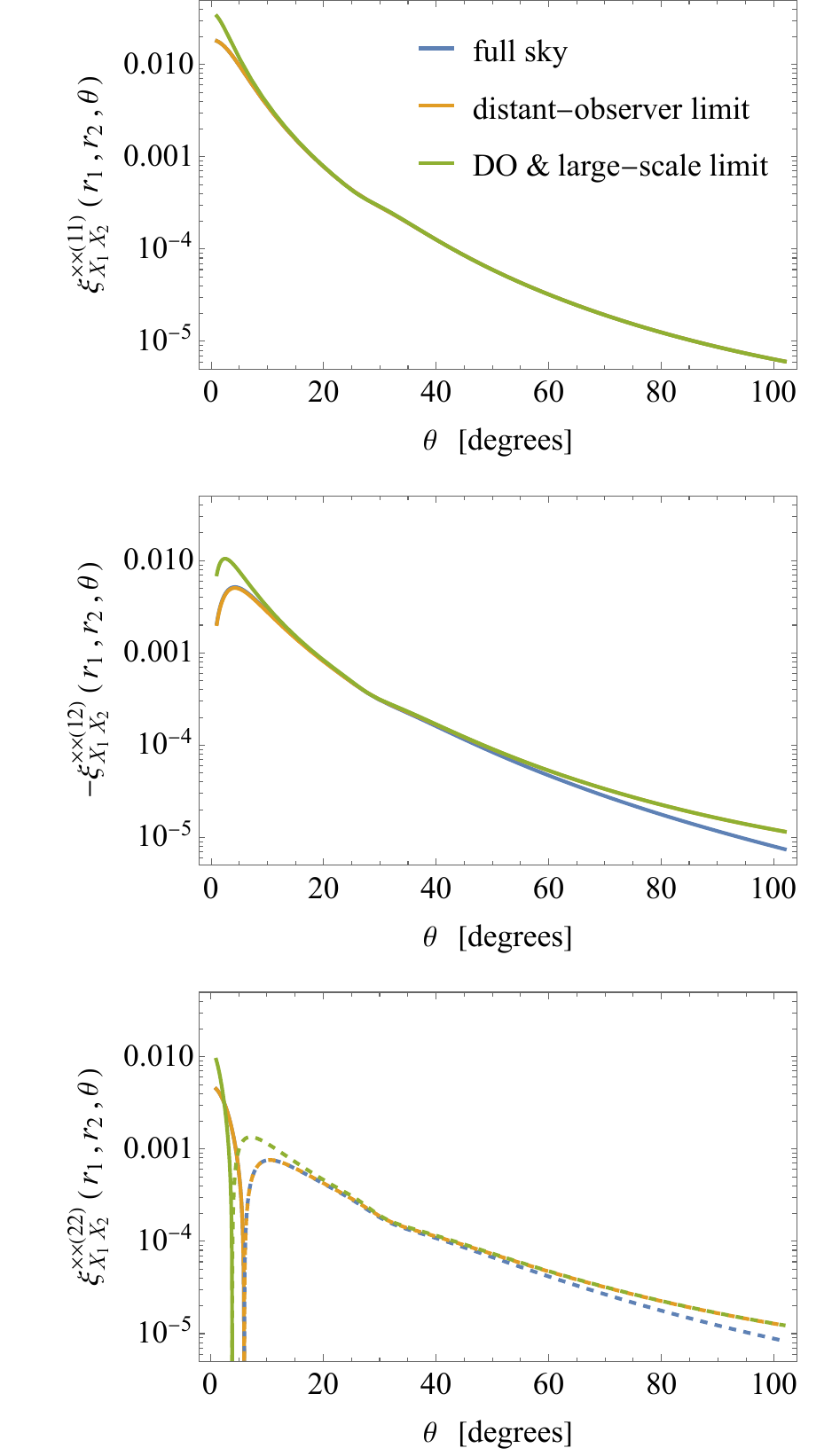}
\caption{\label{fig:5} Same as Fig.~\ref{fig:3}, but for the
  $\times\times$ correlation functions
  $\xi^{\times\times(s_1s_2)}_{X_1X_2}(r_1,r_2,\theta)$. For
  $\xi^{\times\times(12)}_{X_1X_2}$, the values are negative over the
  entire plotted range, the negative values are plotted as solid lines
  in this special case. }
\end{figure}
\begin{figure}
\centering
\includegraphics[width=21pc]{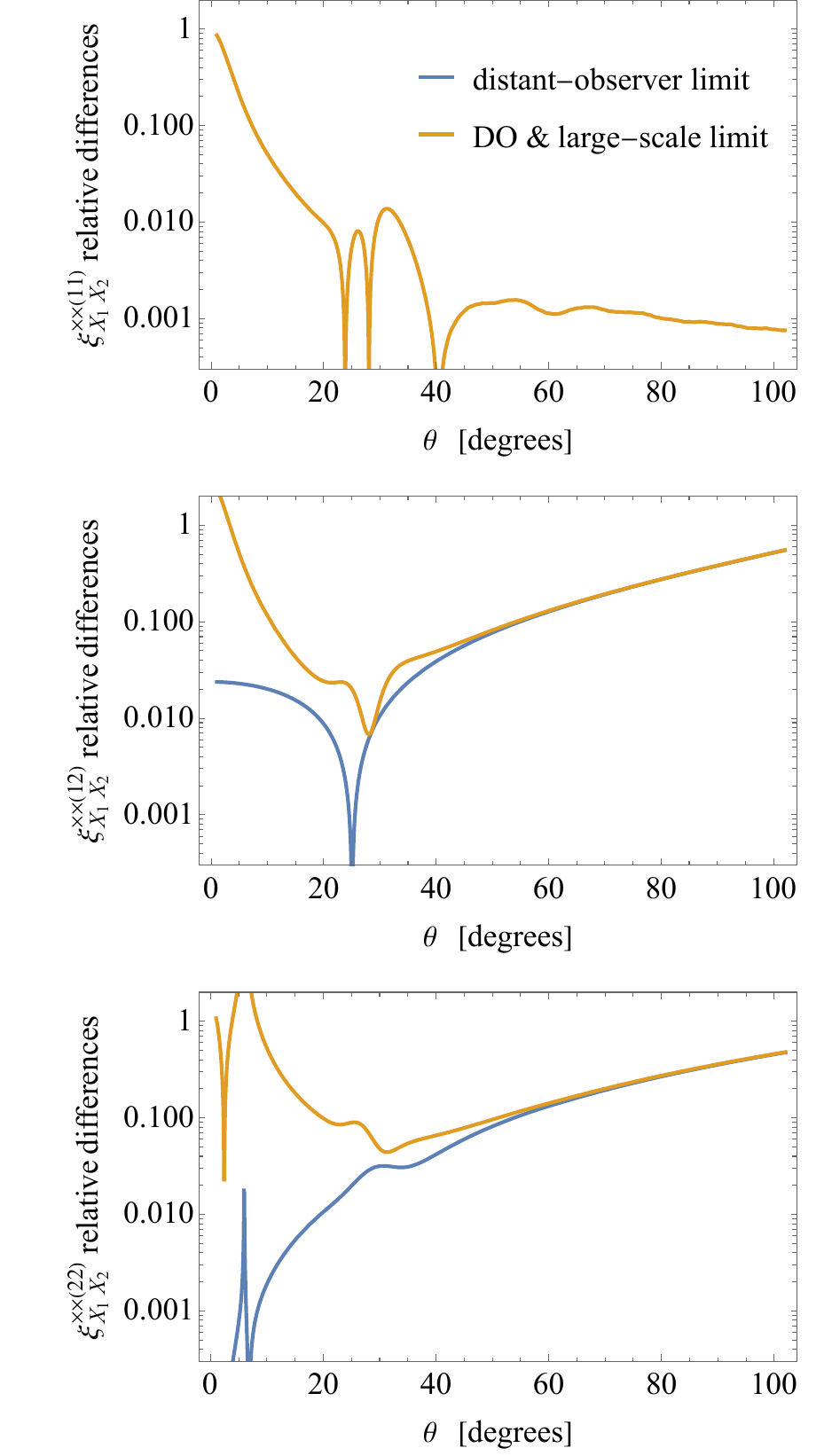}
\caption{\label{fig:6} Relative differences as those in
  Fig.~\ref{fig:4}, but for the $\times\times$ correlation functions
  $\xi^{\times\times(s_1s_2)}_{X_1X_2}(r_1,r_2,\theta)$. For
  $\xi^{\times\times(11)}_{X_1X_2}$, the distant observer limit is the
  same as the full-sky result in this special case.}
\end{figure}

In order to numerically demonstrate how accurately the expressions of
the correlation function in the distant-observer limit can approximate
those with wide-angle effects, we need to relate the relative vector
$\bm{x}$ between two objects in the distant-observer approximation and
variables $(r_1,r_2,\theta)$ of full-sky correlation function. There
is an ambiguity in choosing the direction to the line of sight in the
parametrization of the distant-observer approximation, because there
are two lines of sight in the full-sky formulation. It is natural to
choose the direction of the distant-observer line of sight between the
two. Dividing the angle between two lines of sight by a fraction $p$
with $0\leq p\leq 1$, the geometry is shown in Fig.~\ref{fig:1}. The
vertical arrow in the center labeled by ``LOS'' corresponds to the
line of sight in the distant-observer approximation, given two
positions of 1 and 2 relative to the position of the observer.
Assuming the flat geometry of the Universe with globally zero
curvature, the elementary geometry of triangles in the figure shows
that the separation $x=|\bm{r_1}-\bm{r}_2|$ is given by
Eq.~(\ref{eq:256}), and the angle $\theta_x$ between the separation
vector $\bm{r}_1-\bm{r}_2$ and the direction to line of sight in the
distant-observer approximation is given by
\begin{equation}
  \cos\theta_x = 
  \frac{r_1\cos(p\theta) - r_2 \cos\left[(1-p)\theta\right]}
  {\sqrt{{r_1}^2 + {r_2}^2 - 2r_1 r_2 \cos\theta}},
  \label{eq:271}
\end{equation}
where $(r_1,r_2,\theta)$ is the set of arguments for the full-sky
correlation function. The angles $\tilde{\theta}_1$ and
$\tilde{\theta}_2$ between the separation vector $\bm{r}_1-\bm{r}_2$
and radial vectors $\bm{r}_1$ and $\bm{r}_2$, respectively, are given
by Eqs.~(\ref{eq:257}) and (\ref{eq:258}). They are related to the
angles $\theta_x$ and $\theta$ by
\begin{align}
  &
  \tilde{\theta}_1 = \theta_x - p \theta, \quad
    \tilde{\theta}_2 = \theta_x + (1-p) \theta,
  \label{eq:272}\\
  &
  \theta = \tilde{\theta}_2 - \tilde{\theta}_1, \quad
    \theta_x =
    \frac{\tilde{\theta}_1 + \tilde{\theta}_2 - (1-2p)\theta}{2}.
  \label{eq:273}
\end{align}

A symmetric choice with respect to exchanging the two points,
$1\leftrightarrow 2$, is given by $p=1/2$, and the relations of
Eqs.~(\ref{eq:271})--(\ref{eq:273}) in this case reduces to
\begin{align}
  &
    \cos\theta_x =  \frac{r_1-r_2}{x}\cos\frac{\theta}{2},
  \label{eq:274}\\
  &
    \tilde{\theta}_1 = \theta_x - \frac{\theta}{2}, \quad
    \tilde{\theta}_2 = \theta_x + \frac{\theta}{2},
  \label{eq:275}\\
  &
    \theta = \tilde{\theta}_2 - \tilde{\theta}_1, \quad
    \theta_x = \frac{\tilde{\theta}_1 + \tilde{\theta}_2}{2}.
  \label{eq:276}
\end{align}
We apply the last parametrization with the choice $p=1/2$ to
represent the correlation functions of
Eqs.~(\ref{eq:255})--(\ref{eq:266}), which are considered to be
functions of $x$, $\theta_x$ and $\theta$, i.e.,
$\xi^{++(s_1s_2)}_{X_1X_2}(x,\theta_x;\theta)$. The correlation
functions in the distant-observer limit are given by substituting
$\theta=0$, $\tilde{\theta}_1=\tilde{\theta}_2=\theta_x$ in the
general expressions with wide-angle effects. 

In Figs.~\ref{fig:3}--\ref{fig:6}, the correlation functions in linear
theory are numerically calculated for the same cosmological model and
parameters as for the power spectra in Fig.~\ref{fig:2}. As the radii
$r_1$ and $r_2$ are fixed, the correlation functions in the figures
are functions only of opening angle $\theta$, and the variables $x$,
$\theta_x$, $\tilde{\theta}_1$ and $\tilde{\theta}_2$ all depend
solely on $\theta$. In Figs.~\ref{fig:3} and \ref{fig:5}, the values
of the correlation functions are given in logarithmic plots of
absolute values, where the positive values are represented by solid
lines while negative values are represented by dashed lines. The $++$
correlation functions are plotted in Fig.~\ref{fig:3} and the
$\times\times$ correlations are plotted in Fig.~\ref{fig:5}. The
latter correlation functions involving scalar fields all vanish. As
many parts of lines in the plots of Figs.~\ref{fig:3} and \ref{fig:5}
are overlapped, the relative differences of the limiting cases to the
full-sky results are shown in Figs.~\ref{fig:4} and \ref{fig:6},
corresponding to $++$ and $\times\times$ correlation functions given
in Figs.~\ref{fig:3} and \ref{fig:5}, respectively.

Since the correlation functions represent the clustering in
configuration space, the distant-observer limit and the small-angle
limit more directly correspond to each other, in contrast to the case
of the power spectrum. In Fig.~\ref{fig:3}, the full-sky results (blue
lines) and the distant-observer limits (orange lines) are
indistinguishable by eyes in every case when the opening angle
$\theta$ is small. It is only when the opening angle $\theta$ is
larger than about $35^\circ$--$45^\circ$ that the distant-observer
approximation deviates to some extent from the full-sky results. In
the exceptional case of $\xi^{\times\times(11)}_{X_1X_2}$, the
distant-observer and full-sky results are the same even when the
opening angle is large.

The results with simultaneous approximations of distant-observer and
of large-scale limits are also plotted in Figs.~\ref{fig:3} and
\ref{fig:5} by green lines, analytic expressions of which are derived
from equations in the Appendix, taking distant-observer approximation,
$\theta \rightarrow 0$ fixing the separation $x$. The last
approximation is applied in most literature, where the effect of the
smoothing window function is neglected, and the last effects more or
less should be there on scales of galaxy or halo formation. The
perturbation theory is usually applied to sufficiently large scales,
and the last effects are intentionally neglected. However, the
clustering of halos and galaxies on smaller scales should have been
affected by the effect. While the simple application of the smoothing
window is just an empirical representation of the effect, one can
estimate possible uncertainty regarding such effects of object
formation. In our figures, the lines with the large-scale limit
deviate from those without the same limit on small angular scales. On
large scales, the effects of window function are not seen at all, and
these two cases are indistinguishable in Figs.~\ref{fig:3} and
\ref{fig:5}, while they are different from the full-sky results
because of the wide-angle effect. For scalar correlation functions,
the small-scale effects of the window function are minimal (the
left-top panel of Fig.~\ref{fig:3}). The effects on small scales are
larger in correlation functions of higher spin fields.

\section{\label{sec:Conclusions}%
  Summary and Conclusions
}

\begin{figure*}
\centering
\includegraphics[width=34pc]{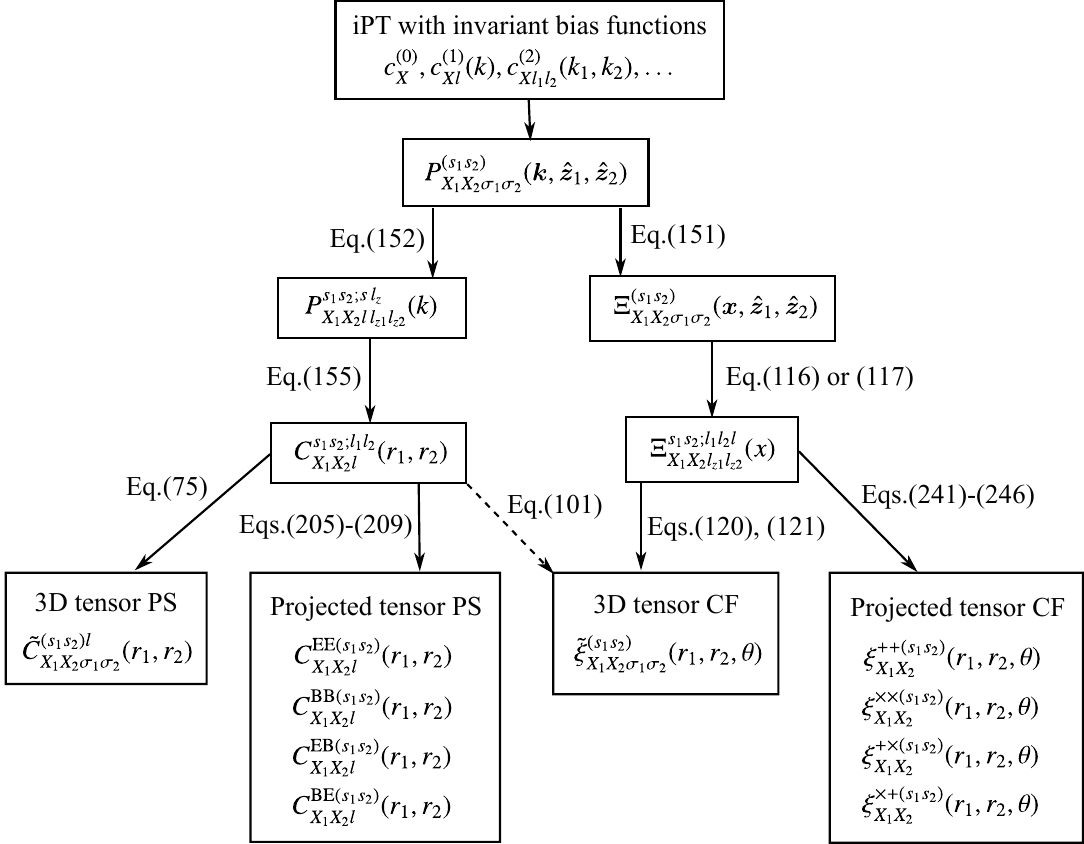}
\caption{\label{fig:7} A flowchart of the procedure for deriving the
  power spectra (PS) and correlation functions (CF) from the iPT.
  Solid arrows correspond to practical routes of derivation, while
  the dashed arrow represents a formal relation. This procedure applies to nonlinear perturbation theory of any order and is
  not necessarily limited to linear theory.}
\end{figure*}

In this paper, we construct a full-sky formulation of the iPT for
cosmological tensor fields. In previous papers of the series,
Papers~I--III, the same theory is formulated in the Cartesian
coordinates system, in which the distant-observer and flat-sky
approximations are assumed. We generalize the previous results to
those without assuming the last approximations, so that one can apply
the iPT in more general situations in which full-sky and wide-angle
effects in the power spectra and correlation functions, and so forth,
are not negligible. Taking the distant-observer limit, it is
explicitly shown that the generalized formalism correctly reproduces
the previously derived results in Papers~I--III. In the course of
showing the last property, the correspondence between the
three-dimensional power spectra introduced in Paper~I with the
distant-observer approximation and the angular power spectra with
fixed radial distances are explicitly derived, i.e.,
Eqs.~(\ref{eq:137}) and (\ref{eq:138}) for three-dimensional tensor
fields, and Eqs.~(\ref{eq:210}) and (\ref{eq:211}) for projected
tensor fields.

With the iPT, nonlinear effects in power spectra and correlation
functions can be systematically evaluated in the present formalism
with full-sky and wide-angle effects included, without assuming
flat-sky and distant-observer approximations. The flowchart of
deriving various power spectra and correlation functions of tensor
fields is shown in Fig.~\ref{fig:7}. For a given set of invariant bias
functions,
$c^{(0)}_X, c^{(1)}_{Xl}(k), c^{(2)}_{Xl_1l_2}(k_1,k_2), \ldots$, the
generalized power spectra
$P^{(s_1s_2)}_{X_1X_2\sigma_1\sigma_2}(\bm{k};\hat{\bm{z}}_1,\hat{\bm{z}}_2)$
are directly calculated to arbitrarily nonlinear orders of
perturbations from the iPT, using the methods presented in Papers~I and
II. From the generalized power spectra, the invariant spectra
$C^{s_1s_2;l_1l_2}_{X_1X_2l}(r_1,r_2)$ are derived through
Eqs.~(\ref{eq:152}) and (\ref{eq:155}), and from which, the power
spectra of three-dimensional tensor fields
$\tilde{C}^{(s_1s_2)l}_{X_1X_2\sigma_1\sigma_2}(r_1,r_2)$ are derived
through Eq.~(\ref{eq:75}), and those of projected tensor fields
$C^{\mathrm{EE}(s_1s_2)}_{X_1X_2l}(r_1,r_2)$,
$C^{\mathrm{BB}(s_1s_2)}_{X_1X_2l}(r_1,r_2)$,
$C^{\mathrm{EB}(s_1s_2)}_{X_1X_2l}(r_1,r_2)$, and
$C^{\mathrm{BE}(s_1s_2)}_{X_1X_2l}(r_1,r_2)$ are derived through
Eqs.~(\ref{eq:205})--(\ref{eq:209}). For correlation functions, the
generalized correlation functions
$\Xi^{(s_1s_2)}_{X_1X_2\sigma_1\sigma_2}(\bm{x};\hat{\bm{z}}_1,\hat{\bm{z}}_2)$
are derived from the generalized power spectra through
Eq.~(\ref{eq:151}), and invariant functions
$\Xi^{s_1s_2;l_1l_2l}_{X_1X_2l_{z1}l_{z2}}(x)$ are derived through
Eqs.~(\ref{eq:116}) or (\ref{eq:117}), and from which the correlation
functions of three-dimensional tensor fields
$\tilde{\xi}^{(s_1s_2)}_{X_1X_2\sigma_1\sigma_2}(r_1,r_2,\theta)$ are
derived through Eqs.~(\ref{eq:120}) and (\ref{eq:121}), and those of
projected tensor fields $\xi^{++(s_1s_2)}_{X_1X_2}(r_1,r_2,\theta)$,
$\xi^{\times\times(s_1s_2)}_{X_1X_2}(r_1,r_2,\theta)$,
$\xi^{+\times(s_1s_2)}_{X_1X_2}(r_1,r_2,\theta)$, and
$\xi^{\times+(s_1s_2)}_{X_1X_2}(r_1,r_2,\theta)$, are derived through
Eqs.~(\ref{eq:241})--(\ref{eq:246}).

With the above procedure, the statistics of tensor fields are
calculated to arbitrarily nonlinear orders of perturbations. In this
paper, however, we mostly focus on deriving explicit results of
applying the lowest-order approximation, or the linear perturbation
theory of gravitational evolution. In this lowest-order approximation,
the analytic expressions of the power spectra and correlation
functions for general spin fields are simple enough and we present
explicit closed-form expressions for statistics of three-dimensional
tensor fields and also of projected tensor fields. The results are
numerically evaluated in simple models of tensor bias just to
illustrate the behaviors of power spectra and correlation functions of
projected fields. The general formulas in this paper with full-sky and
wide-angle effects are compared with those with flat-sky and
distant-observer limits, which are previously derived in literature,
including our Papers~I--III.

The full-sky power spectra of projected fields
$C^{\mathrm{EE}(s_1s_2)}_{X_1X_2l}(r_1,r_2)$ in linear theory are
compared with the corresponding predictions with flat-sky and
distant-observer approximations, as shown in Fig.~\ref{fig:2}. The
power spectra of BB/EB/BE modes in linear theory vanish. Depending on
the values of spins, we see differences to some extent between
full-sky and flat-sky results. The differences are not confined only
to low $l$, but sometimes also distributed for larger $l$. This is
partly because the radial coordinates, $r_1$ and $r_2$, are not
Fourier transformed in the full-sky spectra, while they are in spectra
with flat-sky and distant-observer approximations.

The full-sky correlation functions with wide-angle effects of
projected fields
$\tilde{\xi}^{\mathrm{++}(s_1s_2)}_{X_1X_2}(r_1,r_2,\theta)$ and
$\tilde{\xi}^{\mathrm{\times\times}(s_1s_2)}_{X_1X_2}(r_1,r_2,\theta)$
in linear theory are compared with the corresponding predictions with
flat-sky and distant-observer approximations, as shown in
Figs.~\ref{fig:3} and \ref{fig:4}. The correlation functions of
$+\times$/$\times +$ modes in linear theory vanish. In correlation
functions, full-sky and distant-observer predictions agree with each
other when the opening angle $\theta$ is less than a few tens of
degrees in these examples. The angles of a few tens of degrees are
quite large in practical galaxy surveys to detect correlation signals
as the absolute values of the correlation functions are quite small
compared to those of small angles. Therefore, the distant-observer
approximation is valid in most practical applications of the
perturbation theory.

However, it is plausible to have full-sky formulas of correlation
statistics, as one can estimate to what extent one can rely on the
approximation with flat-sky and distant-observer limits. We found that
the inclusion of full-sky and wide-angle effects does not complicate
the analytic expressions by the perturbation theory. As for the linear
correlation functions of tensor fields, it is enough to add some extra
terms in the expressions for the wide-angle effect. Regardless of
whether the wide-angle effects are small or large, there is no reason
to exclude them in the analysis. The wide-angle effects can sneak
into a given analysis of galaxy survey when the measurements of the
galaxies are precise enough and the survey regions are wide enough or
shallow enough.

Apart from its usefulness in analyzing galaxy surveys, the full-sky
formalism in this paper is mathematically interesting on its own
right. The basic formalism of characterizing the two-point statistics
of irreducible tensor fields, described in the first half of this
paper, Sec.~\ref{sec:TensorFullSky}, relies only on symmetric
properties of the statistics of tensor fields using irreducible
representations of SO(3) rotation group, identifying rotationally
invariant functions which represent physical degrees of freedom of the
statistics. This part does not depend on any dynamics of the structure
formation and therefore is independent of the perturbation theory. In
the second half of this paper, we apply the iPT to this general
formalism and concretely derive various formulas in the lowest-order
approximation, in particular. The iPT is a natural and general method
of perturbation theory to describe the gravitational evolution in
weakly nonlinear regime in the presence of bias, which is not
restricted to be local nor scalar, etc. One can also apply any
theories of gravitational evolutions of the full-sky formalism, and
they do not have to be a perturbation theory, but they can be a fully
nonlinear model of structure formation, etc.

In this paper, we only consider two-point statistics, i.e., power
spectra and two-point correlation functions. The method can be
straightforwardly generalized to the cases of higher-order statistics,
such as bispectra, trispectra, three- and four-point correlation
functions, and so forth. In Paper~I, we consider a generalization to
the bispectra of irreducible tensors of any ranks in the Cartesian
coordinates with flat-sky and distant-observer approximations. The
present formalism with full-sky and wide-angle effects can be
generalized to higher-order statistics as well.

In this paper and previous Papers~I--III, we do not assume any
mechanism of bias for tensor fields, as the formalism of iPT generally
does not need to assume a model of bias from the first place, and
possible degrees of freedom to characterize the bias are all included
in the series of renormalized bias functions. In addition, the
relations between biased fields and the mass density field are not
expanded in multivariate Taylor series of the underlying mass or
gravitational potential fields and their spatial derivatives, so that
the fully nonlinear structures in the relation of biasing can be taken
into account. In some models of bias for the structure formation, such
as the peak theory, halo models, and so on, the relation between the
biased field and underlying density or potential fields cannot be
expanded in multivariate Taylor series, as they contain singular
functions such as the delta function and step function, etc. This is
quite different from other methods in conventional perturbation theory
of biased cosmological fields. As a future direction of work, it would
be interesting to investigate the effects of individual bias models
with singular bias functions, including the peak theory and halo
models, in predicting the correlation statistics by the perturbation
theory.

\begin{acknowledgments}
  I thank T.~Okumura, A.~Taruya, and K.~Akitsu for useful discussions.
  This work was supported by JSPS KAKENHI Grants No.~JP19K03835 and
  No.~21H03403.
\end{acknowledgments}

\newpage

\appendix

\onecolumngrid

\section{\label{app:LargeScaleLimit}%
  Large-scale limits of full-sky formulas of power spectra and correlation functions}

On large scales where the scale dependencies of the first-order
renormalized bias functions $c^{(1)}_{Xs}(k)$ are not effective, they
are considered constants that do not depend on scales $k$. In this
case, the results of the lowest-order expressions of power spectra and
correlation functions derived in the main text are somehow simplified.
The corresponding expressions in the distant-observer approximation
are already derived in Paper~III. For the user's convenience, below we
give generalized expressions with full-sky results in power spectra of
E/B modes and correlation functions of $+/\times$ modes for projected
tensor fields in this paper.

The large-scale limits of the first-order renormalized bias functions
and of Eq.~(\ref{eq:270}) in linear theory are useful quantities for
the purpose. We denote the large-scale limits of the scale-dependent
quantities by
\begin{equation}
  \bar{c}^{(1)}_{Xs} = \lim_{k\rightarrow 0} c^{(1)}_{Xs}(k), \quad
  \bar{b}_X = \lim_{k\rightarrow 0} b_X(k), \quad
  \bar{\beta}_X = \lim_{k\rightarrow 0} \beta_X(k),
  \label{eq:277}
\end{equation}
and thus the large-scale limits of Eq.~(\ref{eq:270}) are given by
\begin{equation}
  \bar{b}_X \equiv c^{(0)}_X + \bar{c}^{(1)}_{X0}, \quad
  \bar{\beta}_X
  \equiv \frac{c^{(0)}_X f}{\bar{b}_X}.
    \label{eq:278}
\end{equation}
Ignoring the scale dependencies on small scales, the integrals of
Eqs.~(\ref{eq:167}) and (\ref{eq:168}) are simply proportional to
Eq.~(\ref{eq:166}), and also the integrals of Eqs.~(\ref{eq:249}) and
(\ref{eq:250}) are simply proportional to Eq.~(\ref{eq:248}):
\begin{equation}
  I_{Xl_1l_2}^{(s)}(r_1,r_2) =
  \bar{c}^{(1)}_{Xs} I_{l_1l_2}(r_1,r_2), \quad
  I_{X_1X_2l_1l_2}^{(s_1s_2)}(r_1,r_2) =
  \bar{c}^{(1)}_{X_1s_1} \bar{c}^{(1)}_{X_2s_2} I_{l_1l_2}(r_1,r_2), \quad
  \xi_{Xl}^{(s)}(x) =
  \bar{c}^{(1)}_{Xs} \xi_l(x), \quad
  \xi_{X_1X_2l}^{(s_1s_2)}(x) =
  \bar{c}^{(1)}_{X_1s_1} \bar{c}^{(1)}_{X_2s_2} \xi_{l}(x).
  \label{eq:279}
\end{equation}

Substituting the above limits into the lowest-order expressions of the full-sky power spectra, Eqs.~(\ref{eq:218})--(\ref{eq:220}), we have
\begin{multline}
  C^{\mathrm{EE}(00)}_{X_1X_2l}(r_1,r_2)
  = \bar{b}_{X_1} \bar{b}_{X_2} (2l+1)
  \Biggl\{
  \left(1 + \frac{\bar{\beta}_{X_1}}{3}\right)
  \left(1 + \frac{\bar{\beta}_{X_2}}{3}\right)
  I_{ll}(r_1,r_2)
  \\
  + \frac{2}{3}
  \sum_{l'} (-1)^{(l'-l)/2} (2l'+1)
  \begin{pmatrix}
    l & l' & 2 \\ 0 & 0 & 0
  \end{pmatrix}^2
  \left[
    \bar{\beta}_{X_1}\left(1 + \frac{\bar{\beta}_{X_2}}{3}\right)
    I_{l'l}(r_1,r_2)
    +
    \bar{\beta}_{X_2}\left(1 + \frac{\bar{\beta}_{X_1}}{3}\right)
    I_{ll'}(r_1,r_2)
  \right]
  \\
  + \frac{4}{9} \bar{\beta}_{X_1}\bar{\beta}_{X_2}
  \sum_{l_1,l_2} (-1)^{(l_1+l_2-2l)/2} (2l_1+1) (2l_2+1)
  \begin{pmatrix} l & l_1 & 2 \\ 0 & 0 & 0 \end{pmatrix}^2
  \begin{pmatrix} l & l_2 & 2 \\ 0 & 0 & 0 \end{pmatrix}^2
  I_{l_1l_2}(r_1,r_2)
  \Biggr\},
  \label{eq:280}
\end{multline}
\begin{multline}
  C^{\mathrm{EE}(s0)}_{X_1X_2l}(r_1,r_2)
  = \bar{c}^{(1)}_{X_1s} \bar{b}_{X_2}
  \frac{(2l+1)A_s}{\sqrt{2s+1}}
  \sum_{l_1} (-1)^{(l_1+s-l)/2} (2l_1+1)
  \begin{bmatrix}
    s & l_1 & l \\ s & 0 & -s
  \end{bmatrix}
  \\ \times
  \left[
    \left(1+\frac{\bar{\beta}_{X_2}}{3}\right)
    I_{l_1l}(r_1,r_2)
    + \frac{2}{3}\bar{\beta}_{X_2}
    \sum_{l_2} (-1)^{(l_2-l)/2} (2l_2+1)
    \begin{pmatrix}
      l & l_2 & 2 \\ 0 & 0 & 0
    \end{pmatrix}^2
    I_{l_1l_2}(r_1,r_2)
  \right],
  \label{eq:281}
\end{multline}
\begin{equation}
  C^{\mathrm{EE}(s_1s_2)}_{X_1X_2l}(r_1,r_2)
  = \bar{c}^{(1)}_{X_1s_1} \bar{c}^{(1)}_{X_2s_2}
  \frac{(2l+1)A_{s_1}A_{s_2}}{\sqrt{(2s_1+1)(2s_2+1)}}
  \sum_{l_1,l_2} (-1)^{(l_1+l_2+s_1+s_2-2l)/2} (2l_1+1)(2l_2+1)
  \begin{bmatrix}
    s_1 & l_1 & l \\ s_1 & 0 & -s_1
  \end{bmatrix}
  \begin{bmatrix}
    s_2 & l_2 & l \\ s_2 & 0 & -s_2
  \end{bmatrix}
  I_{l_1l_2}(r_1,r_2),
  \label{eq:282}
\end{equation}
where $s,s_1,s_2 \geq 1$ in the above equations. Similarly,
correlation functions of Eqs.~(\ref{eq:255})--(\ref{eq:266}) in the
large-scale limit reduce to
\begin{multline}
  \xi^{++(00)}_{X_1X_2}(r_1,r_2,\theta)
  = \bar{b}_{X_1} \bar{b}_{X_2}
  \Biggl\{
    \left[
      1 + \frac{\bar{\beta}_{X_1}}{3} + \frac{\bar{\beta}_{X_2}}{3}
      + \frac{1}{5}\bar{\beta}_{X_1}\bar{\beta}_{X_2}
      \left( 1 - \frac{2}{3} \sin^2\theta \right)
    \right]
    \xi_0(x)
    \\
    -
    \left[
      \bar{\beta}_{X_1}\left(1 + \frac{3}{7}\bar{\beta}_{X_2}\right)
      \left(\cos^2\tilde{\theta}_1 - \frac{1}{3} \right)
      + \bar{\beta}_{X_2}\left(1 + \frac{3}{7}\bar{\beta}_{X_1}\right)
      \left(\cos^2\tilde{\theta}_2 - \frac{1}{3} \right)
      - \frac{2}{21}\bar{\beta}_{X_1}\bar{\beta}_{X_2} \sin^2\theta
    \right]
    \xi_2(x)
  \\
  + \frac{1}{4} \bar{\beta}_{X_1} \bar{\beta}_{X_2}
  \left[
    \frac{12}{35}
    + 2 \cos\tilde{\theta}_1 \cos\tilde{\theta}_2
    \cos\left(\tilde{\theta}_1+\tilde{\theta}_2\right)
    - \frac{5}{7}
    \left(
      \cos^2\tilde{\theta}_1 + \cos^2\tilde{\theta}_2
    \right)
    - \frac{3}{35} \sin^2\theta
  \right]
  \xi_4(x)
  \Biggr\},
  \label{eq:283}
\end{multline}
\begin{multline}
  \xi^{++(10)}_{X_1X_2}(r_1,r_2,\theta)
  = - \frac{\bar{c}^{(1)}_{X_11}\bar{b}_{X_2}}{\sqrt{6}}
  \Biggl\{
  \left[
    \left(1 + \frac{2}{5} \bar{\beta}_{X2} \right)
    \sin\tilde{\theta}_1
    + \frac{\bar{\beta}_{X_2}}{5}
    \sin\left(\tilde{\theta}_1 - 2\tilde{\theta}_2\right)
    \right]
    \xi_1(x)
    \\
    - \frac{\bar{\beta}_{X_2}}{20}
    \left[
      2 \sin\tilde{\theta}_1
      + 5 \sin\left(\tilde{\theta}_1 + 2\tilde{\theta}_2\right)
      + \sin\left(\tilde{\theta}_1 - 2\tilde{\theta}_2\right)
    \right]
    \xi_{3}(x)
    \Biggr\},
  \label{eq:284}
\end{multline}
\begin{multline}
  \xi^{++(20)}_{X_1X_2}(r_1,r_2,\theta)
  = -\frac{\bar{c}^{(1)}_{X_12} \bar{b}_{X_2}}{2\sqrt{5}}
  \Biggl\{
  \frac{2}{15} \bar{\beta}_{X_2} \sin^2\theta\, \xi_0(x)
  -
  \left[
   \left(1+\frac{3}{7}\bar{\beta}_{X_2}\right) \sin^2\tilde{\theta}_1
    - \frac{2}{7}\bar{\beta}_{X_2} \sin^2\tilde{\theta}_2
    + \frac{2}{21}\bar{\beta}_{X_2} \sin^2\theta
  \right]
  \xi_2(x)
  \\
  + \frac{\bar{\beta}_{X_2}}{28}
  \left[
    \frac{3}{2}
    - \frac{7}{2}
    \cos\left(2\tilde{\theta}_1 + 2\tilde{\theta}_2\right)
    - \cos 2\tilde{\theta}_1
    + 3 \cos 2\tilde{\theta}_2
    + \frac{3}{5} \sin^2\theta
  \right]
  \xi_4(x)
  \Biggr\},
  \label{eq:285}
\end{multline}
\begin{align}
  \xi^{++(11)}_{X_1X_2}(r_1,r_2,\theta)  
  &= \frac{\bar{c}^{(1)}_{X_11}\bar{c}^{(1)}_{X_21}}{36}
  \left\{
    2 \cos\theta\, \xi_0(x)
    +
    \left[
    3\cos\left(\tilde{\theta}_1+\tilde{\theta}_2\right)
    - \cos\theta
    \right]
    \xi_2(x)
    \right\},
  \label{eq:286}\\
  \xi^{\times\times(11)}_{X_1X_2}(r_1,r_2,\theta)  
  &= \frac{\bar{c}^{(1)}_{X_11}\bar{c}^{(1)}_{X_21}}{18}
    \left[
    \xi_0(x) + \xi_2(x)
    \right],
  \label{eq:287}
\end{align}
\begin{align}
  \xi^{++(12)}_{X_1X_2}(r_1,r_2,\theta)
  &= \frac{\bar{c}^{(1)}_{X_11}\bar{c}^{(1)}_{X_22}}{20\sqrt{30}}
    \sin\tilde{\theta}_1
    \left\{
    4 \cos\theta\, \xi_1(x)
    +
    \left[
    5\cos\left(\tilde{\theta}_1+ \tilde{\theta}_2\right)
    - \cos\theta
    \right]
    \xi_3(x)
    \right\},
  \label{eq:288}\\
  \xi^{\times\times(12)}_{X_1X_2}(r_1,r_2,\theta)
  &= \frac{\bar{c}^{(1)}_{X_11}\bar{c}^{(1)}_{X_22}}{5\sqrt{30}}
    \sin\tilde{\theta}_1
    \left[
    \xi_1(x) + \xi_3(x)
    \right],
  \label{eq:289}
\end{align}
\begin{multline}
  \xi^{++(22)}_{X_1X_2}(r_1,r_2,\theta)
  = \frac{\bar{c}^{(1)}_{X_12}\bar{c}^{(1)}_{X_22}}{525}
  \Biggl\{
  7\left(1 - \frac{1}{2}\sin^2\theta\right)
  \xi_0(x)
  + \frac{5}{4}
  \left[
    2 + 3\cos 2\tilde{\theta}_1 + 3\cos 2\tilde{\theta}_2
    + \sin^2\theta
  \right]
  \xi_2(x)
  \\
  + \frac{3}{32}
  \left[
    57
    - 30\cos 2\tilde{\theta}_1 - 30\cos 2\tilde{\theta}_2
    + 35
    \cos\left(2\tilde{\theta}_1+2\tilde{\theta}_2\right)
    - 6 \sin^2\theta
  \right] 
  \xi_4(x)
  \Biggr\},
  \label{eq:290}
\end{multline}
and
\begin{equation}
  \xi^{\times\times(22)}_{X_1X_2}(r_1,r_2,\theta)
  = \frac{\bar{c}^{(1)}_{X_12}\bar{c}^{(1)}_{X_22}}{525}
  \Biggl\{
  7 \xi_0(x)
  + \frac{5}{2}
  \left[
    3 \cos\left(\tilde{\theta}_1+\tilde{\theta}_2\right)
    + \cos\theta
  \right]
  \xi_2(x)
  + \frac{3}{2}
  \left[
    5 \cos\left(\tilde{\theta}_1+\tilde{\theta}_2\right)
    - 3 \cos\theta
  \right]
  \xi_4(x)
  \Biggr\},
  \label{eq:291}
\end{equation}
where $\tilde{\theta}_1$, $\tilde{\theta}_2$ and $x$ are functions of
$(r_1,r_2,\theta)$ as given by Eqs.~(\ref{eq:256})--(\ref{eq:258}). It
is a straightforward exercise to show that the above expressions reduce
to the results of Paper~III in the distant-observer approximation,
taking the limits of $\theta \rightarrow 0$ and
$\tilde{\theta}_1 = \tilde{\theta}_2 = \theta_x$.


\twocolumngrid
\renewcommand{\apj}{Astrophys.~J. }
\newcommand{\aap}{Astron.~Astrophys. }
\newcommand{\aj}{Astron.~J. }
\newcommand{\apjl}{Astrophys.~J.~Lett. }
\newcommand{\apjs}{Astrophys.~J.~Suppl.~Ser. }
\newcommand{\apss}{Astrophys.~Space Sci. }
\newcommand{\cqg}{Class.~Quant.~Grav. }
\newcommand{\jcap}{J.~Cosmol.~Astropart.~Phys. }
\newcommand{\mnras}{Mon.~Not.~R.~Astron.~Soc. }
\newcommand{\mpla}{Mod.~Phys.~Lett.~A }
\newcommand{\pasj}{Publ.~Astron.~Soc.~Japan }
\newcommand{\physrep}{Phys.~Rep. }
\newcommand{\ptp}{Progr.~Theor.~Phys. }
\newcommand{\ptep}{Prog.~Theor.~Exp.~Phys. }
\newcommand{\jetp}{JETP }
\newcommand{\jhep}{Journal of High Energy Physics}


\end{document}